\documentclass[a4paper,11pt]{article}
\pdfoutput=1 

\usepackage{jcappub} 

\usepackage[T1]{fontenc} 

\title{\boldmath Probing non-Gaussian Stochastic Gravitational Wave Backgrounds with
 LISA}
 
\def\lsim{\mathrel{\rlap{\lower3pt\hbox{\hskip0pt$\sim$}}
   \raise1pt\hbox{$<$}}}         
\def\gsim{\mathrel{\rlap{\lower4pt\hbox{\hskip1pt$\sim$}}
   \raise1pt\hbox{$>$}}}         

\def\beq{\begin{equation}}
\def\eeq{\end{equation}}
\def\be{\begin{equation}} 
\def\ee{\end{equation}}
\def\bea{\begin{eqnarray}}
\def\eea{\end{eqnarray}}

\usepackage{color}
\definecolor{darkgreen}{RGB}{0, 100, 0}

\author[a,b,c]{Nicola Bartolo\,}
\author[d]{, Valerie Domcke\,}
\author[e,1]{, Daniel G. Figueroa\,\note{Group coordinators: daniel.figueroa@cern.ch, angelo.ricciardone@uis.no}}
\author[f,g]{, Juan Garcia-Bellido\,}
\author[a,h]{, Marco Peloso\,}
\author[g]{, Mauro Pieroni\,}
\author[i,1]{, Angelo Ricciardone\,}
\author[j]{, Mairi Sakellariadou\,}
\author[k]{, Lorenzo Sorbo\,}
\author[l]{, Gianmassimo Tasinato\,}

\affiliation[a]{Dipartimento di Fisica e Astronomia ``G. Galilei'', Universit\`a degli Studi di Padova, via Marzolo 8, 
I-35131, Padova, Italy}
\affiliation[b]{INFN, Sezione di Padova, via Marzolo 8, I-35131, Padova, Italy}
\affiliation[c]{INAF-Osservatorio Astronomico di Padova, Vicolo dell'Osservatorio 5, I-35122 Padova, Italy}
\affiliation[d]{Deutsches Elektronen Synchrotron (DESY), 22607 Hamburg, Germany.}
\affiliation[e]{ Laboratory of Particle Physics and Cosmology Institute of Physics (LPPC), \'Ecole Polytechnique F\'ed\'erale de Lausanne (EPFL), CH-1015 Lausanne, Switzerland.}
\affiliation[f]{Department of Theoretical Physics, CERN, CH-1211 Gen\`eve, Switzerland}
\affiliation[g]{ Instituto de F\'{\i}sica Te\'orica UAM/CSIC, Nicol\'as Cabrera 13, Universidad
Aut\'onoma de Madrid, Cantoblanco 28049 Madrid, Spain}
\affiliation[h]{School of Physics and Astronomy, and Minnesota Institute for Astrophysics,
University of Minnesota, Minneapolis, 55455, USA}
\affiliation[i]{Faculty of Science and Technology, University of Stavanger, 4036, Stavanger, Norway}
\affiliation[j]{Theoretical Particle Physics and Cosmology Group, Department of Physics, King's College London, University of London, Strand, London WC2R 2LS, United Kingdom}
\affiliation[k]{Amherst Center for Fundamental Interactions, Department of Physics, University of Massachusetts,
Amherst, MA 01003 (USA)}
\affiliation[l]{Department of Physics, Swansea University, Swansea, SA2 8PP, UK}

\abstract{The stochastic gravitational wave background (SGWB) contains a wealth of information on astrophysical and cosmological processes. A major challenge of upcoming years will be to extract the information contained in this background and to disentangle the contributions of different sources. In this paper we provide the formalism to extract, from the correlation of three signals in the Laser Interferometer Space Antenna (LISA), information about the tensor three-point function, which characterizes the non-Gaussian properties of the SGWB.   This observable can be crucial to discriminate whether a SGWB has  a primordial or astrophysical origin. Compared to the two-point function, the SGWB three-point function has a richer dependence on the gravitational wave momenta and chiralities. It can be used therefore as a powerful discriminator between different models.  For the first time we provide the response functions of LISA to a general SGWB three-point function. As examples, we study in full detail the cases  of  an equilateral and squeezed SGWB bispectra, and provide the explicit form of the response functions, ready to be convoluted with any theoretical prediction of the bispectrum to obtain the observable signal. We further derive the optimal estimator to compute the signal-to-noise ratio. Our formalism covers general shapes of non-Gaussianity, and can be extended straightaway to other detector geometries. Finally, we provide a short overview of models of the early universe that can give rise to a non-Gaussian SGWB. }

\begin{document}

\begin{flushright}
UMN-TH/3720-18; 
CERN-TH-2018-130; \\
IFT-UAM/CSIC-18-58;
DESY 18-086; \\
KCL-PH-TH/2018-22;
ACFI-T18-08
\end{flushright}
\maketitle
\flushbottom

\section{Introduction}

Many cosmological and astrophysical scenarios predict the existence
of stochastic gravitational wave backgrounds (SGWBs) with a sufficiently
large amplitude to be detectable at interferometer scales, see e.g.~\cite{Maggiore:1900zz,Maggiore:2018sht,Caprini:2018mtu} for reviews. A detection 
of a SGWB signal would provide important information
about its origin, hence it  is essential to characterize  its properties in detail,  and 
to devise observables that will enable us to distinguish among different possible sources
of gravitational waves~(GWs).

The statistical features of a SGWB can offer various  observables that can be measured
with interferometers. While so far the attention has been mainly focussed on the two-point function (power spectrum) of the GW signal (see e.g.~the recent review \cite{Romano:2016dpx}), in this work we study for the first time how LISA~\cite{Audley:2017drz} (and for this matter any other interferometer) can probe the three-point function (bispectrum) of the SGWB. A non-vanishing GW three-point function  is associated with the {\it non-Gaussian} features of the SGWB. Some of the key ingredients of early universe models (in particular inflation and topological defects) predict that the primordial SGWB  can have large non-Gaussianity, whose features can  be calculated from first principles, and can be used to discriminate among different models. On the other hand, a SGWB due to a combination of a large number of uncorrelated astrophysical sources, or due to sub-horizon cosmological processes such as phase transitions or preheating, is Gaussian to a high degree, due to the central limit theorem (it remains to be studied to which degree the central limit theorem applies to the distribution of astrophysical sources that contribute to the stochastic GW background measured by LISA). Thus a detection of a non-Gaussian SGWB would be a distinctive indication of a signal of cosmological origin, and the study of its properties would provide crucial information on the physics of the very early universe.

The Fourier transform of a  GW three-point function defines the  non-Gaussian tensor bispectrum, which depends on the frequencies of the three GWs involved, as well as on their polarization. Its scale and chirality dependence can be very rich, in contrast to the GW two-point function, whose features are very constrained by the symmetries of the underlying background geometry. This  implies that a study of the GW  bispectrum can lead to a large number of new observables that can be used to differentiate among different models of cosmology. So far, the physics of primordial tensor fluctuations has been mainly investigated at  Cosmic Microwave Background (CMB) scales, and indeed the non-Gaussianity of primordial GWs produced in certain models of inflation  have been observationally constrained by the {\it Planck} satellite~\cite{Ade:2015ava}. 
 
In this work, we investigate for the first time the capabilities of interferometers, in particular LISA, to test the non-Gaussian features of a SGWB  via its three-point functions \footnote{Here we are interested in probing the intrinsic non-Gaussian statistics of the SGWB, and not in describing `pop-corn-like' non-Gaussian bursts sometimes present in the unresolved stochastic background due to combining many individual sources. See \cite{Romano:2016dpx} for a detailed review on searches for this kind of non-Gaussian signatures using the SGWB two-point function.}. We do so by developing a formalism to compute the interferometer cubic response function to a non-Gaussian SGWB, connecting the  GW
 bispectrum to the statistics of the actual signal outputs obtained from LISA (see Sections
 \ref{sec:formalism} and \ref{sec-signal}).  We show that cubic correlators of the interferometer signal are sensitive  to the properties of the GW bispectrum -- its 
 dependence on the GW wavenumber and chirality indexes
  -- and also depend on specific features of the interferometer, like its arm configurations. Measurements
 of the cubic interferometer signal can distinguish among different cosmological models for a SGWB, and we show how different 
  examples of primordial bispectra, with distinctive dependence on the momenta of the incoming GWs, lead to qualitatively
 distinguishable  features in the interferometer cubic correlators.  We also demonstrate that any measurement of the GW bispectrum at LISA is invariant under parity.  Nonetheless, we show that in the presence of non-Gaussianity the analysis of LISA data in different frequency intervals can allow to extract nontrivial dependence on the correlators of the left- and right-handed gravitational waves, such as the relative amplitudes of the $(\langle RRR\rangle+\langle LLL\rangle)$ and $(\langle RRL\rangle+\langle LLR\rangle)$ correlation functions.
   
 These results allow us to build the optimal estimator for the GW bispectrum observed at LISA, generalizing results previously developed for studying the GW two-point function \cite{Smith:2016jqs} to the three-point function,
 see Section \ref{sec-estimator}.
 Our final 
 expression for  the  Signal-to-Noise-Ratio  (SNR) is physically transparent, and we apply it to a specific scenario of a primordial non-Gaussian GW signal whose bispectrum is amplified at a specific scale
 to quantitatively 
   demonstrate  the ability of LISA to measure tensor non-Gaussianity, depending on the amplitude of the GW  bispectrum.
   
   Our  results demonstrate that the statistics of the GW signal measured at interferometer is sensitive to various
   distinctive properties of the bispectrum of primordial GWs from the early universe: Section \ref{sec-mod} surveys 
   existing cosmological models   capable of producing  large non-Gaussianity of the SGWB, analyzing the
   features of the corresponding bispectra,  and briefly  discussing  prospects  of detection with the LISA
   interferometer and with CMB experiments, in light of our findings.
   
   Section    \ref{sec-conclusions} contains our conclusions, with a summary of our results and suggestions for
   future studies, while six appendices contain technical results used in the main text.

\section{A formalism for tensor non-Gaussianity} \label{sec:formalism}

In this section we build a  formalism for describing tensor non-Gaussianity, which can be used to conveniently analyze how to probe this observable with interferometers. We consider a stochastic background of gravitational waves  associated with the transverse-traceless metric perturbation $h_{ab} \left(t,\vec{x} \right)$ of the background metric (where   $a,b$ correspond to  spatial indexes, 
while  $\lambda$  denotes tensor polarization). We decompose the tensor modes as 
\begin{eqnarray}
h_{ab} \left( t ,\, \vec{x} \right) &=& \int d^3 k \, {\rm e}^{-2 \pi i \, \vec{k} \cdot \vec{x}} \, \sum_\lambda e_{ab,\lambda} ({\hat k}) 
\left[ {\rm e}^{2 \pi i \, k t} \, h_\lambda ( \vec{k}) +  {\rm e}^{-2 \pi i \, k t} \, h_\lambda^* ( - \vec{k}) \right] \nonumber\\ 
&\equiv&  \int d^3 k \, {\rm e}^{-2 \pi i \, \vec{k} \cdot \vec{x}} \, \sum_\lambda e_{ab,\lambda} ( {\hat k}) \, h_\lambda ( t ,\, \vec{k}) \;, 
\label{GW-classical}
\end{eqnarray}
where $k \equiv \vert \vec{k} \vert$,  $\hat k$ denotes a unit vector in the direction of the vector $\vec{k}$.
In this expression, we sum over the two transverse polarizations $\lambda$ of a GW, with $e_{ab,\lambda} ({\hat k})$ being the polarization operator for the polarization $\lambda$. For a more detailed discussion on the definition and construction of different polarization basis, and the properties they satisfy, we refer the reader to Appendices \ref{app:polarization}, \ref{app:polarizationII}.

As we require that $h_{ab} (t ,\vec{x})$ is real, it must follow that 
\be
\sum_\lambda e_{ab,\lambda}^* ( {\hat k}) \, h_\lambda^* ( t ,\, \vec{k}) = \sum_\lambda e_{ab,\lambda} (-{\hat k}) \, h_\lambda (t,-\vec{k})
\ee
holds. This condition, however, does not fix completely the choice of the polarization tensors $e_{ab,\lambda} ({\hat k})$. Two basis for the GW polarizations are commonly employed in the literature, the $\left\{ + ,\, \times \right\}$ basis, and the $\left\{ R ,\, L \right\}$ chiral basis. In the explicit computations performed in this paper, the chiral basis is used. Furthermore, we choose chiral polarization operators that specifically satisfy $e_{ab,\lambda}^* ( {\hat k}) = e_{ab,\lambda} (-{\hat k})$ [we   discuss the freedom of choosing different polarization operators in Appendices~\ref{app:polarization}, \ref{app:polarizationII}]. This implies that the same property is also satisfied by the momentum space variable $h_\lambda ( t ,\vec{k})$, so that the reality of $h_{ab} ( t ,\vec{x} )$ is ensured. 
Given our choice, we have 
\begin{eqnarray}
e_{ab,\lambda}^* \left(  {\hat k} \right) = e_{ab,\lambda} \left( - {\hat k} \right) = e_{ab,-\lambda} \left(  {\hat k} \right)  
\;\; \;\;  ,\;\; \;\;  e_{ab,\lambda}^* \left( {\hat k} \right) e_{ab,\lambda'} \left( {\hat k} \right) = \delta_{\lambda \lambda'} \;. 
\label{e-properties}
\end{eqnarray} 
In particular, the second equality in the first expression states that $e_{ab,R} \, \leftrightarrow e_{ab,L}$ under a parity transformation. 

The statistical properties of the GW background are controlled by its correlation functions in
Fourier space. In this work we consider the $2-$point correlator $\left\langle h^2 \right\rangle$, and, more in detail, the  $3-$point correlator $\left\langle h^3 \right\rangle$, which is non-vanishing for  a non-Gaussian SGWB. Assuming statistical isotropy, the equal-time momentum-space correlator is given by 
\begin{equation}
\left\langle h_\lambda \left( t ,\, \vec{k} \right) \,  h_{\lambda'} \left( t' ,\, \vec{k}' \right) \right\rangle  = \frac{P_\lambda \left( k \right)}{4 \pi k^3} \, 
\delta_{\lambda \lambda'} \, \delta^{(3)} \left( \vec{k} + \vec{k}' \right) 
\cos \left[ 2 \pi k_1 \left( t - t' \right) \right]  \;, 
\label{PS} 
\end{equation}
where $P_\lambda \left( k \right)$ is the power spectrum of the helicity $\lambda$, and  the numerical factor at the right-hand side has been fixed imposing   that the combination of  eqs.~(\ref{GW-classical}) and (\ref{PS}) leads to the real space correlator 
\begin{equation}
\left\langle h_{ab} \left( t ,\, \vec{x} \right) \,  h_{ab} \left( t ,\, \vec{y} \right) \right\rangle = \int_0^{\infty} \frac{d k}{k} \, \sum_\lambda P_\lambda \left( k \right) \, \frac{\sin \left( 2 \pi k r \right)}{2 \pi k r} \;\;\;,\;\;\; r \equiv \left\vert \vec{x} - \vec{y} \right\vert  \,. 
\label{hx-hy} 
\end{equation} 

\bigskip

For studying the $3-$point function, we use an ansatz
analogous to the one  used for describing the statistics of  primordial scalar perturbations~\cite{Salopek:1990jq,Gangui:1993tt,Komatsu:2001rj,Babich:2004gb,Maldacena:2002vr,Acquaviva:2002ud}, see for example \cite{Scoccimarro:2011pz,Wagner:2010me,Liguori:2003mb}.
 Specifically, we assume a small departure from Gaussianity, so that a tensor mode
   is the sum of a dominant Gaussian component, and its quadratic convolution 
\begin{align}
{h}_\lambda \left( t ,\, \vec{k} \right) = {h}_{\lambda,g} \left( t ,\, \vec{k} \right) + &    \sum_{\lambda',\lambda''} 
 f_{\rm NL}^{\lambda,\lambda',\lambda''} \, 
\int d^3 p \, d^3 q {h}_{\lambda',g} \left( t ,\, \vec{p} \right) \, {h}_{\lambda'',g} \left( t ,\, \vec{q} \right)  \nonumber \\
& \times 
\delta^{(3)} \left( \vec{k} - \vec{p} - \vec{q}  \right) \, K_{\lambda\lambda'\lambda''} \left( \vec{k} ,\, - \vec{p} ,\,  - \vec{q} \right) \,. 
\label{f-NL}
\end{align}
This ansatz is characterized by a kernel  $K_{\lambda\lambda'\lambda''}$, which depends
on the GW momenta and polarizations. We shall see that the kernel defines the properties
of the non-Gaussian tensor bispectrum. 
We assume that, in general, the two different polarizations can be coupled in the convolution. 
The kernel is symmetric in the last two arguments ($\lambda' \leftrightarrow \lambda''$ together with $\vec{p} \leftrightarrow \vec{q}$)
\begin{equation}
f_{\rm NL}^{\lambda,\lambda',\lambda''}  \, K_{\lambda\lambda'\lambda''} \left( \vec{k} ,\, - \vec{p} ,\, - \vec{q} \right) =  
f_{\rm NL}^{\lambda,\lambda'',\lambda'}  \, K_{\lambda\lambda''\lambda'} \left( \vec{k} ,\, - \vec{q} ,\, - \vec{p} \right)  \;.
\label{eq:shape}
\end{equation} 

\smallskip

\noindent
 The transformation of the kernel under a rotation is discussed in Appendix~\ref{app:polarization}. The dependence of the kernel on the magnitude of the three momenta controls the so-called {\it shape} of the non-Gaussianity~\cite{Babich:2004gb},
 namely how the bispectrum changes 
according to different triangular configurations in Fourier space. 
 The simplest form of non-Gaussianity is the so called local shape, enhanced in the squeezed limit of the bispectrum, for which (assuming also that the different helicities are not mixed in the convolution) $K_{\lambda\lambda'\lambda''}  \left( \vec{k} , \, \vec{p} ,\, \vec{q}  \right) \propto \delta_{\lambda \lambda'} \delta_{\lambda \lambda''}$. More in general, we normalize
 $K_{\lambda\lambda'\lambda''}  \left( k_p \hat v_1, \, k_p \hat v_2 ,\, k_p \hat v_3  \right)=1$ for a reference unit triangle formed by $\hat v_1$, $\hat v_2$ and $\hat v_3 = - \hat v_1 - \hat v_2$ (see Eq.~\eqref{v1hat-v2hat}) and some given pivot scale $k_p$.  
 In this way,  the size of non-Gaussianity is controlled by the nonlinear parameter $f_{\rm NL}^{\lambda,\lambda',\lambda''}$. In the concrete example of non scale-invariant non-Gaussianity that we study {in Section~\ref{section-bumpy},
 the pivot scale is chosen to be the scale at which the bispectrum is 
maximal, see eq.~(\ref{bump-ansatz}).  

It is important to note that the mode functions appearing in the relation (\ref{f-NL}) are evaluated today. If the GW has a cosmological origin, we need to account for its evolution. It is conventional to encode this in a cosmological transfer function
\begin{equation}
h_\lambda \left( t_0 ,\, \vec{k} \right) = {\bf T} \left( t_0 ,\, k \right) \, h_\lambda^{\rm pr} ( \vec{k} ) \;, 
\label{eq:T0}
\end{equation}
where $t_0$ indicates the present time, and $h^{\rm pr}$ is the primordial value of the GW mode. For an
adiabatic tensor mode  produced during inflation, $h_\lambda^{\rm pr}$ is constant (time independent) at super-horizon scales. For GW produced inside the horizon after inflation, we take the value of the mode at the end of GW production. Correspondingly, eq.~(\ref{f-NL}) is changed into 
\begin{align}
 {\bf T} \left( t_0 ,\, k \right) \, {h}_\lambda^{\rm pr} ( \vec{k}) = & {\bf T} \left( t_0 ,\, k \right) \, {h}_{\lambda,g}^{\rm pr} (  \vec{k} )  \nonumber \\
 & +   \sum_{\lambda',\lambda''} 
 f_{\rm NL}^{\lambda,\lambda',\lambda''} \, 
\int d^3 p d^3 q
 {\bf T} \left( t_0 ,\, p \right) \, {h}_{\lambda',g}^{\rm pr} \left(  \vec{p} \right) \, {\bf T} \left( t_0 ,\, q \right) \, {h}_{\lambda'',g}^{\rm pr} \left(  \vec{q} \right)   \nonumber \\
 & \quad \times \delta^{(3)} \left( \vec{k} - \vec{p} - \vec{q}  \right) \, K_{\lambda\lambda'\lambda''} \left( \vec{k} ,\, -\vec{p} ,\,  -\vec{q} \right) \,. 
\end{align}
%
%
This leads to the relation 
\begin{equation} 
f_{\rm NL}^{\lambda,\lambda',\lambda'',{\rm pr}} \, K_{\lambda\lambda'\lambda'',{\rm pr}} \left( \vec{k} ,\, \vec{p},\, \vec{q} \right)  \equiv  \frac{{\bf T} \left( t_0 ,\, p \right) \, {\bf T} \left( t_0 ,\, q \right) }{{\bf T} \left( t_0 ,\, k \right) } \,  f_{\rm NL}^{\lambda,\lambda',\lambda''} \, K_{\lambda\lambda'\lambda''} \left( \vec{k} ,\, \vec{p} ,\, \vec{q} \right) \,, 
\label{f-NL-primordial}
\end{equation} 
between the parametrization of non-Gaussianity in terms of the primordial vs.\ the present day GW mode functions.~\footnote{To be precise, both sides in this relation are multiplied by $\delta^{(3)} ( \vec{k} + \vec{p} + \vec{q} )$. Eq. (\ref{f-NL-primordial}) indicates in an unambiguous way how the nonlinear parameter $f_{\rm NL}$ and the kernel function $K$ should be independently rescaled, once we demand that both the present day and the primordial kernel function are normalized to $K_{\lambda\lambda'\lambda''}  \left( k_p \hat v_1,  \, k_p \hat v_2,\, k_p \hat v_3  \right) = K_{\lambda\lambda'\lambda'',{\rm pr}}  \left( k_p \hat v_1,  \, k_p \hat v_2,\, k_p  \hat v_3 \right) = 1$ (see text below eq.~\eqref{eq:shape}).} While the product of the last two factors on the $rhs$ is more directly related to what we measure, the $lhs$ is more immediately connected to the theory that provides the origin of the 
non-Gaussianity. 

From the ansatz (\ref{f-NL}) we obtain the equal-time three-point function, to linear order in the nonlinear parameters, as 
\begin{eqnarray}
&& \!\!\!\!\!\!\!\!  \!\!\!\!\!\!\!\!  \!\!\!\!\!\!\!\!  \!\!\!\!\!\!\!\! 
\left\langle {h}_{\lambda_1} \left( t  ,\, \vec{k}_1 \right)  {h}_{\lambda_2} \left( t  ,\, \vec{k}_2 \right) 
{h}_{\lambda_3} \left( t  ,\, \vec{k}_3 \right) \right\rangle  =   \delta^{(3)} \left( \vec{k}_1+\vec{k}_2+\vec{k}_3 \right)  {\cal B}_{\lambda_1,\lambda_2,\lambda_3} \left( \vec{k}_1 , \vec{k}_2 , \vec{k}_3 \right) \,, 
\label{eq:BS0}
\end{eqnarray} 
with 
\begin{eqnarray} 
{\cal B}_{\lambda_1,\lambda_2,\lambda_3} \left( \vec{k}_1 , \vec{k}_2 , \vec{k}_3 \right) &=& 
 \frac{f_{\rm NL}^{\lambda_1,\lambda_2,\lambda_3}}{8 \pi^2} \, 
 K_{\lambda_1\lambda_2\lambda_3} \left( \vec{k}_1 ,\, \vec{k}_2 ,\, \vec{k}_3 \right) 
\frac{P_{\lambda_2} \left( k_2 \right)}{k_2^3} \, \frac{P_{\lambda_3} \left( k_3 \right)}{k_3^3} \nonumber\\ 
&&  +   \frac{f_{\rm NL}^{\lambda_2,\lambda_1,\lambda_3}}{8 \pi^2} \, 
K_{\lambda_1\lambda_2\lambda_3} \left( \vec{k}_1 ,\, \vec{k}_2 ,\, \vec{k}_3 \right) 
\frac{P_{\lambda_1} \left( k_1 \right)}{k_1^3} \, \frac{P_{\lambda_3} \left( k_3 \right)}{k_3^3} \nonumber\\ 
&&  +   \frac{f_{\rm NL}^{\lambda_3,\lambda_1,\lambda_2}}{8 \pi^2} \, 
K_{\lambda_1\lambda_2\lambda_3} \left( \vec{k}_1 ,\, \vec{k}_2 ,\, \vec{k}_3 \right) 
\frac{P_{\lambda_1} \left( k_1 \right)}{k_1^3} \, \frac{P_{\lambda_2} \left( k_2 \right)}{k_2^3}  \,.
\label{BS} 
\end{eqnarray} 
This expression for the bispectrum can describe the different shapes of non-Gaussianity, and the dependence
on chirality.  The expression for the tensor three-point function \eqref{eq:BS0} evaluated at non-equal times, which will be used in Section \ref{sec-estimator}, is discussed in Appendix \ref{app:non-equal-time}.

\section{LISA signal and response functions } \label{sec-signal}

Our aim 
in this Section is to build the tools to study the primordial tensor three-point functions with LISA: we
 connect 
the theoretical results of the previous section with the actual LISA design. We analyze how 
the signal cubic correlator can be used to probe the  tensor bispectrum
described in Section \ref{sec:formalism},
 deriving for the first
time  the interferometer cubic response function to tensor non-Gaussianity. As we shall see, the interferometer 
response
function  is sensitive to the shape and polarization dependence of the primordial tensor bispectrum.

\smallskip

We first analyze how a SGWB influences the 
time  a photon takes for  traveling along a single arm of an interferometer.
 We then study the signal measured at LISA. We follow the derivation done in \cite{Finn:2008np,Romano:2016dpx} based on the frequency basis (\ref{GW-f}) that we introduce in the appendix~\ref{app:f-basis}, but reformulated in the basis (\ref{GW-classical}).

\subsection{Single arm of an interferometer}

We consider an interferometer arm at rest, with mass $1$ and mass $2$ at the two ends. Mass $1$ and mass $2$ are located, respectively,  at $\vec{x}_1$ and at  $\vec{x}_2 = \vec{x}_1 + L \, {\hat l}_{12}$, where $L$ is the length of the arm, and ${\hat l}_{12}$ the unit vector in the direction from mass $1$ to mass $2$. The change in the light-travel time for a photon emitted at mass $1$ at the time $t_1$ and arriving at mass 2, due to a passing gravitational wave $h_{ab}$ is  given by 
\begin{equation}
\Delta T \left( t_1 \right) =  \frac{1}{2} {\hat l}_{12}^a  {\hat l}_{12}^b \int_{0}^L d s \, h_{ab} \left(  t_1  + s ,\, \vec{x}_1 + s \, {\hat l}_{12} \right) \;. 
\end{equation} 
Inserting the decomposition  (\ref{GW-classical}) and performing the integral, we find  
\begin{align}
 \Delta T \left( t_1 \right) = L \int &  d^3 k \, {\rm e}^{-2 \pi i \vec{k} \cdot \vec{x}_1} \sum_\lambda \, e_{ab,\lambda} \left( {\hat k} \right) \frac{{\hat l}_{12}^a {\hat l}_{12}^b}{2}  \nonumber \\
 \times
& \Bigg[ {\rm e}^{2 \pi i k t_1} {\hat h}_\lambda \left( \vec{k} \right) \, {\cal M} \left(  {\hat l}_{12} \cdot {\hat k} ,\, k \right)       
+  {\rm e}^{-2 \pi i k t_1} {\hat h}_\lambda^* \left( - \vec{k} \right) \, {\cal M}^* \left( - {\hat l}_{12} \cdot {\hat k} ,\, k \right) \Bigg] \;, 
\label{dT-M}
\end{align} 
%
with 
\begin{equation} 
 {\cal M}  \left( {\hat l}_{12} \cdot {\hat n} ,\, k \right)  \equiv  {\rm e}^{ \pi i k L \left[ 1 - {\hat n} \cdot {\hat l}_{12} \right]} \,  \frac{\sin \left( \pi k L \left[ 1 - {\hat n} \cdot {\hat l}_{12} \right] \right)}{\pi k  L \left[ 1 - {\hat n} \cdot {\hat l}_{12} \right] } \,,
\label{calM}
\end{equation} 
where $\hat{n}$ is the direction of propagation of the wave. Let us consider now the signal $s_{12} \left( t ,\, \vec{x}_1 \right)$,  measured at time $t$ by a detector of mass $1$ located at $\vec{x}_1$, as the total change in the light-travel time for a photon emitted at mass $1$ (at the time $t-2L$), arriving to mass $2$ (at the time $t-L$), and then coming back to mass $1$, i.e.,
\begin{equation}
s_{12} \left( t \right) = \Delta T_{12} \left( t - 2 L \right) +  \Delta T_{21} \left( t -  L \right) + n_1 \left( t \right) \;,
\end{equation}
where $n_1 \left( t \right) = n_1 \left(t,\vec x_1 \right)$ is the noise measured by the detector {at the vertex $\vec x_1$}. Using eq.~(\ref{dT-M}) we obtain 
\begin{equation}
s_{12} \left( t ,\, \vec{x}_1 \right) -  n_1 \left( t ,\, \vec{x}_1 \right) =  L \int d^3 k \, {\rm e}^{-2 \pi i \vec{k} \cdot \vec{x}_1} \sum_\lambda 
{\cal G}_\lambda \left( \hat k, \hat l_{12} \right)  \; {h}_\lambda \left( t - L ,\, \vec{k} \right) \, {\cal T} \left( k L ,\,  {\hat k} \cdot {\hat l}_{12} \right) \;,  \label{eq:sarm}
\end{equation}
where the detector transfer function\footnote{The detector transfer function $\cal T$ encodes the response of the detector to a gravitational wave, and should not be confused with the cosmological transfer function introduced in eq.~\eqref{eq:T0}.} is 
\begin{equation}
{\cal T} \left( k L ,\,  {\hat k} \cdot {\hat l}_{12} \right)  \equiv  {\rm e}^{- \pi i k L \left[ 1 + {\hat k} \cdot {\hat l}_{12} \right]} \,  \frac{\sin \left( \pi k L \left[ 1 - {\hat k} \cdot {\hat l}_{12} \right] \right)}{\pi k L \left[ 1 - {\hat k} \cdot {\hat l}_{12} \right] } 
+ {\rm e}^{ \pi i k L \left[ 1 - {\hat k} \cdot {\hat l}_{12} \right]} \,  \frac{\sin \left( \pi k L \left[ 1 + {\hat k} \cdot {\hat l}_{12} \right] \right)}{\pi k L \left[ 1 + {\hat k} \cdot {\hat l}_{12} \right] } \,, 
\end{equation}
and we have also introduced the combination  
\begin{equation}
{\cal G}_\lambda \left( \hat k, \hat l_{12} \right) \equiv e_{ab,\lambda} \left( {\hat k} \right) \frac{{\hat l}_{12}^a {\hat l}_{12}^b}{2} \,. 
\label{calG}
\end{equation}
Using (\ref{PS}), we obtain the $2-$point correlation function of the signal 
\begin{eqnarray} 
\left\langle s_{12}^2 \left( t ,\, \vec{x}_1 \right)  \right\rangle - 
\left\langle n_1^2 \left( t ,\, \vec{x}_1 \right)  \right\rangle  =  \frac{L^2}{4 \pi} 
\int \frac{d^3 k}{k^3} \, \sum_\lambda  P_{\lambda} \left( k \right) 
\left\vert {\cal G}_\lambda \left( \hat k, \hat l_{12} \right) \right\vert^2  \, 
\left\vert {\cal T} \left( k L ,\,  {\hat k} \cdot {\hat l}_{12} \right) \right\vert^2 \,,  
\end{eqnarray}
as well as the  $3-$point correlation function 
\begin{eqnarray} 
&& \!\!\!\!\!\!\!\! 
\left\langle s_{12}^3 \left( t ,\, \vec{x} _1\right) \right\rangle - \left\langle n_1^3 \left( t ,\, \vec{x} _1\right) \right\rangle   = L^3 
\int d^3 k_1 \int d^3 k_2  \int d^3 k_3  \, 
\delta^{(3)} \left( \vec{k}_1 + \vec{k}_2 + \vec{k}_3 \right) {\cal B}_{\lambda_1\lambda_2\lambda_3} \left( k_1 , k_2 , k_3 \right) \nonumber\\ 
&& \quad\quad\quad\quad \quad\quad  \quad\quad \quad\quad \quad \times \; 
\sum_{\lambda_1,\lambda_2,\lambda_3} \, 
{\cal G}_{\lambda_1}  \left( {\hat k}_1 ,\, \hat l_{12} \right) \, 
{\cal G}_{\lambda_2}  \left( {\hat k}_2 ,\, \hat l_{12} \right) \, 
{\cal G}_{\lambda_3}  \left( {\hat k}_3 ,\, \hat l_{12} \right)  \nonumber\\ 
& &  \quad\quad\quad\quad \quad\quad  \quad\quad \quad\quad \quad \times \; 
{\cal T} \left( k_1 L ,\,  {\hat k}_1 \cdot {\hat l}_{12} \right) \, {\cal T} \left( k_2 L ,\,  {\hat k}_2 \cdot {\hat l}_{12} \right) 
{\cal T} \left( k_3 L ,\,  {\hat k}_3 \cdot {\hat l}_{12} \right) \,, 
\end{eqnarray} 
where we have assumed that signal and noise are uncorrelated (i.e., $\left\langle s\,n \right\rangle=\left\langle s\,n^2 \right\rangle=\left\langle s^2 n \right\rangle=0)$.

\subsection{Quadratic and cubic signal correlation functions at LISA}

Equation~\eqref{eq:sarm} describes the signal generated in a single arm of an interferometer. From this result we can construct the response functions of the full instrument, where the phase measurements in the individual arms  are combined to minimize the instrumental noise (see e.g.~\cite{Adams:2010vc}). Combining two arms of the equilateral triangular LISA configuration with a common mass at $\vec x_1$ yields a Michelson interferometer:
\begin{align}
\sigma_X \left( t ,\, \vec{x}_1 \right) & \equiv s_X(t, \vec x_1) - n_X(t, \vec x_1) \nonumber \\
&  = \Delta T_{12}(t - 2 L) +\Delta T_{21}(t - L) -\Delta T_{13}(t - 2 L) - \Delta T_{31}(t - L) \,.
\label{signal-X}
\end{align}

Cyclic permutation of the endpoints $\vec x_1, \vec x_2, \vec x_3$ results in a total of three Michelson interferometers, which we label $X$, $Y$ and $Z$. Specifically 
\begin{eqnarray}
\sigma_1 \equiv \sigma_X &\propto& \left( {\rm time \; } \vec{x}_1 \rightarrow \vec{x}_2 \rightarrow \vec{x}_1 \right) -  \left( {\rm time \; } \vec{x}_1 \rightarrow \vec{x}_3 \rightarrow \vec{x}_1 \right) \nonumber\,,\\ 
\sigma_2 \equiv \sigma_Y &\propto& \left( {\rm time \; } \vec{x}_2 \rightarrow \vec{x}_3 \rightarrow \vec{x}_2 \right) -  \left( {\rm time \; } \vec{x}_2 \rightarrow \vec{x}_1 \rightarrow \vec{x}_2 \right) \nonumber\,,\\ 
\sigma_3 \equiv \sigma_Z &\propto& \left( {\rm time \; } \vec{x}_3 \rightarrow \vec{x}_1 \rightarrow \vec{x}_3 \right) -  \left( {\rm time \; } \vec{x}_3 \rightarrow \vec{x}_2 \rightarrow \vec{x}_3 \right)  \,.  
\label{sigma-123} 
\end{eqnarray}

\begin{figure}[tbp]
\centering 
\includegraphics[width=.35\textwidth]{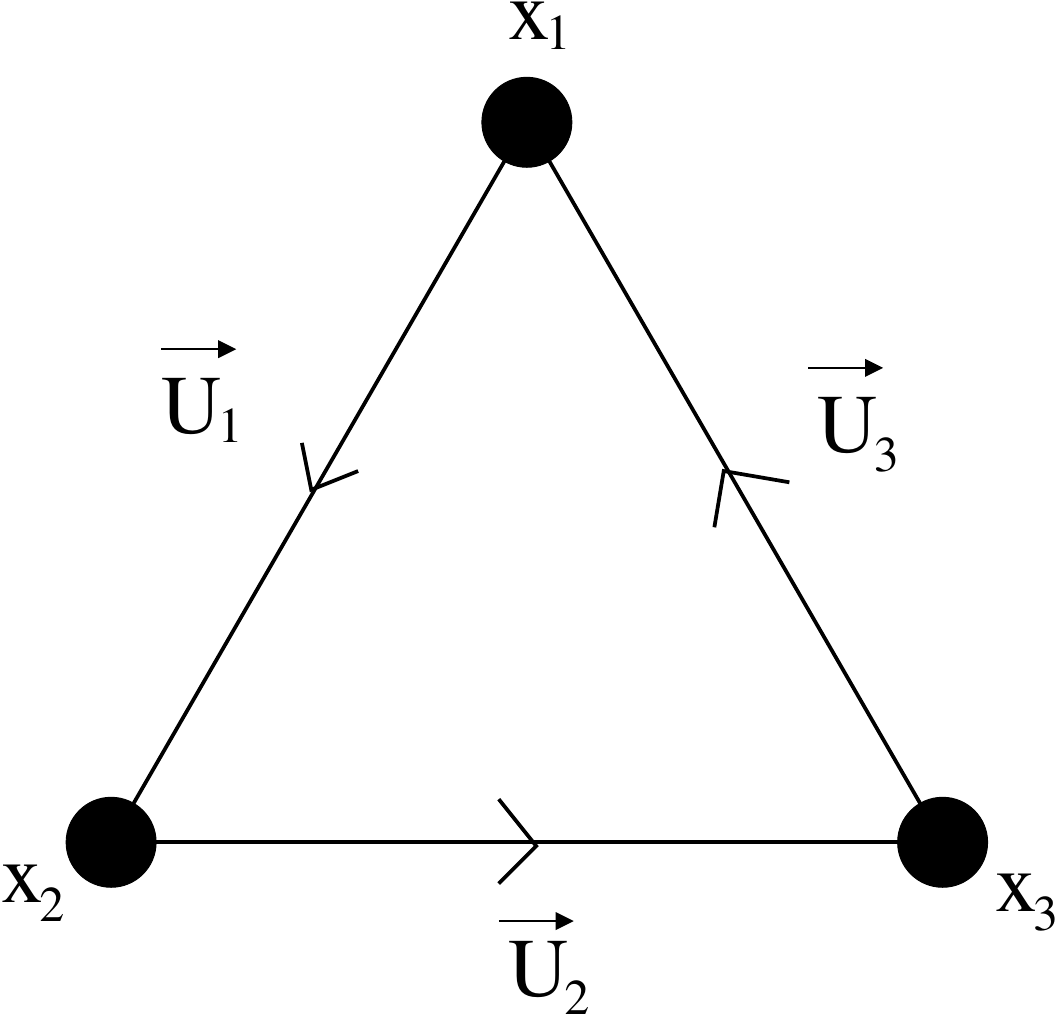}
\caption{\it Labeling of the LISA arms used in this work, cf. eq.  (\ref{LISA-arms}). 
}
\label{fig:LISA}
\end{figure}
Defining the three arm directions as  
\begin{equation} 
(\vec{x}_2 - \vec{x}_1)/L =  \hat l_{21}  \equiv  {\hat U}_1 \;\;\;,\;\;\; 
(\vec{x}_3 - \vec{x}_2)/L =  \hat l_{32}\equiv {\hat U}_2  \;\;\;,\;\;\; 
(\vec{x}_1 - \vec{x}_3)/L =  \hat l_{13}\equiv  {\hat U}_3  \;, 
\label{LISA-arms} 
\end{equation}
(see Figure \ref{fig:LISA}), the three measurements (\ref{sigma-123})  acquire the form 
\begin{equation} 
\sigma_j \left( t \right) = L \, \sum_\lambda \int d^3 k \;  h_\lambda \left( t - L ,\, \vec{k} \right) \frac{1}{2} \, {\rm e}_{ab,\lambda} \left( {\hat k} \right) \; {\cal Q}^j_{ab} \left( \vec{x}_j  ;\, {\hat U}_i  ;\, \vec{k} \right) \;, 
\label{sigma-XYZ}
\end{equation} 
with 
\begin{eqnarray} 
{\cal Q}^1_{ab}  \left( \vec{x}_1  ;\, {\hat U}_i  ;\, \vec{k} \right) &\equiv& 
{\rm e}^{-2 \pi i \vec{k} \cdot \vec{x}_1} \,  \left[  {\cal T} \left( k L ,\, {\hat k} \cdot {\hat U}_1 \right) {\hat U}_1^a {\hat U}_1^b 
-   {\cal T} \left( k L ,\, - {\hat k} \cdot {\hat U}_3 \right) {\hat U}_3^a {\hat U}_3^b \right] \;, \nonumber\\ 
{\cal Q}^2_{ab}  \left( \vec{x}_2  ;\, {\hat U}_i  ;\, \vec{k} \right) &\equiv& 
{\rm e}^{-2 \pi i \vec{k} \cdot \vec{x}_2} \, \left[  {\cal T} \left( k \, L ,\, {\hat k} \cdot {\hat U}_2 \right) \, {\hat U}_2^a  \, {\hat U}_2^b -   {\cal T} \left( k \, L ,\, - {\hat k} \cdot {\hat U}_1 \right) \, {\hat U}_1^a  \, {\hat U}_1^b \right] \;, \nonumber\\ 
{\cal Q}^3_{ab}  \left(   \vec{x}_3 ;\, {\hat U}_i  ;\, \vec{k} \right) &\equiv& 
{\rm e}^{-2 \pi i \vec{k} \cdot \vec{x}_3} \, \left[  {\cal T} \left( k \, L ,\,  {\hat k} \cdot {\hat U}_3 \right) \, {\hat U}_3^a  \, {\hat U}_3^b -   {\cal T} \left( k \, L ,\, - {\hat k} \cdot {\hat U}_2 \right) \, {\hat U}_2^a  \, {\hat U}_2^b \right] \;. \nonumber\\ 
\label{calQ}
\end{eqnarray} 

From these, the standard LISA output channels $A$, $E$ and $T$ are constructed as \cite{Adams:2010vc}
\begin{eqnarray} 
\Sigma_1 \equiv \Sigma_A &=& \frac{1}{3} \left( 2 \sigma_X - \sigma_Y - \sigma_Z \right) \;, 
\nonumber\\ 
\Sigma_2 \equiv \Sigma_E &=& \frac{1}{\sqrt{3}} \left( \sigma_Z - \sigma_Y \right)  \;, 
\nonumber\\ 
\Sigma_3 \equiv \Sigma_T &=&  \frac{1}{3} \left( \sigma_X + \sigma_Y + \sigma_Z \right)  \;. 
\label{signal-AET}
\end{eqnarray} 
We write these through the compact expression 
\begin{equation} 
\Sigma_O = L \sum_\lambda \int d^3 k \;  h_\lambda \left( t - L ,\, \vec{k} \right) \; 
\frac{1}{2} \, {\rm e}_{ab,\lambda} \left( {\hat k} \right) \; c^O_i \, {\cal Q}^i_{ab}  \left( \vec{x}_i  ;\, {\hat U}_j   ;\, \vec{k} \right) \;, 
\label{Sigma-I}
\end{equation} 
where ${ O} = \{ A,E,T \}$ labels the detector channel and  the matrix $c$ is given by 
\begin{equation}
c =  \begin{pmatrix}
\tfrac{2}{3} & - \tfrac{1}{3} & - \tfrac{1}{3}  \\[5pt]
0 & - \tfrac{1}{\sqrt{3}}  &  \tfrac{1}{\sqrt{3}} \\[5pt]
\tfrac{1}{3} & \tfrac{1}{3} & \tfrac{1}{3} 
\end{pmatrix} \;. 
\label{matrix-c}
\end{equation}  

\smallskip
We now have the tools to study the two-point and three-point correlation functions for the signal, and
 analyze how they depend on the statistical properties of the  SGWB, namely
 its power spectrum and bispectrum as described in Section \ref{sec:formalism}.  We derive the quadratic and cubic interferometer response functions, 
 and discuss how the interferometer measurements allow to probe properties of the primordial tensor non-Gaussianity.
 
\subsubsection{The quadratic interferometer response function}

The quadratic auto-correlation of the interferometer signal reads
\begin{equation} 
\left\langle \Sigma_O^2 \left( t \right)  \right\rangle = L^2  \, \sum_{\lambda}   \int \frac{d k}{k} P_\lambda \left( k \right) 
{\cal R}_\lambda^{OO} \left( k \right) \;, 
\end{equation} 
where 
\begin{eqnarray} 
{\cal R}_\lambda^{OO} \left( k \right)  &\equiv& \frac{ 1 }{4 \pi } \int_0^\pi d \theta \, \sin \theta \int_0^{2 \pi} d \phi \, \left\vert \frac{c^O_i}{2} \, {\cal Q}^i_{ab}  \left( \vec{x}_i  ;\, {\hat U}_j  ;\, \vec{k} \right) \,  {\rm e}_{ab,\lambda} \left( {\hat k} \right) \right\vert^2  \;.
\nonumber\\ 
\end{eqnarray} 
Here we use standard spherical coordinates, where $\theta \in \left[ 0 ,\, \pi \right]$ is the polar angle, and $\phi \in \left[ 0 ,\, 2 \pi \right]$ is the azimuthal angle to express the direction of ${\hat k}$. To obtain these 
expressions we have inserted eq.~\eqref{PS} (evaluated at equal times) and used the fact that ${\cal Q}^i_{ab}  {\rm e}_{ab,\lambda} \left( - \vec{k} \right) = {\cal Q}^{i*}_{ab}  {\rm e}_{ab,\lambda}^* \left(  \vec{k} \right) $  [recall that this property is specific to our choice of polarization operators eq.~(\ref{e-properties})]. 

The quantity ${\cal R}^{OO}_\lambda(k)$ gives the (scale-dependent) detector response function to the GW of polarization {$\lambda$}
for the 2-point correlation function of the channel $O$. This is obtained after integrating $ \left\vert \frac{c^O_i}{2} \, {\cal Q}_{ab}^i e_{ab,\lambda} \left( \vec k \right) \right|^2$ over all the directions ${\hat k}$ of the incoming GW.

\begin{figure}[tbp]
\centering 
\includegraphics[width=.45\textwidth]{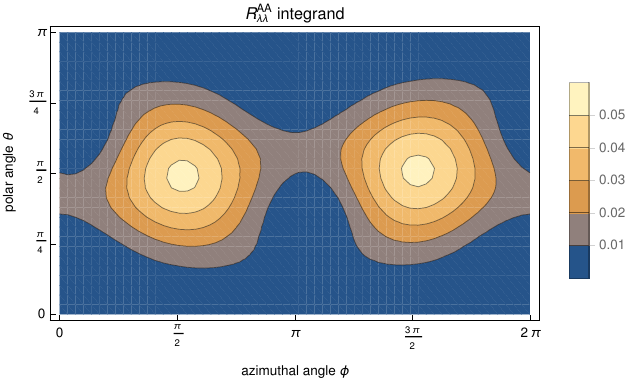}
\hfill
\includegraphics[width=.45\textwidth]{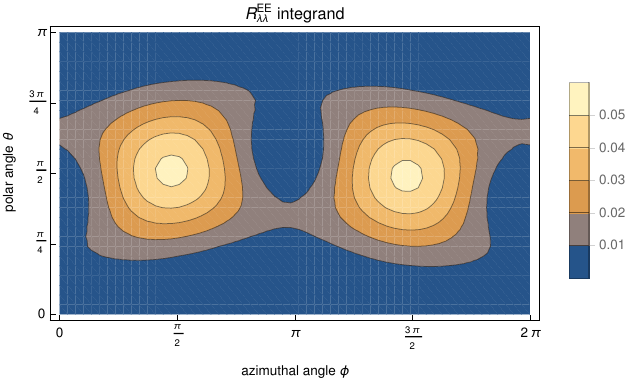}
\caption{\it Directional dependence of the integrand for the response functions ${\cal R}^{AA}_R \left( k \right)$ (left panel) and ${\cal R}^{EE}_R \left(k \right)$. The wavenumber $k$ of the GW has been chosen to be $0.1 / L$. The response functions for the left-handed polarizations are obtained by a parity transformation about the plane of the detector (the $xz$ plane, in our choice), which for our case corresponds to $\phi \rightarrow - \phi$ (as discussed in the text).}
  \label{fig:PSinteg}
\end{figure}

The angle-dependent integrand is shown in Fig.~\ref{fig:PSinteg}, for the specific choice $k = 0.1 / L$ of the GW wavenumber and for the two channels A (left panel) and E (right panel), respectively. We disregard the $T$ channel, as its sensitivity to the GW background is well below that of the A and E channels \cite{Adams:2010vc} (we have verified that this is the case also for the $3-$point functions that we study below).  
For definiteness, we set  the detector in the $xz$-plane, with the three masses, respectively at the $\left\{ x ,\, z \right\}$ locations given by $ \left\{ 0 ,\, 0 \right\} ,\; \left\{ 0 ,\, L \right\}$, and $\left\{ \frac{\sqrt{3}}{2} \, L ,\, \frac{L}{2} \right\}$ (we verified that rotating the plane of the interferometer does not change the final value of the response function obtained after integrating over the angular directions of $\vec{k}$).  As expected GWs arriving from the perpendicular direction to the detector (the $y$ axis, characterized by $\theta = \frac{\pi}{2}$ and $\phi = \frac{\pi}{2} ,\, \frac{3 \pi}{2}$) are those that provide the largest contribution to the detector two-point function. For equal length interferometer arms the {response function of the $A-E$ cross-correlation} 
${\cal R}^{AE}$ vanishes \cite{Adams:2010vc}. 

\begin{figure}
\begin{center}
\includegraphics[width = 0.49 \textwidth]{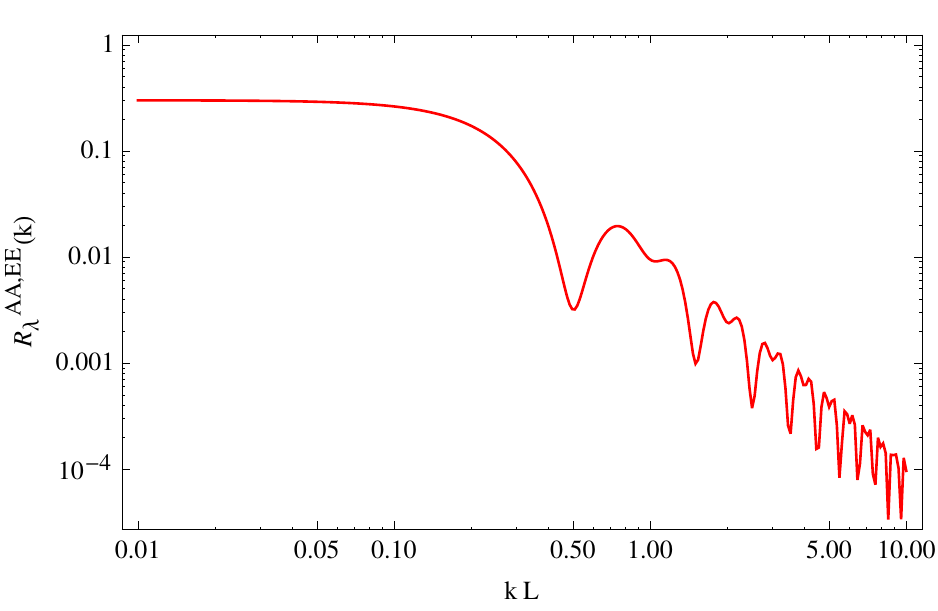}
\caption{\it 
Integrated response function ${\cal R}_\lambda^{OO} \left( k \right)$ for $O = \{A,E\}$ and $\lambda = \{L, R\}$. }
\label{fig:PS}
\end{center}
\end{figure}

In Fig.~\ref{fig:PS} we show the detector response function (after integrating over the GW directions). The same result is obtained for both polarizations (we discuss this below) and in the two channels A and E. The response function is nearly constant up to $k \sim 0.2/L$, and then it strongly decreases with higher $k$'s. By Taylor expanding the relations (\ref{calQ}) in the limit of small $k$, we obtain the analytic result 
\begin{equation}
{\cal R}^{AA}_\lambda \left(k \right) = {\cal R}^{EE}_\lambda \left(k \right) = 
 \frac{3}{10} - \frac{169}{1680} \, \left( \frac{k}{k_*} \right)^2 + {\cal O } \left( \frac{k^4}{k_*^4} \right)  \;\;\;,\;\;\; 
k_* \equiv \frac{1}{2 \pi L} \;. 
\end{equation} 
This agrees with the first of eqs.~(18) in \cite{Cornish:2001bb}. 

The response function for short wavelengths $kL \gg 0.1$, contrary to the long wavelength limit, becomes more supressed the shorter the wavelength. Simple inspection of eq.~(\ref{calQ}) leads to predict that its envelope amplitude scales as $\propto (k_*/k)^2$. This power law suppression at large $k$ can be clearly appreciated in Fig.~\ref{fig:PS}.

\subsubsection{The cubic interferometer response function}
\label{sec-cubicr}

Starting from eq.~(\ref{Sigma-I}), we obtain the three point correlation function of the measurement as 
\begin{align} 
\left\langle \Sigma_O \left( t \right) \Sigma_{O'} \left( t \right) \Sigma_{O''} \left( t \right) \right\rangle &= L^3 \sum_{\lambda_1\lambda_2\lambda_3} 
\int d^3 k_1  d^3 k_2  d^3 k_3 \,\delta^{(3)} \left( \vec{k}_1 + \vec{k}_2 + \vec{k}_3 \right) {\cal B}_{\lambda_1,\lambda_2,\lambda_3} \left( \vec{k}_1 ,\, \vec{k}_2 ,\, \vec{k}_3 \right) \nonumber\\ 
 \times & \,   \frac{c^O_i c^{O'}_j c^{O''}_k }{8} \, {\rm e}_{ab,\lambda_1} \left( {\hat k}_1 \right)  
{\rm e}_{cd,\lambda_2} \left( {\hat k}_2 \right)  {\rm e}_{ef,\lambda_3} \left( {\hat k}_3 \right) 
{\cal Q}^i_{ab}  \left( \vec{x}_i  ;\, {\hat U}_p ;\, \vec{k}_1 \right)  \nonumber\\  \times & \,   {\cal Q}^j_{cd}  \left( \vec{x}_j  ;\, {\hat U}_q ;\, \vec{k}_2 \right) 
{\cal Q}^k_{ef}  \left( \vec{x}_k  ;\, {\hat U}_r ;\, \vec{k}_3 \right) \;. 
\label{3-point-start} 
\end{align} 

With respect to the two-point function, the computation of the three-point function  has the additional complication 
that the bispectrum changes under a rotation of the triangle formed by the three vectors $\vec{k}_i$. {We discuss 
this technical point}  in Appendix~\ref{app:polarization}, in particular see eq.~(\ref{rotate-B}).  
To perform the angular integration, we fix the orientation of a reference triangle, formed by  vectors $\vec{k}_i^*$  satisfying the momentum conservation
condition $\vec k_1^*+\vec k_2^*+\vec k_3^*\,=\,0$, and compute the reference bispectrum ${\cal B}_{\lambda_1,\lambda_2,\lambda_3} \left( \vec{k}_i^* \right)$. For any generic orientation we have then 
\begin{equation} 
{\cal B}_{\lambda_1,\lambda_2,\lambda_3} \left( \vec{k}_1 = R \, \vec{k}_1^* ,\,  \vec{k}_2 = R \, \vec{k}_2^* ,\,  \vec{k}_3 = R \, \vec{k}_3^* \right) = \Phi_{\lambda_i} \left( {\hat k}_i^* ,\, R \right) \, 
{\cal B}_{\lambda_1,\lambda_2,\lambda_3} \left(  \vec{k}_1^* ,\,   \vec{k}_2^* ,\,   \vec{k}_3^* \right) \;, 
\end{equation} 
where $R$ is a rotation matrix, and where $\Phi_{\lambda_i} \left( {\hat k}_i^* ,\, R \right) $ is the combination of phases given in eq.~(\ref{rotate-B}). We note that this phase is eliminated by the corresponding opposite phase in the transformation of the helicity operators, cf. eqs.~(\ref{rotate-e}) and  (\ref{rotate-h}). Namely, 
\begin{align} 
& {\cal B}_{\lambda_1,\lambda_2,\lambda_3} \left( \vec{k}_1 ,\, \vec{k}_2 ,\, \vec{k}_3  \right)  {\rm e}_{ab,\lambda_1} \left( {\hat k}_1 \right)  
{\rm e}_{cd,\lambda_2} \left( {\hat k}_2 \right)  {\rm e}_{ef,\lambda_3} \left( {\hat k}_3 \right) \big\vert_{\vec{k}_i =  R \vec{k}_i^*}  \nonumber\\ 
& \quad  =  R_{aA} R_{bB} R_{cC} R_{dD} R_{eE} R_{fF} \; 
{\cal B}_{\lambda_1,\lambda_2,\lambda_3} \left( \vec{k}_1^* ,\, \vec{k}_2^* ,\, \vec{k}_3^*  \right)   {\rm e}_{AB,\lambda_1} \left( {\hat k}_1^* \right)  {\rm e}_{CD,\lambda_2} \left( {\hat k}_2^* \right)  {\rm e}_{EF,\lambda_3} \left( {\hat k}_3^* \right) \;. 
\nonumber\\ 
\end{align} 
The disappearance of $\Phi$ from the product is due to the fact that the measurement is proportional to $h_{ab} h_{cd} h_{ef}$, which is a product of three tensors. 

Among the possible choices for the orientation of the reference vectors $\vec{k}_i^*$,  in what follows we adopt this one: 
\begin{equation}
\vec{k}_1^* = k_1 \, {\hat v}_1 \;\;\;\;,\;\;\;\; \vec{k}_2^* = k_2 \, {\hat v}_2 \;\;\;\;,\;\;\;\;  \vec{k}_3^* = - \vec{k}_1^* - \vec{k}_2^* \;, 
\label{starframe}
\end{equation}
where 
\begin{eqnarray} 
{\hat v}_1 = \left( \begin{array}{c} 1 \\ 0 \\ 0 \end{array} \right) \;\;,\;\; 
{\hat v}_2 = \left( \begin{array}{c} \frac{k_3^2-k_1^2-k_2^2}{2 k_1 k_2} \\  \sqrt{1-\left( \frac{k_3^2-k_1^2-k_2^2}{2 k_1 
k_2} \right)^2} \\ 0  \end{array} \right) \;. 
\label{v1hat-v2hat} 
\end{eqnarray} 
Starting from the configuration (\ref{starframe}), a generic orientation of the bispectrum is obtained through the rotation matrix 
\begin{equation} 
R \left[ \theta_1 ,\, \phi_1 ,\, \phi_2 \right] = 
\left( \begin{array}{ccc} 
\sin \theta_1 \cos \phi_1  & \cos \theta_1  \cos \phi_1  & - \sin \phi_1   \\ 
\sin \theta_1  \sin \phi_1  & \cos \theta_1  \sin \phi_1  & \cos \phi_1    \\  
\cos \theta_1  & - \sin \theta_1  & 0 
\end{array} \right)  \cdot 
\left( \begin{array}{ccc} 
1 & 0 & 0 \\ 
0 & \cos \phi_2 & - \sin \phi_2 \\ 
0 & \sin \phi_2 & \cos \phi_2  \\ 
\end{array} \right) \;. 
\label{rotmafc}
\end{equation} 

The angles in this rotation matrix can be used to parametrize the independent angular integrals in eq.~(\ref{3-point-start}). Namely, we note that the angle between $\vec{k}_1$ and $\vec{k}_2$ is fixed by the fact that 
\begin{equation} \label{k3prop1}
\vec{k}_3 = - \vec{k}_1 - \vec{k}_2 \;\;\; \Rightarrow \;\;\; \cos \, \theta_{\vec{k}_1 ,\, \vec{k}_2} = \frac{k_3^2-k_1^2-k_2^2}{2 k_1 k_2} \,. 
\end{equation} 
As a rotation preserves angles between vectors, this angle has been imposed in eq.~(\ref{v1hat-v2hat}).  The right matrix in 
eq.~(\ref{rotmafc}) then represents a rotation (by a generic angle $\phi_2$) of $\vec{k}_2$ around the axis of $\vec{k}_1$, at a fixed value of $\theta_{\vec{k}_1 ,\, \vec{k}_2}$. (We note that this matrix leaves ${\hat v}_1$ unchanged.) The left matrix in eq.~(\ref{rotmafc}) then represents a common rotation of $\vec{k}_1$ and $\vec{k}_2$. Therefore, the  matrix 
$R \left[ \theta_1 ,\, \phi_1 ,\, \phi_2 \right] $ parametrizes the most generic orientation for $\vec{k}_1$ and $\vec{k}_2$, compatible with the fact that the three wavevectors close into a triangle. No additional independent angle is required to parametrize $\vec{k}_3$, as this vector is simply $-\vec{k}_1-\vec{k}_2$. 

With this in mind, eq.~(\ref{3-point-start}) can be rewritten as 
\begin{align}
\left\langle \Sigma_O \left( t \right) \Sigma_{O'} \left( t \right) \Sigma_{O''} \left( t \right) \right\rangle\,= & \,
%
 L^3 \sum_{\lambda_1\lambda_2\lambda_3} 
\int_0^\infty d k_1 d k_2 d k_3 k_1 k_2 k_3  
{\cal B}_{\lambda_1,\lambda_2,\lambda_3} \left( \vec{k}_1^* ,\, \vec{k}_2^* ,\, \vec{k}_3^* \right)  \nonumber \\
& \times 
\frac{c^O_i c^{O'}_j c^{O''}_k }{8} \, {\rm e}_{AB,\lambda_1} \left( {\hat k}_1^* \right)  
{\rm e}_{CD,\lambda_2} \left( {\hat k}_2^* \right)  {\rm e}_{EF,\lambda_3} \left( {\hat k}_3^* \right)  \nonumber\\ 
& \times
\int_0^\pi d \theta_1 \sin \theta_1 \int_0^{2 \pi} d \phi_1 \int_0^{2 \pi} d \phi_2 \; 
R_{aA} R_{bB} R_{cC} R_{dD} R_{eE} R_{fF} \; 
\nonumber\\ 
& \times 
{\cal Q}^i_{ab}  \left( \vec{x}_i  ;\, {\hat U}_p  ;\, R \vec{k}_1^* \right)  
 {\cal Q}^j_{cd}  \left( \vec{x}_j  ;\, {\hat U}_q ;\, R \vec{k}_2^* \right) 
{\cal Q}^k_{ef}  \left( \vec{x}_k  ;\, {\hat U}_r ;\, R \vec{k}_3^* \right) \,, 
\label{3-point-middle} 
\end{align} 
%
where $\sin \theta_1$ is the Jacobian of the transformation from the integrals in eq.~(\ref{3-point-start}) to those in 
(\ref{3-point-middle}) (after the $d^3 \vec{k}_3$ integration has been eliminated through the $ \delta^{(3)} \left( \vec{k}_1 + \vec{k}_2 + \vec{k}_3 \right) $ function). 

We then note that 
\begin{eqnarray} 
R_{aA} R_{bB} {\cal Q}^i_{AB}  \left( \vec{x}_i  ;\, {\hat U}_p  ;\, R \vec{k}^* \right)  &=&  {\cal Q}^i_{ab}  \left( \vec{x}_i^*  ;\, {\hat U}_p^*  ;\,  \vec{k}^* \right)  \;, \nonumber\\ 
\end{eqnarray} 
where 
\begin{equation} 
\vec{x}_i^{\,*} \equiv R^{-1} \, \vec{x}_i \;\;\;,\;\;\; {\hat U}_p^* \equiv R^{-1} \, {\hat U }_p  \;. 
\end{equation} 
Therefore, eq.~\eqref{3-point-middle} can be written as 
\begin{align} \label{forstar}
\left\langle \Sigma_O \left( t \right) \Sigma_{O'} \left( t \right) \Sigma_{O''} \left( t \right) \right\rangle =& L^3 \sum_{\lambda_1\lambda_2\lambda_3} 
\int_0^\infty d k_1 \, d k_2 \, d k_3 \, k_1 \, k_2\,  k_3  \nonumber \\
& \times {\cal B}_{\lambda_1\lambda_2\lambda_3} \left( \vec{k}_1^* ,\, \vec{k}_2^* ,\, \vec{k}_3^* \right) \; 
{\cal R}^{OO'O''}_{\lambda_1\lambda_2\lambda_3} \left( \hat{k}_i^* ;\, k_i  \right) \;, 
\end{align} 
with the three-point response function 
\begin{eqnarray} 
{\cal R}^{OO'O''}_{\lambda_1\lambda_2\lambda_3} \left( {\hat k}_i^* ;\, k_i \right)  &\equiv& 
 \frac{c^O_i c^{O'}_j c^{O''}_k }{8} \, {\rm e}_{AB,\lambda_1} \left( {\hat k}_1^* \right)  
{\rm e}_{CD,\lambda_3} \left( {\hat k}_2^* \right)  {\rm e}_{EF,\lambda_3} \left( {\hat k}_3^* \right)  \nonumber\\ 
& &  \!\!\!\!\!\!\!\!  \!\!\!\!\!\!\!\!  \!\!\!\!\!\!\!\! 
\times \int_0^\pi d \theta_1 \sin \theta_1 \int_0^{2 \pi} d \phi_1 \int_0^{2 \pi} d \phi_2 \; 
{\cal Q}^i_{AB}  \left( \vec{x}_i^*  ;\, {\hat U}_p^*  ;\,  \vec{k}_1^* \right)  
\nonumber\\ 
&&  \!\!\!\!\!\!\!\!  \!\!\!\!\!\!\!\!  \!\!\!\!\!\!\!\! 
\times  \, {\cal Q}^j_{CD}  \left(  \vec{x}_j^* ;\, {\hat U}_q^*  ;\,  \vec{k}_2^* \right)  
{\cal Q}^k_{EF}   \left(   \vec{x}_k^* ;\, {\hat U}_r^*  ;\,  \vec{k}_3^* \right) \;.  
\label{R3}
\end{eqnarray} 

We use this expression to numerically evaluate the response function. To verify that the response function is real, it is convenient to rewrite it as 
\begin{align} 
{\cal R}^{OO'O''}_{\lambda_1\lambda_2\lambda_3} \left( {\hat k}_i^* ;\, k_i \right)  = &  
\frac{c^O_i c^{O'}_j c^{O''}_k  }{8} \, 
\int_0^\pi d \theta_1 \sin \theta_1 \int_0^{2 \pi} d \phi_1 \int_0^{2 \pi} d \phi_2 \; \, 
{\rm e}^{2 i \lambda_1 \gamma \left[ {\hat k}_1^* ,\, R \right]}  {\rm e}_{ab,\lambda_1} \left( R {\hat k}_1^* \right)   \nonumber\\ 
& \times
{\rm e}^{2 i \lambda_2 \gamma \left[ {\hat k}_2^* ,\, R \right]}  {\rm e}_{cd,\lambda_2} \left( R {\hat k}_2^* \right)  
{\rm e}^{2 i \lambda_3 \gamma \left[ {\hat k}_3^* ,\, R \right]}  {\rm e}_{ef,\lambda_3} \left( R {\hat k}_3^* \right)
\nonumber\\ 
& \times
{\cal Q}^i_{ab}  \left( \vec{x}_i ;\, {\hat U}_p  ;\, R \vec{k}_1^* \right)  
{\cal Q}^j_{cd}  \left( \vec{x}_j ;\, {\hat U}_q ;\, R \vec{k}_2^* \right) 
{\cal Q}^k_{ef}  \left( \vec{x}_k ;\, {\hat U}_r ;\, R \vec{k}_3^* \right) \,,  
\end{align} 
%
where the quantities $\gamma$ are defined in (\ref{gamma}). One can verify by direct inspection that all terms in this integrand go  to their complex conjugate under parity (namely, under 
$\vec{k}_i \rightarrow - \vec{k}_i$, corresponding to $\theta_1 \rightarrow \pi - \theta_1$, $\phi_1 \rightarrow \pi + \phi_1$, $\phi_2 \rightarrow \pi - \phi_2$). Therefore, the response function is real. 

As we will discuss in Section \ref{subsec:propertiesR3}, measurements of correlation functions by LISA are invariant under parity. This implies that ${\cal R}_{RRR} = {\cal R}_{LLL}$ and  ${\cal R}_{RRL} = {\cal R}_{LLR}$, for any choice of channels and momenta.  For this reason, we show the response function in the RRR and RRL cases only. In the same Section, we will also demonstrate that the only nonvanishing 3-point response functions are the one among three $E$ channels, and the one among two $A$ channels and one {$E$ channel.}
We will show that they are opposite to each other, namely $\left\langle EEE \right\rangle = - \left\langle AAE \right\rangle $ (and permutations). 

We now use the expression (\ref{R3}) to evaluate the response function numerically.  We do so for two cases:

\begin{figure}[tbp]
\centering 
\includegraphics[width=.45\textwidth]{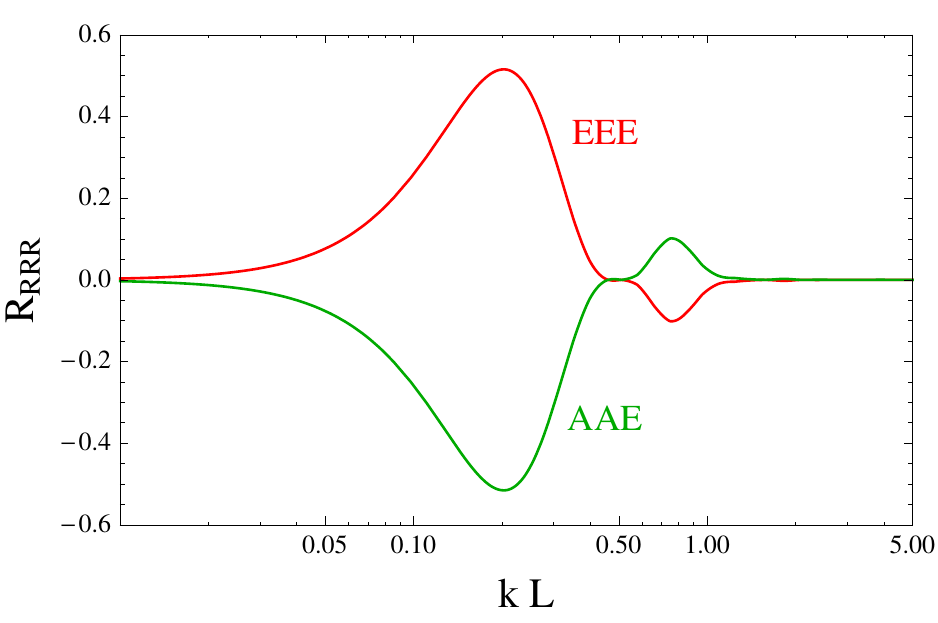}
\hfill
\includegraphics[width=.45\textwidth]{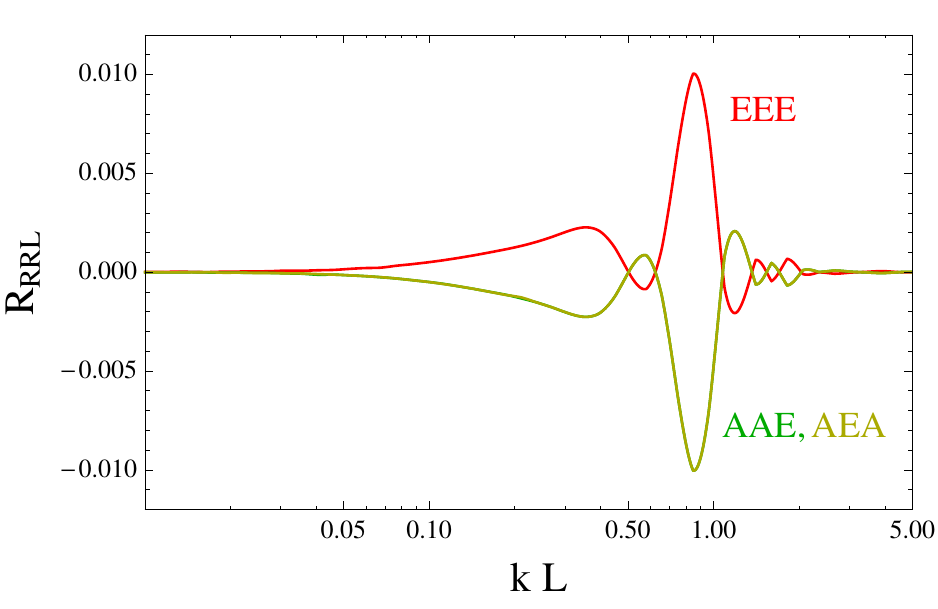}
\caption{\it 
Three point response function for equilateral configurations, $k_1 = k_2 = k_3 \equiv k$. The left (resp., right) panel shows the functions induced by three right-handed (resp, two right-handed and one left-handed) GWs. The same result is obtained from three left-handed (resp. two left-handed and one right-handed). Different lines correspond to different channels, as specified next to each line. In each panel, we show the only two distinct nonvanishing response functions, noting that they are of equal magnitude and opposite sign. 
}
\label{fig:BS-eq}
\end{figure}

\bigskip
{\it -- An equilateral configuration}. This is the case of equal wavenumbers, $k_1 = k_2 = k_3 \equiv k$. In Figure 
\ref{fig:BS-eq} we show the nonvanishing response functions for the RRR (left panel) and {RRL} 
(right panel) cases. The figure confirms that $\left\langle EEE \right\rangle = - \left\langle AAE \right\rangle $. We also see that the response functions vanish both in the $k L \gg 1 $ and $k L \ll 1$ limit. In the large $k$ limit, the same behavior is obtained for the two-point function, cf. Figure \ref{fig:PS}, and it is due to the inability of the interferometer to resolve scales much smaller than its size. The small $k$ limit is instead studied in Appendix \ref{app:R3k0}.

\bigskip
{\it --  A squeezed isosceles configuration}. For the isosceles squeezed case, $k_3 \ll k_1 = k_2$, eqs.~(\ref{v1hat-v2hat}), together with ${\hat v}_3 = - \frac{k_1 \, {\hat v}_1 + k_1 \, {\hat v}_1}{k_3}$, give 
\begin{equation}
{\hat v}_1 = \left( 1,\, 0,\, 0 \right)  \;\;\;,\;\;\; 
{\hat v}_2 = \left( -1,\, 0,\, 0 \right)  \;\;\;,\;\;\; 
{\hat v}_3 = \left( 0,\, -1,\, 0 \right) \;. 
\end{equation} 

{Starting from eq.~(\ref{3-point-middle}), we find that, in this limit, the combinations $R_{aA} R_{bB}   e_{AB,\lambda_1} \left( \hat v_1 \right)$ and $R_{cC} R_{dD}   e_{CD,\lambda_2} \left( -\hat v_1 \right)$ can be expressed as $ {\rm e}^{\mp 2 i \lambda_1 \phi_2} $ times a factor that is independent of $\phi_2$, respectively. We further note that in this limit, }
the quantities ${\cal Q}^i_{ab}  \left( \vec{x}_i  ;\, {\hat U}_p  ;\, k_1 R  {\hat v}_1 \right)$ and ${\cal Q}^j_{cd}  \left( \vec{x}_j  ;\, {\hat U}_q   ;\, - k_1 R  {\hat v}_1 \right)$ are independent of $\phi_2$, and that 
\begin{eqnarray} 
\lim_{k_3 \rightarrow 0 } c^{O''}_k {\cal Q}^k_{ef}  \left( \vec{x}_k  ;\, {\hat U}_r  ;\, k_3 R  {\hat v}_3 \right) 
&=& 2 c^{O''}_1 \left( {\hat U}_1^e {\hat U}_1^f - {\hat U}_3^e {\hat U}_3^f  \right) 
+ 2 c^{O''}_2 \left( {\hat U}_2^e {\hat U}_2^f - {\hat U}_1^e {\hat U}_1^f  \right) \nonumber\\ 
&& + 2 c^{O''}_3 \left( {\hat U}_3^e {\hat U}_3^f - {\hat U}_2^e {\hat U}_2^f  \right) \;, 
\end{eqnarray} 
 is also $\phi_2-$independent (the matrix $c^{O''}_k$ in the formula above is given in eq.~(\ref{matrix-c}). The $\phi_2$ integration then gives 
(all other quantities are $\phi_2$ independent, and factorize out of it) 
\begin{equation} 
\int_0^{2 \pi} d \phi_2 \; 
e_{EF,\lambda_3} \left( \hat v_3 \right) \,R_{eE} R_{fF} \,  {\rm e}^{2 i \left( \lambda_2 - \lambda_1 \right) \phi_2} 
\equiv  \delta_{\lambda_1\lambda_2} F_{ef}(\theta_1,\phi_1)\;, 
\end{equation} 
namely we find that the result is nonvanishing only for $\lambda_1 = \lambda_2$, in which case the $\lambda_{1,2}$ dependence drops, and it is also independent of $\lambda_3$. Specifically, we find 
\begin{eqnarray}
F_{11} &=& \frac{1}{8} \pi  \left(-6 \cos ^2(\phi_1) \cos (2 \theta_1)+3 \cos (2 \phi_1)-1\right) \;, \nonumber\\ 
F_{12} = F_{21}  &=& \frac{3}{2} \pi  \sin (\phi_1) \cos (\phi_1) \sin ^2(\theta_1) \;, \nonumber\\ 
F_{13} = F_{31} &=& \frac{3}{2} \pi  \cos (\phi_1) \sin (\theta_1) \cos (\theta_1) \;, \nonumber\\ 
F_{22} &=& -\frac{1}{8} \pi  \left(6 \sin ^2(\phi_1) \cos (2 \theta_1)+3 \cos (2 \phi_1)+1\right) \;, \nonumber\\ 
F_{23} = F_{32} &=& \frac{3}{2} \pi  \sin (\phi_1) \sin (\theta_1) \cos (\theta_1) \;, \nonumber\\ 
F_{33} &=& \frac{1}{4} \pi  (3 \cos (2 \theta_1)+1) \;. 
\end{eqnarray} 

The remaining quantities collect into 
\begin{align} 
{\cal R}_{OO'O'',\lambda_1\lambda_2\lambda_3}^{\rm squeezed} & \left( {\hat v}_1 ,\, - {\hat v}_1 ,\, {\hat v}_3 \right) =   \nonumber \\
& =  \delta_{\lambda_1 \lambda_2} 
\left[   \left( c^{O''}_1 - c^{O''}_2 \right) {\hat U}_1^e {\hat U}_1^f 
+ \left( c^{O''}_2 - c^{O''}_3  \right) {\hat U}_2^e {\hat U}_2^f 
+ \left( c^{O''}_3 - c^{O''}_1 \right)  {\hat U}_3^e {\hat U}_3^f 
\right] \nonumber\\ 
&\times
 \frac{c^O_i c^{O'}_j}{4}  e_{AB,\lambda_1} \left( {\hat v}_1 \right) e_{CD,\lambda_1} \left( - {\hat v}_1 \right)  \; 
\int_0^\pi d \theta_1 \int_0^{2 \pi} d \phi_1 \sin \theta_1 F_{ef}(\theta_1,\phi_1) \nonumber\\ 
& \times 
{\cal Q}^i_{AB}  \left( \vec{x}_i^* ;\, {\hat U}_p^*  ;\, k_1  {\hat v}_1 \right) 
{\cal Q}^j_{CD}  \left( \vec{x}_j^* ;\, {\hat U}_q^*  ;\, - k_1  {\hat v}_1 \right) \;. 
\end{align} 
%
Apart from the $F_{ef}(\theta_1,\phi_1) $ factor, the last two lines in this expression are the two-point response function ${\cal R}_{\lambda}^{OO'}$ times $4 \pi$. This is a consequence of the fact that the squeezed bispectrum is nothing but the power spectrum of the short wavelength modes modulated by the long wavelength mode. Numerical evaluation of this expression results in the response functions shown in Figure \ref{fig:BS-sq}. As for the equilateral case, we find the only nonvanishing response functions shown in the figure, and we find that the response function for two $A$ channels and one $E$ channel is of equal magnitude but opposite sign compared to the one for 
three $E$ channels. The reason for this is explained in the next Subsection.

\begin{figure}[tbp]
\centering 
\includegraphics[width=.45\textwidth]{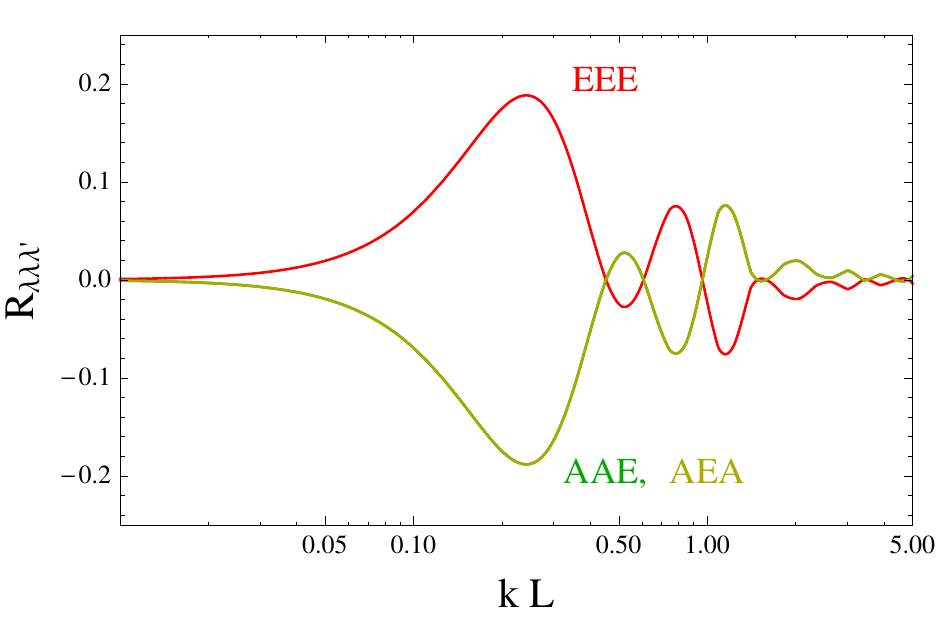}
\hfill
\caption{\it 
Squeezed isosceles bispectrum, for $\lambda_1 = \lambda_2 \equiv \lambda$, and $\lambda_3 \equiv \lambda'$ (the same result is obtained for all helicities). The third channel is taken in the squeezed limit, $k_3 L \rightarrow 0$, while $k_1 = k_2 \equiv k$ is varied in the figure. 
}
\label{fig:BS-sq}
\end{figure}

\subsubsection{Three-point response functions and LISA geometry}
\label{subsec:propertiesR3}

In this Subsection we describe how the geometrical features of the LISA configuration are sufficient to 
characterise some of the properties of the interferometer three point function.

  \smallskip
 
 \noindent
 {\bf Behaviour of LISA response functions under a parity transformation}
 
 \smallskip
 
 \noindent
 Besides the dependence on the shape and orientation of the triangle formed by the three momenta, 
  an additional feature 
 that  can characterize the tensor bispectrum is its behaviour under a parity transformation.
  Certain models of early universe cosmology (see Section \ref{sec-mod}), predict
  {tensor bispectra which violate parity, and values of the bispectrum components which}
are not invariant under a parity transformation. We now show that LISA -- and any planar interferometer -- cannot distinguish among components
of the bispectrum that differ only by a parity transformation, and consequently cannot detect parity breaking effects.
 Reference~\cite{Smith:2016jqs} noted that the LISA response function for the two-point function is the same for the left and right polarizations of the GW, namely ${\cal R}^{OO}_{L}  = {\cal R}^{OO}_{R}$, 
 due to a mirror symmetry with respect to  the plane of the interferometer. As we show now, this result can be readily extended also to the three-point function, implying the equalities ${\cal R}_{RRR} = {\cal R}_{LLL}$ and ${\cal R}_{RRL} = {\cal R}_{LLR}$. To understand this, consider a triangular configuration of {test masses in} 
  the $xz$ plane. The mirror symmetry then corresponds to changing the $y$ component of a vector. We would like to understand what this implies for the response functions. Denoting by $V_\parallel$ and by $V_\perp$ the component of a vector $V$, respectively, within and perpendicular to $xz$ plane, we note that 
the GW momentum, and the basis vectors expressed by eqs.~(\ref{eq:CanonicalOrthonormalBasis})  change as 
\begin{eqnarray}
&& {\hat k}_{\parallel} \Rightarrow \hat k_{\parallel} \;\;\;,\;\;\;
{\hat k}_{\perp} \Rightarrow - \hat k_{\perp} \;, \nonumber\\ 
&& {\hat u}_{\parallel} \Rightarrow - \hat  u_{\parallel} \;\;\;,\;\;\;
{\hat u}_{\perp} \Rightarrow \hat  u_{\perp} \;, \nonumber\\ 
&& {\hat v}_{\parallel} \Rightarrow \hat v_{\parallel}  \;\;\;,\;\;\;
{\hat v}_{\perp} \Rightarrow - \hat v_{\perp} \;. 
\label{kuv-mirror}
\end{eqnarray} 
As the unit vectors in eq.~(\ref{LISA-arms})  (expressing the displacements between the test masses) lie in the plane of the instrument, only the parallel components contribute to the ${\cal Q}^i_{ab} e_{ab,\lambda}$ contraction. Due to this, and thanks to the properties of  (\ref{kuv-mirror}), we see that ${\hat U}_a \, {\hat U}_b \, e_{ab,\lambda}$ goes to ${\hat U}_a \, {\hat U}_b \, e_{ab,-\lambda}$  under this mirror transformation (and analogously for the contractions with ${\hat V}$ and ${\hat W}$). On the other hand, the transfer function ${\cal T}$ and the phase in the first term in front of the square brackets in eqs.~(\ref{calQ})  are invariant under this transformation. This implies that  ${\rm e}_{ab,\lambda} {\cal Q}^j_{ab}  \left( k_\parallel ,\, k_\perp \right) =  {\rm e}_{ab,-\lambda} {\cal Q}^j_{ab}  \left( k_\parallel ,\, -k_\perp \right) $  in eq.~(\ref{sigma-XYZ}), and analogously for the linear combinations in eq.~(\ref{signal-AET}).  These combinations form the response functions, which, therefore, satisfy
\begin{equation}
{\cal R}_{\lambda}^{OO'}  \left( k_\parallel ,\, k_\perp \right) = {\cal R}_{-\lambda}^{OO'}  \left( k_\parallel ,\, - k_\perp \right) \;\;\;,\;\;\; 
{\cal R}_{\lambda \lambda' \lambda''}^{OO'O''}  \left( k_{i,\parallel} ,\, k_{i,\perp} \right) = 
{\cal R}_{-\lambda -\lambda' -\lambda''}^{OO'O''}  \left( k_{i,\parallel} ,\, - k_{i,\perp} \right) \;. 
\end{equation} 
After integrating over the angles, this implies the equalities ${\cal R}_{RR} = {\cal R}_{LL}$ for the two-point function found in \cite{Smith:2016jqs}, as well as the equalities ${\cal R}_{RRR} = {\cal R}_{LLL}$ and ${\cal R}_{RRL} = {\cal R}_{RLL}$ for the three-point function.  This implies that parity violation  in the tensor bispectrum can not be detected
using  a planar interferometer, like LISA.

 \smallskip
 
 \noindent
 {\bf Symmetries of the LISA response functions under exchange of the interferometer vertices}

 \smallskip
 
 \noindent
In eqs.~\eqref{sigma-123} we described how the three Michelson
interferometer signals $X$, $Y$, $Z$ associated with the interferometer vertices are constructed. Since LISA forms an equilateral triangle, we should expect that physical results are {\it independent} on the 
vertex labelling, and should be invariant exchanging vertices. For example,  exchanging the second and third vertices, one finds
\be
\label{imex23}
\sigma_X\,\to\,-\sigma_X \hskip1cm, \hskip1cm \sigma_Y\,\to\,-\sigma_Z
 \hskip1cm, \hskip1cm \sigma_Z\,\to\,-\sigma_Y\,,
\ee
and physical results are invariant under this transformation. Analogously, 
exchanging the first and second vertex, one has 
\be
\label{imex12}
\sigma_Z\,\to\,-\sigma_Z \hskip1cm, \hskip1cm \sigma_X\,\to\,-\sigma_Y
 \hskip1cm, \hskip1cm \sigma_Y\,\to\,-\sigma_X\,,
\ee
and lastly, exchanging first and third vertex, one finds invariance of results under the simultaneous
transformations 
\be
\label{imex13}
\sigma_Y\,\to\,-\sigma_Y \hskip1cm, \hskip1cm \sigma_X\,\to\,-\sigma_Z
 \hskip1cm, \hskip1cm \sigma_Z\,\to\,-\sigma_X\,. 
\ee

We can derive some consequences of these relations for the bispectra. First, the previous formulas imply that 
\be
\langle \sigma_X^3\rangle\,=\, \langle \sigma_Y^3\rangle\,=\,\langle \sigma_Z^3\rangle\,=\,\langle \sigma_X
\sigma_Y \sigma_Z
\rangle
\,=\, 0\,.
\ee
Moreover,  {\bf only one} component of the {three-point function} 
is independent, 
say $\langle \sigma_X \sigma_Y^2 \rangle$, and all the remaining ones coincide with this correlator up to a sign. We can write these relations as (we use the notation: $\sigma_1 \equiv \sigma_X \;,\; \sigma_2 \equiv \sigma_Y \;,\; \sigma_3 \equiv \sigma_Z $) 
\begin{equation}
\left\{ 
\begin{array}{l} 
\left\langle \sigma_1 \sigma_2 \sigma_3 \right\rangle = 0 \;\;\;, \\ \\ 
\left\langle \sigma_i \sigma_j \sigma_j \right\rangle = 
\epsilon_{ijk} \, C \;\;,\;\; {\rm where} \;\; C \equiv \left\langle \sigma_X \sigma_Y^2 \right\rangle 
\,.
\end{array} \right.
\end{equation} 

This also implies that {three-point functions} 
computed in terms of $E$, $A$ channels are particularly simple to express. One
finds 
\be
\label{imex46}
\langle \Sigma_E^3\rangle\,=\,-\langle \Sigma_E \Sigma_A^2\rangle\,=\,-\frac{2}{\sqrt{3}}\,
\langle \sigma_X \sigma_Y^2\rangle\,,
\ee
while all the other {three-point functions}  
vanish. We note that the identity $\langle \Sigma_E^3\rangle\,=\,-\langle \Sigma_E \Sigma_A^2\rangle$ is  confirmed by the numerical computations shown in Figures \ref{fig:BS-eq} and \ref{fig:BS-sq}.  Among the vanishing correlators, we note that $\langle \Sigma_A^3\rangle\,=\,0$. We also note that all the bispectra involving the $T$ channel vanish. For example 
\be
\langle \Sigma_E \Sigma_T^2\rangle\,=\,\langle \Sigma_E \Sigma_T \Sigma_A\rangle\,=\,0
\,.
\ee
On the one hand, this prediction for the null-channel, together with the relation \eqref{imex46} provides a non-trivial consistency check of the underlying assumptions of our analysis, such as statistical isotropy and Gaussian instrument noise. On the other hand, these relations predict that for any given frequency interval, (cross-)correlating different interferometer channels only allows us to extract a single measurement of the bispectrum, a serious obstacle in measuring its different helicity contributions. However, with mild assumptions on the frequency dependence of the bispectrum, the different frequency dependences of the ${\cal R}_{RRR} = {\cal R}_{LLL}$ and ${\cal R}_{RRL} = {\cal R}_{LLR}$ response functions (see Fig.~\ref{fig:BS-eq}) can be used to measure the chirality structure by comparing measurements in different frequency intervals.


\section{The optimal signal-to-noise ratio}
\label{sec-estimator}

In this section we construct a frequency-dependent estimator of the stochastic gravitational wave bispectrum. Our procedure 
extends  the arguments developed in 
 \cite{Smith:2016jqs} to estimate the optimal signal-to-noise ratio in the gravitational wave power spectrum to the case of a signal three-point function. 
 {In most of this section we  assume for generality that all the correlators 
$\langle AAA\rangle$, $\langle AAE \rangle$, $\langle AEE\rangle$ and $\langle EEE\rangle$ can be nonvanishing and independent from each other; only at the end
of Section \ref{sec-optest} we  specialize to the case of an equilateral LISA configuration, where only the correlators $\langle EEE\rangle\,=\,-\langle AAE\rangle$ are nonvanishing.}

\subsection{The signal in frequency space} 

Since our goal is to generate a frequency-dependent estimator, our first task is to compute the three-point function of the signal 
 in frequency space, defined as (for a return flight along the first arm) 
\begin{align}
\tilde{s}_{12} \left( f \right)&\equiv\int dt\,e^{-2\pi i f t}\,{s}_{12} \left( t\right)  \nonumber\\
& =  L \int d^3 k \, {\rm e}^{-2 \pi i \vec{k} \cdot \vec{x}_1} \sum_\lambda 
{\cal G}_\lambda \left( \hat k, \hat l_{12} \right) 
\int dt\,e^{-2\pi i f t}\,{h}_\lambda \left( t - L ,\, \vec{k} \right) \, {\cal T} \left( k L ,\,  {\hat k} \cdot {\hat l}_{12} \right) +  \tilde{n}_{12} \left( f \right),  \label{eq:stilde}
\end{align}
where $\tilde{n}_{12} \left( f \right)$ denotes the corresponding noise in frequency space. Defining
\begin{align}\label{eq:def_tildesigma}
\tilde{\Sigma}_O(f)=\tilde{s}_O \left( f \right) -  \tilde{n}_O \left( f \right) \;, 
\end{align}
where now $O$ denotes a LISA channel, we obtain the following three-point function
\begin{align}
\langle \tilde{\Sigma}_{O_1}(f_1)\,\tilde{\Sigma}_{O_2}(f_2)\,\tilde{\Sigma}_{O_3}(f_3)\rangle&= L^3 \int \left(\prod_{i=1}^3 dt_i\,e^{-2\pi i \,f_it_i}\,d^3k_i\sum \frac{c_{{O_i} \, j_i}}{2} {\cal Q}_{a_i b_i}^{j_i} \left( \vec{k}_i \right) e_{a_i b_i,\lambda_i} \left( {\hat k}_i \right)  \right)\nonumber\\
&\times\langle h_{\lambda_1}(t_1-L,\,\vec{k}_1)\, h_{\lambda_2}(t_2-L,\,\vec{k}_2) h_{\lambda_3}(t_3-L,\,\vec{k}_3)\rangle\;, 
\end{align}
the calculation of which requires the knowledge of the time correlator for the gravitational waves at unequal times. From the ansatz (\ref{f-NL}), and the 2-point correlator (\ref{PS}), one obtains 
\begin{align} 
\left\langle {h}_{\lambda_1} \left( t_1 ,\, \vec{k}_1 \right)  {h}_{\lambda_2} \left( t_2 ,\, \vec{k}_2 \right)  {h}_{\lambda_3} \left( t_3 ,\, \vec{k}_3 \right) \right\rangle 
& =   \frac{\delta^{(3)} \left( \vec{k}_1 + \vec{k}_2 + \vec{k}_3 \right)}{8 \pi^2}  \nonumber\\ 
&  \!\!\!\!\!\!\!\!  \!\!\!\!\!\!\!\!  \!\!\!\!\!\!\!\!  \!\!\!\!\!\!\!\!  \!\!\!\!\!\!\!\!  \!\!\!\!\!\!\!\!  \!\!\!\!\!\!\!\!  \!\!\!\!\!\!\!\!   \!\!\!\!\!\!\!\!  \!\!\!\!\!\!\!\!   \!\!\!\!\!\!\!\!   \!\!\!\!\!\!\!\!  \times 
 \Bigg\{   
f_{\rm NL}^{\lambda_1,\lambda_2,\lambda_3} \, K_{\lambda_1  \lambda_2 \lambda_3} \left( \vec k_1 , \vec k_2 , \vec k_3 \right) 
\frac{P_{\lambda_2} \left( k_2 \right)}{k_2^3} \, \frac{P_{\lambda_3} \left( k_3 \right)}{k_3^3} \, 
\cos \left[ 2\pi  k_2 \left( t_2 - t_1 \right) \right] \cos \left[ 2\pi k_3 \left( t_3 - t_1 \right) \right]  \nonumber\\ 
&  \!\!\!\!\!\!\!\!  \!\!\!\!\!\!\!\!  \!\!\!\!\!\!\!\!  \!\!\!\!\!\!\!\!  \!\!\!\!\!\!\!\!  \!\!\!\!\!\!\!\!  \!\!\!\!\!\!\!\!  \!\!\!\!\!\!\!\!  \!\!\!\!\!\!\!\!  \!\!\!\!\!\!\!\!   \!\!\!\!\!\!\!\! 
+ f_{\rm NL}^{\lambda_2 \lambda_1 \lambda_3}\, K_{\lambda_2  \lambda_1  \lambda_3} \left(\vec  k_2 , \vec k_1 , \vec k_3 \right) 
\frac{P_{\lambda_1} \left( k_1 \right)}{k_1^3} \, \frac{P_{\lambda_3} \left( k_3 \right)}{k_3^3}\, 
\cos \left[ 2\pi  k_1 \left( t_1 - t_2 \right) \right] \cos \left[ 2\pi  k_3 \left( t_3 - t_2 \right) \right] \nonumber\\ 
&  \!\!\!\!\!\!\!\!  \!\!\!\!\!\!\!\!  \!\!\!\!\!\!\!\!  \!\!\!\!\!\!\!\!  \!\!\!\!\!\!\!\!  \!\!\!\!\!\!\!\!  \!\!\!\!\!\!\!\!  \!\!\!\!\!\!\!\!  \!\!\!\!\!\!\!\!   \!\!\!\!\!\!\!\!   \!\!\!\!\!\!\!\! 
+ f_{\rm NL}^{\lambda_3,\lambda_1,\lambda_2} \, K_{\lambda_3  \lambda_1 \lambda_2} \left( \vec k_3 , \vec k_1 , \vec k_2 \right) 
\frac{P_{\lambda_1} \left( k_1 \right)}{k_1^3} \, \frac{P_{\lambda_2} \left( k_2 \right)}{k_2^3}  \, \cos \left[2\pi  k_1 \left( t_1 - t_3 \right) \right] \cos \left[2\pi k_2 \left( t_2 - t_3 \right) \right]
 \Bigg\} \,. 
\label{h-h-h-FT}
\end{align} 

This leads  to 
\begin{align}
&\int dt_1\,dt_2\,dt_3\,e^{-2\pi i(f_1t_1+f_2t_2+f_3t_3)}\left\langle {h}_{\lambda_1} \left( t_1-L ,\, \vec{k}_1 \right)  {h}_{\lambda_2} \left( t_2-L ,\, \vec{k}_2 \right)  {h}_{\lambda_3} \left( t_3-L ,\, \vec{k}_3 \right) \right\rangle\nonumber\\
&=\frac{\delta(f_1+f_2+f_3)}{32 \pi^2}\,\delta^{(3)} \left( \vec{k}_1 + \vec{k}_2 + \vec{k}_3 \right)  \nonumber\\ 
& \times 
 \Bigg\{   
f_{\rm NL}^{\lambda_1,\lambda_2,\lambda_3} \, K_{\lambda_1;\lambda_2,\lambda_3} \left( \vec k_1 ; \vec k_2 ,\vec k_3 \right) 
\frac{P_{\lambda_2} \left( k_2 \right)}{k_2^3} \, \frac{P_{\lambda_3} \left( k_3 \right)}{k_3^3} \, 
\delta(k_2-|f_2|)\,\delta(k_3-|f_3|)  \nonumber\\ 
&  
+ f_{\rm NL}^{\lambda_2,\lambda_1,\lambda_3}\, K_{\lambda_2;\lambda_1,\lambda_3} \left(\vec k_2 ; \vec k_1 , \vec k_3 \right) 
\frac{P_{\lambda_1} \left( k_1 \right)}{k_1^3} \, \frac{P_{\lambda_3} \left( k_3 \right)}{k_3^3}\, 
\delta(k_1-|f_1|)\,\delta(k_3-|f_3|) \nonumber\\ 
&  
+ f_{\rm NL}^{\lambda_3,\lambda_1,\lambda_2} \, K_{\lambda_3;\lambda_1,\lambda_2} \left( \vec k_3 ;  \vec k_1 , \vec  k_2 \right) 
\frac{P_{\lambda_1} \left( k_1 \right)}{k_1^3} \, \frac{P_{\lambda_2} \left( k_2 \right)}{k_2^3}  \, \delta(k_1-|f_1|)\,\delta(k_2-|f_2|) 
 \Bigg\} \nonumber\\
 &\equiv \delta(f_1+f_2+f_3)\,\delta^{(3)} \left( \vec{k}_1 + \vec{k}_2 + \vec{k}_3 \right)\,\tilde{\cal B}_{\lambda_1\lambda_2\lambda_3}(\vec k_1,\,\vec k_2,\,\vec k_3;\,f_1,\,f_2,\,f_3)\,, 
\label{B-tilde} 
\end{align}
so that  we can express the result in terms of quantities evaluated for a reference closed triangle with sides 
 $\vec k_i^*$, as discussed in Section \ref{sec-cubicr} (see in particular eq.~\eqref{forstar})
\begin{align} 
\langle \tilde{\Sigma}_{O_1}(f_1)\,\tilde{\Sigma}_{O_2}(f_2)\,\tilde{\Sigma}_{O_3}(f_3)\rangle& =\delta(f_1+f_2+f_3)\delta^{(3)} \left( \vec{k}_1 + \vec{k}_2 + \vec{k}_3 \right) \nonumber\\
  \quad\quad   \quad\quad   \quad\quad  \times L^3 
   \,\int \Bigg( \prod_{i=1}^3 &  d^3k_i\, 
   \frac{c^{O_i}_{j_i}}{2} {\cal Q}_{a_i b_i}^{j_i} \left( \vec{k}_i \right) e_{a_i b_i,\lambda_i} \left( {\hat k}_i \right)  \Bigg)\,\,\tilde{\cal B}_{\lambda_1\lambda_2\lambda_3}(\vec k_1,\,\vec k_2,\,\vec k_3;\,f_1,\,f_2,\,f_3)\nonumber\\
\quad\quad & =\delta(f_1+f_2+f_3)\,L^3\,\int k_1\,dk_1\,k_2\,dk_2\,k_3\,dk_3\, \nonumber \\
& \quad \times \tilde{\cal B}_{\lambda_1\lambda_2\lambda_3}(\vec k^*_1,\,\vec k^*_2,\,\vec k^*_3;\,f_1,\,f_2,\,f_3)\,{\cal R}^{O_1O_2O_3}_{\lambda_1\lambda_2\lambda_3}(\vec k^*_1,\,\vec k^*_2,\,\vec k^*_3) \;, 
\end{align}
%
where ${\cal R}^{O_1O_2O_3}_{\lambda_1\lambda_2\lambda_3}(k_i,\hat k_i)$ is the  signal
three-point response function computed in Section~\ref{sec-cubicr}.

\subsection{The estimator and the optimal SNR}\label{sec-optest}

Following \cite{Smith:2016jqs} we define a frequency-dependent estimator for the three-point function as
\begin{align}
\hat{\cal F}(f_1,\,f_2,\,f_3)\equiv \sum_{ijk} W^{ijk}(f_1,\,f_2,\,f_3)\,\tilde{s}_{i}(f_1)\,\tilde{s}_{j}(f_2)\,\tilde{s}_{k}(f_3) \;, 
\end{align}
where the filter function $W^{ijk}(f_1,\,f_2,\,f_3)$ is totally symmetric  under simultaneous permutations of both the indices and the arguments, $W^{ijk}(f_1,\,f_2,\,f_3)=W^{jik}(f_2,\,f_1,\,f_3)=...$, and satisfies the reality condition $W^{ijk}(f_1,\,f_2,\,f_3)^*=W^{ijk}(-f_1,\,-f_2,\,-f_3)$. Since only the signals $O=\{A,\,E\}$ are relevant for our analysis, only four filter functions are in principle independent: $W^{AAA}$, $W^{AAE}$, $W^{AEE}$ and  $W^{EEE}$. The frequency integrated estimator $\hat{\cal F}$ reads
\begin{align}
\hat{\cal F}\equiv  \sum_{ijk}\int df_1\,df_2\,df_3W^{ijk}(f_1,\,f_2,\,f_3)\,\tilde{s}_{i}(f_1)\,\tilde{s}_{j}(f_2)\,\tilde{s}_{k}(f_3) \;, 
\end{align}
with expectation value
\begin{align}
\langle\hat{\cal F}\rangle=\sum_{ijk}\int df_1\,df_2\,df_3W^{ijk}(f_1,\,f_2,\,f_3)\,\langle\tilde{s}_{i}(f_1)\,\tilde{s}_{j}(f_2)\,\tilde{s}_{k}(f_3)\rangle \;. 
\end{align}

Under the assumptions that the noise is Gaussian\footnote{This turned out to be the case, for a broad range of frequencies around $1$~mHz, in LISA Pathfinder, and is being considered as a working assumption in the LISA Data Challenge
(Carlos Fern\'andez Sopuerta, private communication).}  (so that its three-point function vanishes) and uncorrelated with the signal, one has $\langle\tilde{s}_{i}(f_1)\,\tilde{s}_{j}(f_2)\,\tilde{s}_{k}(f_3)\rangle=\langle\tilde{\Sigma}_{i}(f_1)\,\tilde{\Sigma}_{j}(f_2)\,\tilde{\Sigma}_{k}(f_3)\rangle$, where $\tilde{\Sigma}_{i}(f)$ is defined in eq.~(\ref{eq:def_tildesigma}). This implies that the expectation value of the estimator is
\begin{align}
\langle\hat{\cal F}\rangle &=\sum_{ijk}\int df_1\,df_2\,df_3W^{ijk}(f_1,\,f_2,\,f_3)\,\langle\tilde{\Sigma}_{i}(f_1)\,\tilde{\Sigma}_{j}(f_2)\,\tilde{\Sigma}_{k}(f_3)\rangle\nonumber\\
&=\sum_{ijk}\int df_1\,df_2\,df_3W^{ijk}(f_1,\,f_2,\,f_3)\,\delta(f_1+f_2+f_3)\,S^{ijk}_{s}(f_1,\,f_2,\,f_3) \;, 
\end{align}
where we have defined the real quantity 
%
%
\begin{align}
S^{ijk}_{s}(f_1,\,f_2,\,f_3)\equiv & \,   L^3\,\sum_{\lambda_1\lambda_2\lambda_3}\int  k_1\,dk_1\,k_2\,dk_2\,k_3\,dk_3\, \nonumber \\
& \quad \times \tilde{\cal B}_{\lambda_1\lambda_2\lambda_3}(\vec k^*_1,\,\vec k^*_2,\,\vec k^*_3;\,f_1,\,f_2,\,f_3)\,{\cal R}^{ijk}_{\lambda_1\lambda_2\lambda_3}(\vec k^*_1,\,\vec k^*_2,\,\vec k^*_3) \;. 
\label{S-ijk}
\end{align}

We next compute the variance of $\hat{\cal F}$ assuming that the signal is noise dominated, with 
\begin{equation}
\left\langle n_i \left( f_1 \right) \,n_j \left( f_2 \right) \right\rangle = \delta_{ij} \, P^i_n \left( \vert f_1 \vert \right) \, \delta \left( f_1+f_2 \right) \,,\qquad \qquad {\rm (no\ sum\ on\ }i{\rm )}\;, 
\label{noise}
\end{equation}
(for brevity, 
from now on 
 we omit the absolute value  in the frequency dependence of the noise) so that
\begin{align}
\langle \hat{\cal F}^2\rangle=&\sum_{ijk}\int df_1\,df_2\,df_3\, P^i_n(f_1)\,P^j_n(f_2)\,P^k_n(f_3) \nonumber \\
& \times \left[6\,W^{ijk}(f_1,\,f_2,\,f_3)\,W^{ijk}(f_1,\,f_2,\,f_3)^*+9\,W^{ijj}(f_1,\,f_2,\,-f_2)\,W^{ikk}(f_1,\,f_3,\,-f_3)^*\right]\,,
\end{align}
%
where the factors of $6$ and $9$ originate from the symmetry properties of $W^{ijk}$. The signal-to-noise ratio (SNR) is given by $\langle\hat{\cal F}\rangle/\sqrt{\langle \hat{\cal F}^2\rangle}$. 

We note that for any given pair of objects $A^{ijk}(f_1,\,f_2,\,f_3)$ and $B^{ijk}(f_1,\,f_2,\,f_3)$ with the properties of our filter function $W^{ijk}$ we can define a scalar product 
\begin{align}\label{eq:def_scalarprodab}
\left(A^{ijk},\,B^{ijk}\right)& =\sum_{ijk}\int df_1\,df_2\,df_3\,  P^i_n(f_1)\,P^j_n(f_2)\,P^k_n(f_3) \nonumber \\ 
 \times & \, \left[6\,A^{ijk}(f_1,\,f_2,\,f_3)\,B^{ijk}(f_1,\,f_2,\,f_3)^*+9\,A^{ijj}(f_1,\,f_2,\,-f_2)\,B^{ikk}(f_1,\,f_3,\,-f_3)^*\right]\,,
\end{align}
%
so that we can write the SNR as
\begin{align}\label{eq:snr_theo}
{\rm SNR}=\frac{1}{6}\frac{\left(W^{ijk},\,\delta(f_1+f_2+f_3)\,\frac{S^{ijk}_{s}(f_1,\,f_2,\,f_3)}{P^i_n(f_1)\,P^j_n(f_2)\,P^k_n(f_3)}\right)}{\sqrt{\left(W^{ijk},\,W^{ijk}\right)}}\,.
\end{align}
In writing this formula we make the hypothesis that $W^{ijk}(0,\,f_2,\,f_3)=0$, whose validity will be checked in short, which implies that $\sum_{ijk}\int df_1\,df_2\,df_3 W^{ijj}(f_1,\,f_2,\,-f_2)\,\delta(f_1+f_2-f_2)\,S^{ikk}_{s}(f_1,\,f_2,\,-f_2)=0$, i.e., the second term in the scalar product~(\ref{eq:def_scalarprodab}) does not contribute to the numerator in eq.~(\ref{eq:snr_theo}). The SNR is thus maximized for 
\begin{align}
W^{ijk}(f_1,\,f_2,\,f_3)\propto \delta(f_1+f_2+f_3)\,\frac{S^{ijk}_{s}(f_1,\,f_2,\,f_3)}{P^i_n(f_1)\,P^j_n(f_2)\,P^k_n(f_3)}\,,
\end{align}
up to a multiplicative constant that cancels out from the expression of SNR. This last equation shows that our  hypothesis $W^{ijk}(0,\,f_2,\,f_3)=0$ is valid, since the noise diverges at low frequencies.

To sum up, for this optimal estimator the SNR is given by
\begin{align}
{\rm {SNR}}=\left[\frac{T}{6}\sum_{ijk}\int df_1\,df_2\,df_3\,\delta(f_1+f_2+f_3)\frac{S^{ijk}_{s}(f_1,\,f_2,\,f_3)^2}{P^i_n(f_1)\,P^j_n(f_2)\,P^k_n(f_3)}\right]^{1/2}\,,
\label{SNR}
\end{align}
where $T$ is the duration of the experiment, the indices $i$, $j$ and $k$ can take only the values $A$ and $E$ and the sum will contain $8$ terms. 

This relation simplifies in the case in which the channels involved in the correlation have the same noise. The noises of the measurement  (\ref{signal-X}) at the vertex
$X$ and at the two other vertices $Y$ and $Z$ of the instrument satisfy 
\begin{equation}
\left\langle n_i \left( f_1 \right)  n_j \left( f_2 \right) \right\rangle = P_n^{ij} \left(  f_1  \right) \delta \left( f_1 + f_2 \right) \;\;,\;\; \left\{ i,\, j \right\} = \left\{ X,Y,Z \right\} \;. 
\end{equation} 
where, due to symmetry\footnote{The cross-correlation arises because the different interferometers share one common arm.}, 
\begin{eqnarray}
&& P_n^{XX} \left(  f  \right) = P_n^{YY} \left(  f  \right) = P_n^{ZZ} \left(  f  \right) \equiv P_{n,{\rm self}} \left(  f  \right) \;, \nonumber\\ 
&& P_n^{XY} \left(  f  \right) = P_n^{XZ} \left(  f  \right) = P_n^{YZ} \left(  f  \right) \equiv P_{n,{\rm cross}} \left(  f  \right) \;. 
\label{noise-XYZ} 
\end{eqnarray} 
The linear combinations (\ref{signal-AET}) have diagonal noise (\ref{noise}), with spectral dependence 
\begin{eqnarray}
&& P_n^A \left(  f  \right) = P_n^E \left(  f  \right) = \frac{2}{3} \left[  P_{\rm n,self} \left(  f  \right) -  P_{\rm n,cross} \left(  f  \right)\right] \equiv \left( 2 L \right)^2 P_n \left( f \right) \;, \nonumber\\ 
&& P_n^T \left(  f  \right) =  \frac{1}{3} \left[  P_{\rm n,self} \left(  f  \right) + 2 \, P_{\rm n,cross} \left(  f  \right) \right] \;, 
\label{sA-sE-sT} 
\end{eqnarray} 
which shows that the $A$ and $E$ channels have indeed identical noise. The factor $\left( 2 L \right)^2$ has been inserted in the definition of $P_n \left( f \right) $ so to convert from time displacement to strain, by dividing $\Delta T$ by the round-trip light travel distance $2 L$ \cite{Cornish:2018dyw}. In this way,  $P_n \left( f \right) $ has the dimension of time. 

Using these two channels, the relation (\ref{SNR}) can be explicitly written as 

%
\begin{eqnarray}  
&& \!\!\!\!\!\!\!\!  \!\!\!\!\!\!\!\!  \!\!\!\!\!\!\!\! 
 {\rm SNR } = \Bigg\{ \frac{T}{6} \int \frac{d f_1  \, d f_2 \, d f_3}{\left( 4 L^2 \right)^3 P_n \left(  f_1  \right) P_n \left(  f_2  \right) P_n \left(  f_3  \right) }  \, \delta \left( f_1 + f_2 + f_3 \right) \Bigg[  S_s^{AAA} \left( f_1 ,\, f_2 ,\, f_3 \right)^2  \nonumber\\ 
&& \quad\quad \quad\quad   
 + 3 \, S_s^{AAE} \left( f_1 ,\, f_2 ,\, f_3 \right)^2 + 3 \, S_s^{AEE} \left( f_1 ,\, f_2 ,\, f_3 \right)^2 + 
S_s^{EEE} \left( f_1 ,\, f_2 ,\, f_3 \right)^2  \Bigg] \Bigg\}^{1/2} \;, 
\\
&=&
\left\{
 \frac{2\,T}{3} \int \frac{d f_1  \, d f_2 \, d f_3}{\left( 4 L^2 \right)^3 P_n \left(  f_1  \right) P_n \left(  f_2  \right) P_n \left(  f_3  \right) }  \, \delta \left( f_1 + f_2 + f_3 \right) \, S_s^{EEE} \left( f_1 ,\, f_2 ,\, f_3 \right)^2
 \right\}^{1/2}
\label{SNR-AE}
\end{eqnarray} 
where we used the results of Section \ref{subsec:propertiesR3}, in which  we found  that for an equal arm LISA configuration $S_s^{AAE}\,=\,-S_s^{EEE}$ and $S_s^{AAA}\,=\,{S_s^{AEE}} 
\,=\,0$.   

\subsection{SNR for a non-Gaussian signal enhanced at a fixed scale}
\label{section-bumpy}
As an explicit example of application of these results, 
in this subsection we evaluate the SNR (\ref{SNR-AE}) for a specific shape of non-Gaussian bispectrum.
 We make the following choice of kernel entering   in eq.~(\ref{f-NL}), depending on a reference  closed triangle $\vec k_i^*$ and
 a reference scale $\bar k$: 
\begin{equation}
K_{\lambda_1;\lambda_2,\lambda_3} \left(\vec k_1^*;  \vec k_2^*,\, \vec k^*_3 \right) = 
{\rm e}^{-\frac{1}{2 \sigma^2} \left[ \left( |\vec k_1^*| - \bar k \right)^2 + \left( |\vec k_2^*| - \bar k \right)^2  +  \left( |\vec k_3^*| - \bar k \right)^2  \right]} 
\delta_{\lambda_1 L} \, \delta_{\lambda_2 L} \, \delta_{\lambda_3 L}   \;. 
\label{bump-ansatz} 
\end{equation} 
where we assume that $\sigma \ll \bar  k$, and 
normalize the shape function to one at its maximum. This localized, chiral
  bump in non-Gaussianity well approximates
   the predictions of certain models of early universe cosmology, 
   for instance the signal obtained in the model of \cite{Namba:2015gja} that we will review in the  subsection \ref{sec:axion}. For later convenience it is also useful to introduce the energy density of the universe per logarithmic wavenumber interval. Using the expression $\rho_{\rm GW} = \frac{M_p^2}{4} \, \left\langle \dot{h}_{ij} \dot{h}_{ij}  \right\rangle $ for the energy density in GW, one finds 
\begin{equation}
\Omega_{\rm GW} \left( k \right) \equiv \frac{1}{3 H_0^2 M_p^2} \, \frac{\partial \rho_{\rm GW}}{\partial \ln k} 
= \frac{\pi^2}{3} \, \frac{k^2}{H_0^2} \, \sum_\lambda P_\lambda \left( k \right) \;.  
\end{equation} 
where $H_0 \simeq 3.24 \times 10^{-18} \, h \, {\rm Hz}$ is the present  value of the Hubble rate.

At this point we can use the kernel function of Eq. \eqref{bump-ansatz} and the expressions for the SNR we developed in Section \ref{sec-estimator}. We relegate  technical steps of the calculations to Appendix
\ref{app:tech}, and write here the final expression for the SNR for an equilateral configuration: 
\begin{eqnarray} 
{\rm SNR } 
&\simeq&  \frac{ f_{\rm NL}^{LLL} }{ 128 \, \pi^2 } \, \frac{\sqrt{2} \pi \, \sigma^2}{\bar{k}^3}   \, P_L^2 \left( \bar{k} \right) 
\, \sqrt{\frac{T}{P_n^2 \left( \bar{k} \right) \, P_n \left( 2 \bar{k} \right)} } \, 
|{\cal R}_{LLL}^{EEE} \left( \bar{k} \, \hat{k}_1^*,\, \bar{k} \, \hat{k}_2^*,\, \bar{k} \, \hat{k}_3^*\right)| \, ,
\label{SNR-bump} 
\end{eqnarray} 
where $\hat{k}_i^*$ denote the unit vector along the direction of $\vec{k}_i^*$ and equilateral transfer functions are plotted in the left panel of Fig.~\ref{fig:BS-eq}. 

 For reference we consider the optimal scale $ \bar{k} L\,=\,0.028$ at which the ratio 
\begin{equation}
   \frac{|{\cal R}_{LLL}^{EEE} \left( \bar{k} \, \hat{k}_1^* ,\, \bar{k} \, \hat{k}_2^* ,\, \bar{k} \, \hat{k}_3^*  \right)|}{ \bar{k}^4 \, P_n \left( \bar{k} \right) \, \sqrt{P_n \left( 2 \bar{k} \right)} } 
 \end{equation} 
 is maximum (for the expressions of the noise power spectrum $P_n$ see \cite{Cornish:2018dyw}). For this value we find $|{\cal R}_{LLL}^{EEE}|\, \simeq \,0.025$ and thus substituting into the previous expression, we find 
%
\begin{eqnarray}
{\rm SNR} &\simeq&  8.9\times 10^{-5}  \, f_{\rm NL}^{LLL}  \, \frac{\sigma^2}{\bar{k}^3}   \, P_L^2 \left( \bar{k} \right)  \, \frac{|{\cal R}_{LLL}^{EEE} \left( \bar{k} \, \hat{k}_1^* ,\, \bar{k} \, \hat{k}_2^* ,\, \bar{k} \, \hat{k}_3^* \right)|}{0.025}  \,  \left\{ \frac{T}{P_n \left( \bar{k}  \right)  P_n \left( \bar{k}  \right) P_n \left( 2 \bar{k}  \right) }  \right\}^{1/2}   \nonumber\\ 
&\simeq& 47 \, f_{\rm NL}^{LLL} \, \frac{\sigma^2}{ \bar{k}^2} \, \left(\frac{3.4  \times 10^{-3} \, {\rm Hz} }{\bar{k}} \right)^5  \, \left( \Omega_{\rm GW} \left( \bar{k} \right) h^2  \right)^2 \, \frac{|{\cal R}_{LLL}^{EEE} \left( \bar{k} \, \hat{k}_1^* ,\, \bar{k} \, \hat{k}_2^* ,\, \bar{k} \, \hat{k}_3^* \right)|}{0.025} \nonumber \\ 
& &\qquad  \frac{7 \times 10^{-41} \, {\rm Hz}^{-1}}{P_n \left(  \bar{k}  \right)} \, \sqrt{\frac{T}{{3} \, {\rm yrs}} \,  \frac{3.9 \times 10^{-41} \, {\rm Hz}^{-1}}{P_n \left( 2 \bar{k} \right)} }     \;. 
 \label{SNR-fNL0} 
\end{eqnarray} 
where in the second line we normalize the various quantities to useful reference values, to be able to more easily appreciate the relevance of the result. For definiteness, we normalize $ \bar{k}$ to $3.4 \times  10^{-3} \, {\rm Hz}$, which is the frequency corresponding to $\bar{k}$; we then normalize the noise functions $P_n \left( k \right)$ and $P_n \left( 2k \right)$ to $7 \times 10^{-41} \, {\rm Hz}^{-1}$ and $3.9 \times 10^{-41} \, {\rm Hz}^{-1}$ respectively, which are (parametrically) the values obtained at $\bar{k}$ (see for instance Fig.~3 of \cite{Cornish:2018dyw}, where the square root of $P_n$ is shown; we note that the noise power spectrum $P_n$ is related to the sensitivity $S_n$ by $P_n = {\cal R}_2 \; S_n$, where ${\cal R}_2$ is the two-point response function shown in Fig.~\ref{fig:PS}.).  We normalize the time to the nominal LISA mission time of $4$ years times a $75\%$ duty factor.

As  discussed  in Section \ref{sec:formalism},  the value of the non-linear parameter entering in this relation is the present one, related to the primordial one by eq.~(\ref{f-NL-primordial}). A detailed discussion of the transfer functions entering in this relation can be found for instance in \cite{Kuroyanagi:2014nba}. The precise behavior is not needed for the present  estimate. When they are well outside the horizon, the GW have constant amplitude; instead the amplitude decreases as the inverse power of the scale factor while it is well inside the horizon. For the present estimate can simply take ${\bf T} \left( t_0 ,\, k \right) = a_k$, leading to 
\begin{eqnarray} 
{\rm SNR } 
& \simeq & 47 \, \frac{f_{\rm NL}^{LLL,{\rm primordial}}}{a_{\bar k}} \, \frac{\sigma^2}{ \bar{k}^2}  \, \left( \frac{3.4 \times 10^{-3} \, {\rm Hz} }{\bar{k}} \right)^5  \, \left( \Omega_{\rm GW} \left(  \bar{k} \right) h^2  \right)^2 \frac{|{\cal R}_{LLL}^{EEE} \left( \bar{k} \, \hat{k}_1^* ,\, \bar{k} \, \hat{k}_2^* ,\, \bar{k} \, \hat{k}_3^* \right)|}{0.025} \nonumber \\
&& \qquad \qquad \qquad  \times \frac{7 \times 10^{-41} \, {\rm Hz}^{-1}}{P_n \left(   \bar{k}  \right)} \, \sqrt{\frac{T}{3 \, {\rm yrs}} \,  \frac{3.9 \times 10^{-41} \, {\rm Hz}^{-1}}{P_n \left( 2  \bar{k} \right)} }     \;. 
\end{eqnarray} 

We recall that, in the case of GWs produced during inflation, $a_{k_*}$ is the value of the scale factor when the mode of frequency $k_*$ re-entered the horizon in the Friedmann-Lema\^itre-Robertson-Walker (FLRW) stage after inflation (we normalize the scale factor to one today). The same relation applies for GW modes that are produced when their size is comparable to the horizon side during radiation domination. Assuming that the universe is radiation dominated at that moment, and that radiation domination continues until the recent stage matter-radiation equality (at $z_{\rm eq} \sim 
3,400$) we find (see Appendix \ref{app:astar}) 
\begin{eqnarray} 
{\rm SNR } \Big\vert_{\rm from \, inflation} 
&\simeq& 3.2 \times 10^{-7} \, f_{\rm NL}^{LLL,{\rm primord}} \, \frac{\sigma^2}{ \bar{k}^2}  \, \left( \frac{3.4 \times 10^{-3} \, {\rm Hz} }{\bar{k}} \right)^4  \, \left( \frac{\Omega_{\rm GW} \left(  \bar{k} \right) h^2 }{10^{-13}} \right)^2 \times  \nonumber \\
&& \frac{|{\cal R}_{LLL}^{EEE} \left( \bar{k} \, \hat{k}_1^* ,\, \bar{k} \, \hat{k}_2^* ,\, \bar{k} \, \hat{k}_3^* \right)|}{0.025}   \frac{7 \times 10^{-41} \, {\rm Hz}^{-1}}{P_n \left( \bar{k} \right)}    \sqrt{\frac{T}{3 \, {\rm yrs}} \,  \frac{3.9 \times 10^{-41} \, {\rm Hz}^{-1}}{P_n \left( 2 \bar{k} \right)} }.  \nonumber\\ 
\end{eqnarray} 
Notice for instance that for $f_{\rm NL}^{LLL,{\rm primordial}} = 10^{3}$ this equation implies that a detection of non-Gaussianities can only take place for $ \Omega_{\rm GW} \left(  \bar{k} \right) h^2 \gtrsim 10^{-11}$. If however, $ \Omega_{\rm GW} \left(  \bar{k} \right) h^2 \simeq 10^{-13}$ this implies that a detection of non-Gaussianities requires at least an $f_{\rm NL}^{LLL,{\rm primordial}} \gtrsim 10^{7}$. As we discuss in Appendix  \ref{app:astar}, modes that have presently the frequency $\bar{k} = 3.4 \times 10^{-3} \, {\rm Hz}$ re-entered the horizon 
at the temperature of $\sim 50 \, {\rm TeV}$. A different value for $a_*$ can be obtained if we make instead the unconventional hypothesis that the universe had a different equation of state than radiation, before, but close, to Big-Bang Nucleosynthesis. 

A different value for $a_*$ is instead obtained if we assume that GW are produced inside the horizon by a sudden episode that took place during radiation domination at the temperature $T_*$. In this case we find  (see Appendix \ref{app:astar}) 
\begin{eqnarray} 
{\rm SNR } \Big\vert_{\rm  prod. \; inside \; horizon } 
&\simeq& 1.8 \times 10^{-10} \, f_{\rm NL}^{LLL,{\rm primordial}} 
\, \frac{T_*}{100 \, {\rm GeV}}  \, \frac{\sigma^2}{\bar k^2} \, \left( \frac{3.4 \times 10^{-3} \, {\rm Hz} }{\bar{k}} \right)^5  \, \times \nonumber \\
&& \left( \frac{\Omega_{\rm GW} \left(  \bar{k} \right) h^2 }{10^{-13}} \right)^2 \times \frac{|{\cal R}_{LLL}^{EEE} \left( \bar{k} \, \hat{k}_1^* ,\, \bar{k} \, \hat{k}_2^* ,\, \bar{k} \, \hat{k}_3^* \right)|}{0.0158}  \times  \nonumber \\
&&  \frac{7 \times 10^{-41} \, {\rm Hz}^{-1}}{P_n \left(  \bar{k} \right)} \, \sqrt{\frac{T}{3 \, {\rm yrs}} \,  \frac{3.9 \times 10^{-41} \, {\rm Hz}^{-1}}{P_n \left( 2  \bar{k} \right)} }     \;. 
\end{eqnarray} 
The previous result is recovered if we take $T_* = 50 \, {\rm TeV}$. 

In both these relations, $\Omega_{\rm GW} \left( \bar k\right) h^2$ has been normalized to the threshold level that can be detected in the 2-point correlation function at LISA. Finally, we recall that these results assume $\bar k L = 0.028 $, as this scale maximizes the SNR for a flat spectrum $\Omega_{GW}(k)$. In order to evaluate the SNR at arbitrary scale, all the $k$ dependent quantities (in particular $| {\cal R}_{LLL}^{EEE} |$, and $\Omega_{GW}(k)$ if it is scale dependent) must be evaluated consistently.

\section{Tensor non-Gaussianity and Early Universe Cosmology}\label{sec-mod}

In the previous sections, we developed
 techniques for investigating the  non-Gaussianity of
a SGWB by studying the three-point correlation function of 
signals detectable  with LISA. 
 We have shown that
 LISA is in principle able to
   measure specific properties of   the SGWB bispectra, as 
   the dependence on the amplitude of the momentum and polarization
   of the GW signal.
    
 In this  section we review and discuss the current theoretical understanding 
 of non-Gaussian features of a primordial SGWB sourced
 by   Early Universe physics, in particular   inflation and cosmological defects. Our
 aim is to provide  theoretical motivations  for the  results derived in  the previous Sections, showing that 
 they can be used   to distinguish different
 sources for primordial 
  SGWBs. Besides analyzing models, we also briefly review current and perspective constraints
  on tensor non-Gaussianity from the physics of CMB, that is able to probe primordial SGWBs at
  frequency scales much smaller than interferometers.

\subsection{Primordial tensor non-Gaussianity and inflationary physics}
Cosmological inflation predicts the existence of a stochastic background of tensor modes, produced by 
quantum fluctuations of the metric spin-2 degrees freedom during the phase of inflationary
expansion. CMB experiments  constrain the amplitude of the primordial SGWB power spectrum  at large 
CMB scales in terms of the tensor-to-scalar ratio $r$:  the current upper bound from BICEP2/Keck and Planck
 is $r\,<\,0.07$ at 95\% confidence level 
 \cite{Array:2015xqh,Ade:2015lrj} (assuming the consistency relation $r=-8 n_{T}$), 
 and future CMB polarization experiments \cite{Hazumi:2012gjy,Finelli:2016cyd,Abazajian:2016yjj} can lower this bound down to around $10^{-3}$ in
 absence of a detection.

Tensor  self-interactions 
 are expected to make  the non-Gaussianity  of  the primordial SGWB relatively large, already within Einstein
 gravity. 
  Moreover, as we shall review next, there are models of inflation which exploit specific couplings between fields present
  during the inflationary era, in order to 
    enhance the amplitude of the  tensor power spectrum at  scales
   that can be probed by the   future generation of gravitational interferometers. 
   This gives the opportunity to probe inflation  at scales much smaller than
   CMB scales: in these scenarios, the large couplings 
   among the fields involved 
   can enhance tensor non-Gaussianity, 
  making
   it a very useful observable for distinguishing among different scenarios.  
We briefly survey the topic of inflationary tensor non-Gaussianity, considering in succession models of
   increasing complexity. 
   
   \subsubsection*{General Relativity in pure de Sitter space}

The simplest situation to investigate -- as a toy 
model for inflation --  is pure General Relativity  (GR)  in de Sitter space.  
 The  metric  for de Sitter space  can be expressed as $d s^2\,=\,- d t^2+ e^{2 H t} d \vec x^2$, with $H$ the constant Hubble
 parameter. 
  As we have seen in Section \ref{sec:formalism}, the transverse-traceless spin-2 
  tensor fluctuations can be decomposed in two 
  helicity modes, $\lambda=L$ (left) and $\lambda=R$ (right). 
  The Einstein-Hilbert action 
 \be
S_{\rm EH}\,=\,\int d \tau\,d^3 x\,\sqrt{-g}\,\left[ 
\frac{M_{Pl}^2}{2} \,R-\,\Lambda 
\right] \;, 
 \ee
 can be straightforwardly expanded  around the de Sitter background up to third order
 in fluctuations \cite{Maldacena:2002vr,Maldacena:2011nz}. 
 From the  second order action one can compute 
    the two-point function for the Fourier transform of the tensor fluctuations, which  defines the primordial
     tensor power spectrum 
    \bea
    \left\langle h_\lambda \left( t ,\, \vec{k} \right) \,  h_{\lambda'} \left( t ,\, \vec{k}' \right) \right\rangle  &=& \frac{P_\lambda \left( k \right)}{4 \pi k^3} \, 
\delta_{\lambda \lambda'} \, \delta^{(3)} \left( \vec{k} + \vec{k}' \right)  \;, 
\label{PSin1}
\\
P_\lambda (k)&=& \frac{1}{\pi^2}\left(\frac{H}{M_{Pl}} \right)^2\,.\hskip1cm 
\label{PSin2}
   \eea
From the third order action one obtains the three-point function for the fluctuations  \cite{Maldacena:2002vr,Maldacena:2011nz}, which defines the primordial tensor bispectrum 
 \bea
 \left\langle {\hat h}_{\lambda_1} \left( t  ,\, \vec{k}_1 \right)  {\hat h}_{\lambda_2} \left( t  ,\, \vec{k}_2 \right) 
{\hat h}_{\lambda_3} \left( t  ,\, \vec{k}_3 \right) \right\rangle  =   \delta^{(3)} \left( \vec{k}_1+\vec{k}_2+\vec{k}_3 \right)  {\cal B}_{\lambda_1,\lambda_2,\lambda_3} \left(\vec  k_1 , \vec k_2 , \vec k_3 \right) \,, 
\label{eq:BS0a}
 \eea
 and  can be obtained using for example the in-in formalism.
 The tensor bispectrum is the lowest order statistics providing information on 
 non-Gaussianity of tensor fluctuations.
  Its  dependence on the three wave-vectors  characterizes  the shape of tensor non-Gaussianity, one of the properties
  that
   allows one to distinguish among different models. 
 We list here some properties of the tensor bispectrum and the corresponding tensor non-Gaussianities for 
 GR  in de Sitter space: 

\begin{enumerate}
\item The amplitude of tensor non-Gaussianity -- parameterized by the 
ratio between the tensor bispectrum and the square of the tensor power spectrum -- are of order one in pure GR (but not larger). For example, one finds for equilateral
configurations 
\cite{Agrawal:2018gzp} 
\be \label{amprrr}
\frac{| {\cal B}_{RRR}|
}{P_R^2}\,\simeq\,3.6\,.
\ee
This is a difference with respect to  scalar curvature fluctuations in single field vanilla models of inflation, where
scalar  non-Gaussianities are suppressed by slow-roll parameters, and are then at most of order a few percent \cite{Maldacena:2002vr,Salopek:1990jq,Gangui:1993tt,Acquaviva:2002ud}.
This is due to the fact that 
Einstein gravity is a non-linear theory, and cubic interactions are not suppressed with respect to quadratic 
ones by small (e.g. slow-roll) parameters.
 In Einstein gravity the shape of tensor bispectra is peaked in squeezed configurations. 
\item The amplitude of tensor bispectra depends on 
chirality, that is on 
the polarization indexes $L,R$, and in general bispectra
 characterized by  different indexes have different amplitudes. For example, in the case of GR in de Sitter space one finds 
  \cite{Agrawal:2018gzp,Gao:2011vs}
 \be \label{grdifc}
|{\cal B}_{RRL}|
 \,=\,\frac{ |{\cal B}_{RRR}|
 }{81}\,.
 \ee
  This is different with respect to the power spectrum, where $L$ and $R$ modes have the same
  amplitude (see eq.~\eqref{PSin2}) and the cross correlation  vanishes, 
  $\left\langle h_L  \,  h_{R}  \right\rangle =0$. This has interesting phenomenological consequences
  since we have many independent bispectrum components
  we can use 
  to build  observables for distinguishing  among different models of inflation. As we have
  learned in Section \ref{sec-signal}, LISA can in principle probe
  different chiral components of primordial bispectra.
  \item Pure Einstein gravity around de Sitter space {\it preserves parity}, in the sense that the amplitudes of power spectrum and bispectrum
  components obtained interchanging all the $L,R$ indexes is the same. For example, $P_L(\vec k)\,=\,P_R(-\vec k)$, ${\cal B}_{LLL}(\vec k_i) = {\cal B}_{RRR}(-\vec k_i)$,
  ${\cal B}_{LLR}(\vec k_i)\,=\,{\cal B}_{RRL}(-\vec k_i)$ etc. This is potentially not true in more complex inflationary scenarios, as we are going to discuss in the following subsection.
  
\end{enumerate}

We can go beyond a pure Einstein-Hilbert action, and investigate non-Gaussian tensor fluctuations around de Sitter 
space in theories of gravity  including higher order curvature invariants \cite{Maldacena:2011nz,Soda:2011am,Shiraishi:2011st},  schematically denoted as  $\int \,W^3$ and 
$\int \tilde{W} W^2$ ($W_{\mu\nu\rho\sigma}$ being the Weyl tensor). Such perturbative contributions 
 modify the third-order tensor action,  and lead to new, parity preserving  shapes for the tensor bispectra. However,
 their amplitudes can not be (much) larger than the ones one finds within General Relativity,  due to unitarity constraints. 

   \subsubsection*{Single field slow-roll inflation}

An enhanced amplitude of tensor bispectra can be obtained in single field inflation, with
the time variable controlled by a scalar field, the inflaton. 
The power spectrum of tensor modes in single-field slow-roll models of  inflation is straightforward to compute, and one finds
the same result as in de Sitter space, but with an additional scale dependence. In the simplest models of inflation
 the tensor spectrum is red, with a spectral tilt 
given by
\be \label{redsp}
n_{T}\,=\,-r/8\,,
\ee
with $r$ the tensor-to-scalar ratio~\footnote{Equation \eqref{redsp} holds at leading order in slow-roll  for   inflationary 
 models with unit tensor sound speed. See \cite{Guzzetti:2016mkm} for a systematic discussion
of more general scenarios that can  change  this relation. 
}.
The analysis of the tensor bispectrum is instead less straightforward:
 the scalar can have non-minimal derivative
couplings with the metric -- see e.g. the model  \cite{Kobayashi:2011nu} built in terms of the Horndeski theory  -- which can
 complicate the analysis. 
  The complete classification of tensor non-Gaussianity in
 single field models of inflation based on Horndeski scalar-tensor
  theories of gravity has been done in \cite{Gao:2011vs},  finding that besides the Einstein-Hilbert part
\cite{Maldacena:2002vr},
 the third order action for tensor modes acquires a parity-preserving contribution proportional to the cube 
  of time derivatives of tensor fluctuations:
  \be \label{s3new}
  S_{(3)}^{new}\,=\,\int d t\,d^3 x\,G(a,\,\phi)\, \dot h _{ik} \dot h_{kl} \dot h_{li} \,,
  \ee
  where $G$ is a function of the scale factor and of the homogeneous profile of the  scalar field and its first time derivatives (it is related with the function  $G_5$ controlling the quintic Horndeski action). Interestingly, there are no obvious unitarity constraints on the
  size of this correction to the Einstein-Hilbert action, hence -- depending on the inflationary model and the profile for the scalar
  field -- this contribution can lead to a tensor bispectrum whose amplitude is parametrically larger than the GR result (although it might be hard to find an explicit model satisfying all the CMB constraints in the scalar sector).  The shape
  of the tensor bispectrum  has been investigated in  \cite{Gao:2011vs},  finding that it is maximized for equilateral configurations, hence
  it is different from the GR one, maximized in the squeezed limit.
   In the equilateral limit, the bispectrum components associated with
  action \eqref{s3new} satisfy the relation
   \be
   |B^{(new)}_{RRL}(k, k,k)|\,=\,\frac{|B^{(new)}_{RRR}  (k, k, k)|}{9}.
  \ee
  This is different from the GR result of eq \eqref{grdifc}:  we learn that the chiral structure of the tensor bispectra
   -- i.e. a distinctive hierarchy for the amplitudes of the tensor bispectra depending on the polarization indexes $L,\,R$ -- might
 be used to build observables to distinguish among different models.  
  
  Tensor non-Gaussianity in single field models of inflation based on generalization of Horndeski actions \cite{Zumalacarregui:2013pma,Gleyzes:2014dya,Langlois:2015cwa,Langlois:2015skt,Crisostomi:2016czh,BenAchour:2016fzp} 
  is a topic still under investigation, see e.g.~\cite{Akita:2015mho}, and also \cite{Creminelli:2014wna, Bordin:2017hal} for an analysis carried on using techniques based
  on effective field theory for inflation. Also,
  tensor non-Gaussianity in parity-violating  scalar-tensor theories 
  \cite{Lue:1998mq,Jackiw:2003pm} 
     that spontaneously break Lorentz  invariance \cite{Crisostomi:2017ugk} is particularly interesting to motivate a search
     of  parity 
   violating effects in the tensor bispectrum: an analysis
    of the tensor bispectrum in  parity violating gravitational theories can be found in \cite{Soda:2011am,Shiraishi:2011st,Bartolo:2017szm}, 
    although their
    generalization to the extended set-up introduced  in \cite{Crisostomi:2017ugk} 
    is still an open question. Finally an important contribution to the non-Gaussian SGWB  could be induced by primordial second order scalar perturbations~\cite{Mollerach:2003nq,Saito:2008jc,Baumann:2007zm, Ananda:2006af}.

     \subsubsection*{Beyond single field inflation}

Since the simplest  single field models of inflation predict a red tensor tilt, the amplitude of the primordial SGWB is too small to be detected
with interferometers.  On the other hand, 
there are more complex inflationary scenarios that might allow one to enhance the tensor  spectrum at interferometer scales \cite{Bartolo:2016ami,Guzzetti:2016mkm}, and that may
lead to  large  tensor non-Gaussianity with
distinctive features. 
 We briefly discuss here two frameworks, whose predictions for the tensor bispectrum have been studied so far:

\bigskip

      \noindent
      {\it Coupling the inflaton  with additional scalars and vector fields}

\bigskip

\noindent
If other sources of GWs are present during inflation, the primordial SGWB can have richer features. A possibility
to source primordial GWs is to couple fields driving inflation with
 additional scalars \cite{Cook:2011hg,Senatore:2011sp,Carney:2012pk,Biagetti:2013kwa,Biagetti:2014asa,Goolsby-Cole:2017hod}, 
 $U(1)$ gauge vectors~\cite{Sorbo:2011rz,Anber:2012du,Barnaby:2010vf,Barnaby:2012xt},  
 non-Abelian vector fields~\cite{Maleknejad:2011jw, Dimastrogiovanni:2012ew, Adshead:2013qp, Adshead:2013nka, Obata:2014loa, Maleknejad:2016qjz, Dimastrogiovanni:2016fuu, Agrawal:2017awz, Adshead:2017hnc, Caldwell:2017chz,Agrawal:2018mrg}, or Standard Model fields~\cite{Espinosa:2018eve}. 
 These mechanisms  usually exploit  instabilities for the additional source fields during inflation.
 Such instabilities feed in the evolution of tensor modes through higher order  contributions to the anisotropic stress, 
and affect both  power spectra and bispectra of fluctuations: see e.g. \cite{Namba:2015gja,Peloso:2016gqs,Garcia-Bellido:2016dkw,Shiraishi:2016yun}. Given the nature of the couplings, the resulting tensor
power spectrum and bispectra can be parity violating, and the bispectra are usually enhanced in equilateral
 configurations.
   The tensor power spectrum profile can acquire a feature (a `bump' profile)
   and can  also
get sufficiently enhanced at small scales to be detected with interferometers \cite{Bartolo:2016ami}. Models involving couplings
between  (pseudo)scalars and gauge fields  are theoretically well understood, and  their observational prospects 
in CMB polarization  and interferometer experiments are being
developed in great detail, in particular for what respect the parity violating features  of power spectra and bispectra, see
 e.g. the recent papers \cite{Thorne:2017jft,Agrawal:2017awz} and references  therein.

\bigskip

      \noindent
      {\it Breaking space-time symmetries during inflation}

\bigskip

\noindent
Another possibility for enhancing the bispectrum of 
tensor modes   is to break space-time symmetries during inflation. One
way to do so are scenarios of (super)solid inflation: see e.g.
 \cite{Endlich:2012pz,Bartolo:2015qvr,Ricciardone:2016lym,Ricciardone:2017kre,Domenech:2017kno,Ballesteros:2016gwc,Cannone:2015rra,Lin:2015cqa,Cannone:2014uqa,Akhshik:2014bla}, where the vacuum expectation value  of additional scalar fields spontaneously breaks space diffeomorphisms during inflation,
 or more in general
 in models where spatial diffeomorphisms
 are broken spontaneously during inflation.  Tensor bispectra  are maximized  in squeezed configurations,
and their amplitude can be parametrically larger than in Einstein gravity. In the scenarios
explored so far in the literature, power spectrum and bispectrum  components  are all parity preserving.

Alternatively, space-time symmetries can be broken {\it explicitly} as for example a violation of Lorentz symmetry in Ho\v rava-Lifshitz gravity: in this case inflationary tensor fluctuations are automatically chiral \cite{Takahashi:2009wc} and the
tensor bispectra can be chiral and  enhanced with respect
to GR results \cite{Zhu:2013fja,Huang:2013epa}.

For these systems, some additional mechanisms (as growth of perturbations due to instabilities)
might be necessary to amplify the tensor spectrum at interferometer scales.

\subsection{An explicit example: tensor non-Gaussianity and axion  inflation}
\label{sec:axion}

After the general survey of the previous subsection,
we describe in detail the predictions of a concrete model able to produce a signal detectable at interferometers with large
tensor non-Gaussianity.  We assume that during inflation, which is 
driven by a scalar $\phi$,  an auxiliary axion field $\chi$ interacts with a gauge field $A_\mu$: the Lagrangian 
density for the system is 
\begin{equation}\label{int}
\mathcal{L} = - \frac{1}{ 2 }  \left( \partial \phi \right)^2 - V \left( \phi \right) - \frac{1}{4} F_{\mu \nu}F^{\mu \nu} - 
\frac{\chi}{f}\,F_{\mu\nu}\,\tilde{F}^{\mu\nu}\,,
\end{equation}
where $1/f$ is a coupling constant with the dimension of a length, and $F_{\mu \nu}$ ($\tilde F^{\mu \nu}$) is the  field (dual) strength tensor.
 GW non-Gaussianities in different helicity channels become larger than those of the scalar perturbations by different orders of magnitudes.  This can be understood easily noticing that both scalar and tensors are sourced with similar efficiency (and source) but the scalars are suppressed by the helicity conservation, while the tensors can be large and also chiral, meaning that the contribution of the different  correlator are different for different combinations of the helicity.\\
In such a model the gauge field $A_\mu$ acts as a source for both scalar and tensor metric perturbations, and, for the latter, this translates into a source term in the equation of motion of $h_{ij}$
\begin{equation}\label{eqh}
h_{ij}''+2\,\frac{a'}{a}\,h_{ij}'-\Delta\,h_{ij}=\frac{2}{M_P^2}\,\Pi_{ij}{}^{lm}\,T^{EM}_{lm}\,,
\end{equation}
where $\Pi_{ij}{}^{lm}=\Pi^i_l\,\Pi^j_m-\frac{1}{2}\Pi_{ij}\,\Pi^{lm}$ is the transverse traceless projector, with $\Pi_{ij}=\delta_{ij}-\partial_i\,\partial_j/\Delta$ and where $T^{EM}_{lm}$ represents the spatial part of the stress-energy tensor of the gauge field.
The source term arises not from the $F\tilde{F}$ term but from the $FF$ term in eq.~(\ref{int}). Parity-violating information of the gauge field is transmitted through this source term to the tensors that acquire a chiral component. The GW solution for such equation can be found using the Green function method, see~\cite{Cook:2013xea,Garcia-Bellido:2017aan}.

Using the same procedure as described in~\cite{Sorbo:2011rz,Cook:2013xea} we can compute both the two-point function and the three-point function at super-horizon scales ($k\tau\rightarrow 0$)  for the tensor modes. 
The GW two-point function receives contributions from the parity conserving amplification of vacuum fluctuations and from  the excited electromagnetic modes which source the parity violating parts. These two contributions are uncorrelated and the overall right- and left-handed power spectra are
\begin{eqnarray}
P_R&=&\frac{H^2}{\pi^2 M_{Pl}^2}\left(1+8.6\times10^{-7}\frac{H^2}{M_{Pl}^2}\frac{e^{4\pi\xi}}{\xi^{6}}\right) \,, \nonumber\\
P_L&=&\frac{H^2}{\pi^2 M_{Pl}^2}\left(1+8.6\times10^{-9}\frac{H^2}{M_{Pl}^2}\frac{e^{4\pi\xi}}{\xi^{6}}\right)\,, 
\end{eqnarray}
where the parameter $\xi$ encodes the velocity of the field $\chi$, $\xi = \dot \chi / (2 f H)$.
For the three-point function, at late times ($k\tau\rightarrow 0$),  we have 
\begin{eqnarray}
\label{3ptten}
\langle h_{\lambda_1}({\vec{k}_1})\,h_{\lambda_2}({\vec{k}_2})\,h_{\lambda_3}({\vec{k}_3})\rangle&=&
4\,\delta^{(3)}({\vec{k}_1}+{\vec{k}_2}+{\vec{k}_3})\int d^3{q} \,F_{\lambda_1}({\vec{k}_1},\,{\vec{q}})\,F_{\lambda_2}({\vec{k}_2},\,-{\vec{q}})\,F_{\lambda_3}({\vec{k}_3},\,{\vec{q}}-{\vec{k}_1})\nonumber\\&& +({\vec{k}_2}\leftrightarrow{\vec{k}_3})\,,
\end{eqnarray}
where the function $F^{lm}_{\lambda}({\vec{k}},\,{\vec{q}})$ is the  same as in~\cite{Cook:2013xea}. The largest contribution to the three-point function of the GW, given by the $\langle h_{R} h_{R} h_{R}\rangle$ correlator,  was already computed in~\cite{Cook:2013xea, Shiraishi:2013kxa}, where was also shown that the shape of such a bispectrum peaks in the equilateral limit ($|{\bf{k}_1}|=|{\bf{k}_2}|=|{\bf{k}_3}|=\,k$). Here we extend such computations to estimate the amplitude of the other mixed contributions to the GW bispectrum, which are
\begin{eqnarray}
\langle {h}_R({\vec{k}}_1^*)\,{h}_R({\vec{k}}_2^*)\,{h}_R({\vec{k}}_3^*)\rangle_{\rm {equil}}&\simeq& 5.7\times 10^{-10}\,\frac{H^6}{M_{Pl}^6}\,\frac{e^{6\pi\xi}}{\xi^9}\,\frac{\delta^{(3)}({\vec{k}_1^*}+{\vec{k}_2^*}+{\vec{k}_3^*})}{k^6}\,,\nonumber\\
\langle {h}_L({\vec{k}}_1^*)\,{h}_R({\vec{k}}_2^*)\,{h}_R({\vec{k}}_3^*)\rangle_{\rm {equil}}&\simeq& 9.0\times 10^{-12}\,\frac{H^6}{M_{Pl}^6}\,\frac{e^{6\pi\xi}}{\xi^9}\,\frac{\delta^{(3)}({\vec{k}_1^*}+{\vec{k}_2^*}+{\vec{k}_3^*})}{k^6}\,,\nonumber\\
\langle {h}_L({\vec{k}}_1^*)\,{h}_L({\vec{k}}_2^*)\,{h}_R({\vec{k}}_3^*)\rangle_{\rm {equil}}&\simeq& -1.9\times 10^{-15}\,\frac{H^6}{M_{Pl}^6}\,\frac{e^{6\pi\xi}}{\xi^9}\,\frac{\delta^{(3)}({\vec{k}_1^*}+{\vec{k}_2^*}+{\vec{k}_3^*})}{k^6}\,,\nonumber\\
\langle {h}_L({\vec{k}}_1^*)\,{h}_L({\vec{k}}_2^*)\,{h}_L({\vec{k}}_3^*)\rangle_{\rm {equil}}&\simeq& 1.2\times 10^{-14}\,\frac{H^6}{M_{Pl}^6}\,\frac{e^{6\pi\xi}}{\xi^9}\,\frac{\delta^{(3)}({\vec{k}_1^*}+{\vec{k}_2^*}+{\vec{k}_3^*})}{k^6}\,.
\end{eqnarray}
Note that the signs and reality condition are specific to our choice of chiral polarization operators obeying the properties in eq.~(\ref{e-properties}).

We find that the mixed helicity components are non-zero and in Section \ref{sec-estimator} we have seen how LISA is sensitive to such components.  The shape of the GW
three-point functions is close to equilateral and this can be explained by the fact that the mechanism of generation of perturbations happens at sub-horizon scales.

\subsection{Primordial tensor non-Gaussianity at CMB scales }

To put our investigation in a wider context, we discuss in this subsection the current and perspective
constraints on tensor non-Gaussianity from CMB physics.
CMB observations  at large scales impose stringent constraints 
on the non-Gaussianity of scalar curvature (density) perturbations,  and have the potential
to constrain tensor non-Gaussianity as well, but at frequencies much smaller than interferometer
scales. The tightest constraints come from the {\it Planck} collaboration~\cite{Ade:2015ava}, setting strong limits on scalar non-Gaussianity for a variety of well-motivated models. For example, for the ``standard"  local, equilateral and orthogonal shapes the analysis of {\it Planck} temperature (T) and E-mode polarization data constrain the level of these types of non-Gaussianity to $f_{NL}^{local}=0.8\pm5.0$, $f_{NL}^{equil}=-4\pm 43$ and $f_{NL}^{ortho}=-26\pm 21$ ($68\%$ CL, statistical) from the measurements of the angular CMB bispectrum.  These values point into the direction of consistency with the predictions of 
the standard cosmological model based on single-field models of slow-roll inflation, meaning that the structure that we observe today have been sourced by (almost) Gaussian seed perturbations.
However there are also much less studied observables that could open up a new window into the physics of the early universe in the near future. Among these there are non-Gaussian signatures  
coming from tensor (gravitational waves) 3-point correlations, and also non-Gaussianity coming from mixed (tensor-scalar) 3-point correlators. Interesting bounds on these kind of signals already exist.  
A first observational limit on the tensor non-linearity parameter $f_{NL}^{tens}$ has been obtained by {\it Planck} (from temperature data): $f_{NL}^{tensor}/10^2= 4\pm15\; (68\%CL)$, which is consistent with a previous WMAP data analysis~\cite{Shiraishi:2014ila}. The tensor bispectra analyzed derive from some models of inflation predicting a specific parity violation in the tensor sector. 
However, future observations of CMB will try to improve constraints on tensor fluctuations focusing on B-mode polarization.  The reason is that in this case primordial tensor modes are not hidden by scalar perturbations. 
The main contamination to the B mode polarization signal is polarized dust~\cite{Flauger:2014qra} and lensing of the $E$-modes to $B$-modes~\cite{Seljak:2003pn}. However, precise B-mode measurements have just started and the prospects of improvement are large. For example the constraints on the $\langle BTT \rangle$ CMB bispectrum arising from mixed correlators tensor-scalar-scalar can improve by an order of magnitude with a CMB-Stage IV mission w.r.t the ones achievable from temperatura data alone (see, e.g.,~\cite{Meerburg:2016ecv}). Observational constraints on such a primordial tensor-scalar-scalar bispectrum already exist and have been obtained recently in \cite{Shiraishi:2017yrq} using WMAP temperature data finding a constraint $g_{tss}=-48\pm28$ ($68\%$CL) on the amplitude of such interactions. The inclusion of polarization information will certainly improve such a constraint.
On the other hand it might be crucial to have the possibility to probe such signals on different scales, like the ones which are probed by direct measurements from interferometers. This might serve not only as a possible consistency check, but also to test models of inflation that eventually produce a relevant bispectrum signature only at interferometric scales.

\subsection{Other early Universe mechanisms sourcing gravitational waves}

So far in Section~\ref{sec-mod}, we have considered GW backgrounds from inflation which can lead to a significant deviation from Gaussian statistics. Inflation, however, is not the only source of GWs from the early Universe. Various post-inflationary mechanisms can be also responsible for the generation of GW backgrounds having today a large amplitude. Non-linear field dynamical processes after inflation can generate non-trivial quadrupolar field distributions, resulting in an active and efficient source of GWs. The most representative processes are non-perturbative particle production~\cite{Easther:2006gt,GarciaBellido:2007dg,GarciaBellido:2007af,Dufaux:2007pt,Dufaux:2008dn,Dufaux:2010cf,Figueroa:2017vfa,Adshead:2018doq}, strong first order phase transitions~\cite{Kamionkowski:1993fg,Caprini:2007xq,Huber:2008hg,Hindmarsh:2013xza,Hindmarsh:2015qta,Caprini:2015zlo,Cutting:2018tjt} and cosmic defect networks~\cite{Vachaspati:1984gt,Sakellariadou:1990ne,Damour:2000wa,Damour:2001bk,Damour:2004kw,Figueroa:2012kw,Blanco-Pillado:2017oxo}. For a recent review on early Universe GW sources see~\cite{Caprini:2018mtu}.
 
The stochastic background of GWs from strong first order phase transitions is expected to exhibit a single ``bump'' spectrum, with a rather large amplitude, depending on the efficiency of the dynamical processes involved in the emission of GWs. In particular, if the electroweak phase transition is sufficiently strong due to the presence of beyond the Standard Model particle 
physics, the peak frequency of the associated GW background lies naturally within the LISA frequency window. For a discussion on the ability of LISA to measure a GW background from first order phase transitions (and in particular from the electroweak phase transition) see~\cite{Caprini:2015zlo,Caprini:2018mtu}. The stochastic background of GWs from non-perturbative phenomena like periodic~\cite{Figueroa:2013vif,Figueroa:2017vfa}, tachyonic~\cite{Dufaux:2008dn,Dufaux:2010cf} or other~\cite{Dufaux:2009wn,Adshead:2018doq, Tranberg:2017lrx} excitation of fields (typically expected in preheating), is also characterized by a single ``bump'' spectrum, with an amplitude that can be typically large but peaked at very high frequencies (unless extreme fine tuned couplings are considered). In both cases, phase transitions and preheating, the production of GWs is due to a causal process. This implies that today's background is formed by the superposition of billions of independent signals (corresponding to the GWs emitted from the many uncorrelated regions at the time of the background generation), and hence it must have necessarily a Gaussian distribution, due to the central limit theorem. The angular resolution needed to probe any of the individual signals, beyond their effective stochastic Gaussian nature, is far beyond the reach of any reasonable GW detector. GW backgrounds from preheating and phase transitions are therefore expected to be highly Gaussian, and hence their 3-point function is expected to vanish at the data of any interferometer. For a discussion on this aspect, see Section 3.1 of Ref.~\cite{Caprini:2018mtu}.

The case of GW emission from cosmic defect networks is however very different. The GW signal expected from cosmic defects can be the sum of two components. The first component is the emission of GWs produced around the horizon at each time $t$, by the anisotropic stress of the network~\cite{Krauss:1991qu,JonesSmith:2007ne,Fenu:2009qf,Adshead:2009bz,Figueroa:2012kw}. This first component is expected to be emitted by any network of cosmic defects in $scaling$, independently of the topology and origin of the defects~\cite{Figueroa:2012kw}. The second contribution is only expected for a network of cosmic strings -- local gauge one-dimensional topological defects -- and is given by the superposition of GWs emitted from sub-horizon strings chopped off from the string network all along cosmic history, as well as from super-horizon strings with superimposed small-scale structure~\cite{Sakellariadou:1990ne} due to string interconnections  leading to kink formation. For a cosmic string network, this second contribution is expected to be the dominant one. However, the argument of causality discussed above, remains valid, even if the GWs are continuously emitted during several Hubble times.
Thus, this background cannot be resolved beyond its stochastic nature, and is expected to be Gaussian\footnote{In reality, on top of the continuous stochastic Gaussian background from cosmic strings, there can be individual bursts emitted by nearby strings or a ``popcorn'' discontinuous noise. The former results from the fact that at small redshifts there are not many sources and events do not overlap, thus the time interval between events is longer than the duration of a single event. The latter is a noise generated by rare bursts at high redshifts leading to unresolved GW sources. Considering the number of sources at a given frequency as a Poisson process, one may thus expect either to get a signal or to get a superposition of signals, leading to what is called a popcorn-like noise~\cite{Regimbau:2011bm}. These signals due to bursts represent therefore, in a sense, a temporal deviation from Gaussianity, that can be measured from the two-point function. However they do not correspond to the type of non-Gaussianity that we discuss in this paper, as they do not form a continuous stochastic background.}. We therefore expect that any non-Gaussianity in the continuous stochastic background sourced by a defect network, can only be due to the first contribution (even if this is sub-dominant in the case of cosmic strings, in terms of the power spectrum). 

Let us thus focus on the first contribution mentioned above, produced by the anisotropic stress of a defect network. This contribution represents an irreducible emission of GWs sourced by any type of  viable defect network that reached scaling, meaning that the energy density in the defect network is constant fraction of the overall energy density of the universe. Such defects can be either one-dimensional topological defects --- global, Abelian, non-Abelian, semi-local strings --- or (viable) global defects  --- domain walls, monopoles or non-topological textures~\cite{Figueroa:2012kw}. Indeed,  in the case of local strings, scaling is reached by the emission of gravitational waves.  Global defects reach scaling mainly through  long-range interactions and emission of Goldstone bosons~\cite{Vilenkin:2000jqa}. A defect network in scaling produces an irreducible background of GWs emitted continuously around the horizon scale at every moment of the cosmic history. Its spectrum is predicted to be exactly scale-invariant for the modes that are emitted during radiation domination~\cite{Figueroa:2012kw}. At the level of the power spectrum, this background mimics therefore the shape of the inflationary background due to quantum fluctuations. The irreducible emission of GWs from a defect network is however expected to be highly non-Gaussian. This is simply due to the fact that the source of the GWs is the (transverse-traceless part of the) energy-momentum of the network, and this is a function bilinear in the amplitude (modulo derivatives) of the fields that the cosmic defects are made of. Since the GW source is bilinear in field amplitudes, this implies automatically that any correlator of an odd number of tensor perturbations will be characterized by the correlation of an even product of fields. This is non-vanishing even if the fields were Gaussian. Thus, odd tensor correlation functions are non-vanishing, i.e.~the GW background is not Gaussian. As the GWs are continuously emitted around the horizon scale at every moment, we cannot apply now the previous causal argument based on the superposition of many domains from the past. 

Even though the shape of the power spectrum of the irreducible GW background from defects is well understood theoretically, its ultimate amplitude depends on the fine details of the so called {\it unequal-time-correlator} of the network's energy-momentum tensor. Unfortunately, this correlator that can only be obtained accurately from sufficiently fine lattice simulations of defect networks. It is therefore difficult to assess at this point whether this background can be detectable with LISA, much less whether its 3-point function can be measured. The GW signal can be however estimated analytically in a simplified case, known as the large $N$ limit of a global phase transition due to the spontaneous symmetry breaking of $O(N)$ into $O(N-1)$. This kind of phase transition leads to the formation of so called ``self-ordering scalar field'' configurations, which correspond to non-topological defects (textures). The scale-invariant GW power spectrum due to the dynamics of such global defects has been estimated, in the limit $N \gg 1$, by Refs.~\cite{JonesSmith:2007ne,Fenu:2009qf}, whereas the 3-point function (in the equilateral configuration) has been presented in Ref.~\cite{Adshead:2009bz}. Order of magnitude calculations in the large $N$ limit have found a GW bispectrum peaked in the equilateral configuration~\cite{Adshead:2009bz} 
\begin{eqnarray}
k^6| \mathcal{B}(k,k,k)| \sim C_{NL}(k^{3/2}\mathcal{P}_h(k))^2\,,~~~~~~{\rm with}~C_{NL} \sim \frac{3.6}{\sqrt{N}}\,,
\end{eqnarray}
where $\mathcal{P}_h$ is the total power spectrum (summing over the two polarizations) and $N \gg 1$ is the number of components of the symmetry-breaking field. Even though this is a rough estimate, it clearly indicates that we should expect a large departure from Gaussianity for the irreducible GW background from any defect network. A proper assessment of the ability of LISA to detect the power spectrum and bispectrum of this stochastic background, requires however further work; namely lattice simulations of defect networks with a large dynamical range.

Let us finally comment on {\sl string gas cosmology}~\cite{Brandenberger:1988aj,Sakellariadou:1995vk} a string theory motivated early universe scenario, which does not involve a period of cosmological inflation.  The model is based upon T-duality, making use of (fundamental)
string oscillatory modes and winding modes. In this scenario, the universe may have started in a
a quasi-static Hagedorn phase, during which thermal fluctuations of closed (fundamental) strings generate the density perturbations. The obtained spectrum of cosmological perturbations \cite{Nayeri:2005ck}, and of GWs \cite{Brandenberger:2006xi} are nearly scale invariant. The power spectrum of scalar metric fluctuations has a slight red tilt, while the spectrum of gravitational waves  has a slight blue tilt. The string gas model has one free parameter and one free function, the former being the ratio of the string to the Planck length, and the latter the wavenumber dependence of the temperature. In this scenario, the spectrum of cosmological fluctuations may have large non-Gaussianities  due to the thermal origin of the initial perturbations leading to strong three-point correlations~\cite{Chen:2007js}. Such thermal effects have been shown to be important for some inflationary models, such as {\sl chain} inflation~\cite{Feldstein:2006hm} or {\sl warm} inflation~\cite{Berera:1995wh}. In string gas cosmology, the non-Gaussianities of the spectrum of cosmological fluctuations depend linearly on the wavenumber, while the amplitude depends sensitively on the string scale. In slow-roll inflation,  to leading order in perturbation theory, matter fluctuations do not couple to tensor ones, whereas in string gas cosmology matter fluctuations induce both scalar and tensor fluctuations.
Hence, one may expect also non-Gaussianities in the tensor modes, for which the formalism discussed in the previous sections may be applicable.

\section{Conclusions}  \label{sec-conclusions} 

In this work we studied the three-point correlation function of the SGWB expected to be measured by LISA, and we developed the formalism required for this analysis. The three-point correlation function is a key observable to gain information on the statistics of this background and crucial in order to study its departure from Gaussianity. This can  be an important  discriminant for a signal of cosmological origin, since the stochastic background due to a large number of uncorrelated astrophysical sources is Gaussian to a very high degree, due to the central limit theorem. The three-point correlator (also known as bispectrum), if present in the data, is a  richer observable than the standard 
two-point correlator (the power spectrum), due to several reasons. 
 Firstly, this is due to the frequency of the three modes present in the correlator. The wave-vectors of the modes entering in the correlator must add up to zero, namely $\vec{k} + \vec{k}' = 0$ for the power spectrum, and $\vec{k} + \vec{k}' + \vec{k}'' = 0$ for the bispectrum, implying that the momentum vectors form a closed triangle. For a statistically isotropic SGWB, 
the power spectrum only depends on the signal wavenumber ({\it i.e.} $\vert \vec{k} \vert = \vert \vec{k}' \vert$), while the bispectrum depends on the momentum orientation (see eq.~(\ref{rotate-B})), on an overall scaling, and on two relative sizes. This last dependence is known as the ``shape'' of the bispectrum. Secondly, additional combinations of LISA channels can be measured using the three-point signal, some of which may be useful as null tests for non-Gaussianity from unresolved astrophysical point sources.
 Thirdly, the planar nature of the instrument (together with the assumption of statistical isotropy) results in the insensitivity of the two-point function to the polarization of the two GW modes. This situation is ameliorated for the 
three-point function, since more combinations of GW polarizations can be considered (see below). 

In eq.~(\ref{f-NL}) we provided a very general parametrization for the case of a mildly non-Gaussian SGWB. This ansatz is borrowed from the wide literature on the non-Gaussianity of the primordial density perturbations probed through the fluctuations of the CMB radiation. This gives rise to a bispectrum of rather general shape (namely, the dependence on the wavenumbers of the three GW involved in the correlation, and, ultimately, on the frequencies of the three measured signals), see eq.~(\ref{BS}). Different cosmological mechanisms produce non-Gaussianity of different shapes, so that this study can potentially be a powerful discriminant between them. In Section \ref{sec-mod} we provided a survey of several possibilities present in the literature. 

In Section \ref{sec-signal} we computed all the non-vanishing three-point LISA response functions ${\cal R}_{\lambda\lambda;\lambda''}^{OO'O''} \left( \vec k ,\, \vec k' ,\,\vec  k'' \right)$ which, together with the GW correlation function $\langle h_\lambda ( \vec k ) h_{\lambda'} (\vec k' ) h_{\lambda''} ( \vec k'' ) \rangle$, provide the signal $\left\langle s_O s_{O'} s_{O''} \right\rangle$ measured by the different LISA channels (we study the response for all possible combinations of the $A$ and $E$ channels). We provided the formulae to compute the response functions for generic wavenumbers, and we then provided the explicit results for two cases of particular relevance (as they probe distinct physical mechanisms for the generation of non-Gaussianity): the equilateral configuration $k=k'=k''$  and the isosceles squeezed configuration $k \ll k' = k''$. As we already mentioned, the planarity of the instrument and the fact that it has equal arms  results in identities between different response functions, in the case of an isotropic SGWB. It is most immediate to discuss this in terms of circular polarizations, as they change into each other under a parity transformation. In the $2-$point functions we have ${\cal R}_{LL}^{OO'} = {\cal R}_{RR}^{OO'} $ (where $L$ and $R$ stand, respectively, for the left handed and the right handed circular polarization), resulting effectively in an inability to probe a chiral GW signal. 
On the other hand, for the three-point function we have  ${\cal R}_{LLL}^{OO'O''} = {\cal R}_{RRR}^{OO'O''} $, and  ${\cal R}_{LLR}^{OO'O''} = {\cal R}_{RRL}^{OO'O''} $, but these two objects have different frequency dependences. Hence by comparing measurements of the signal three-point function in different frequency ranges, one can hope to discriminate between a chiral and a non-chiral SGWB. 

 Moreover, in Section \ref{sec-estimator} we constructed the frequency-dependent estimator for the SGWB bispectrum that maximizes the signal-to-noise ratio in the measurement.  As an example, we evaluated the general formula for the estimator for the specific case of a non-Gaussian signal narrowly localized at one given scale, $k = k' = k'' = k_*$. 

\smallskip

To summarize, we have provided the complete formalism to compute the three-point correlation function of the signal measured at LISA for any given theoretical SGWB three-point correlator. This formalism can be readily extended to any other combination of GW interferometers. 
  Moreover, if the primordial bispectrum is amplified
on  a squeezed shape -- hence coupling modes of different frequencies -- we can use the  three-point function to correlate
signals from different experiments probing distinct frequency ranges, as LISA, LIGO, or  PTA.  This can allow us to `break' the planarity condition of a single interferometer, and measure effects of parity violation also in small frequency ranges that 
cannot be probed by other means (see also the discussion in Ref.~\cite{Smith:2016jqs}).
If non-vanishing,  measurements of three-point correlation functions  can provide a wealth of information on different cosmological mechanisms, due to its dependence on the scale, the shape, and, possibly, the polarization of the SGWB. Even a null measurement can set limits on specific models that would be otherwise be unconstrained. We leave this study to future work.

\section{Note added} 

In this paper we have computed the three-point function $\left\langle h^3 \right\rangle$ of the GW field using the ansatz \eqref{f-NL}. However, after the publication of this paper it was pointed out in \cite{Bartolo:2018evs,Bartolo:2018rku}  that this is generically largely suppressed and not observable in interferometers due to the decoherence of the phase of the GW wave-function $h$ induced by the GW propagation in the perturbed universe, and due to the finite duration of the measurement. Possible ways out have been proposed to avoid this problem:  the simplest one is to consider the angular three-point function of the SGWB energy density which is not affected by these fast phases \cite{Bartolo:2019oiq}. Alternatively one could also consider very specific shapes of the bispectrum for which the phases cancel \cite{Dimastrogiovanni:2019bfl,Powell:2019kid}.}
\vspace{5mm}

\vspace{0.5cm} 

\noindent {\bf Acknowledgments} \\
\\
\noindent 
We thank Raphael Flauger, Michele Liguori, Sabino Matarrese, and Germano Nardini for useful discussions. We thank the Mainz Institute for Theoretical Physics (MITP)
for hosting the IV LISA workshop, where this work has started.
N.B acknowledges partial financial support by ASI Grant No. 2016-24-H.0. The work of D.G.F was supported by the Swiss National Science Foundation.
J.G-B thanks the Theory Department at CERN for their hospitality during his sabbatical year at CERN. His work is supported by the Research Project FPA2015-68048-C3-3-P (MINECO-FEDER), the Centro de Excelencia Severo Ochoa Program SEV-2016-0597, and the Salvador de Madariaga Program Ref. PRX17/00056. 
The work of M. Pe was supported in part by the DOE grant de-sc0011842 at the University of Minnesota. M. Pi acknowledges the support of the Spanish MINECO's Centro de Excelencia Severo Ochoa Program SEV-2016-0597.
This project has received funding from the European Unions Horizon 2020
research and innovation programme under the Marie Sk\l{}odowska-Curie
grant agreement No 713366.
The work of M.S is supported in part by
the Science and Technology Facility Council (STFC), UK, under the research grant ST/P000258/1.
The work of L.S is partially supported by the US-NSF grant PHY-1520292. G.T
is partially funded by the STFC grant ST/P00055X/1.

\appendix
\setcounter{equation}{0}
\renewcommand{\theequation}{\thesection\arabic{equation}}

\section{Transformation of polarization operators and  bispectrum under a rotation of wave vectors}
\label{app:polarization} 

Let us first of all reproduce for clarity the tensor decomposition given in Eq.~(\ref{GW-classical}), 
\begin{eqnarray}
h_{ab} ( t ,\vec{x} ) = \int d^3 k \, {\rm e}^{-2 \pi i \, \vec{k} \cdot \vec{x}} \, \sum_\lambda e_{ab,\lambda} ( {\hat k}) \, h_\lambda (t,\vec{k}) \;, 
\label{GW-classicalv2}
\end{eqnarray}
expressed in terms of an arbitrary polarization basis $e_{ab,\lambda} ( {\hat k})$. A basis for the polarization operators often considered in the literature is the $\left\{ + ,\, \times \right\}$ basis. To define it one can introduce an auxiliary unit vector, e.g.~in the z-direction of our frame of coordinates, i.e.~${\hat e}_z$, and introduce the orthogonal unit vectors 
\begin{equation}
{\hat u} \left( {\hat k} \right) \equiv \frac{ {\hat k} \times {\hat e}_z  }{\vert  {\hat k} \times {\hat e}_z \vert}  \;\;\;\;,\;\;\;\; 
{\hat v} \left( {\hat k} \right) \equiv {\hat k} \times {\hat u} \left( {\hat k} \right) =   \frac{ \left( {\hat k} \cdot {\hat e}_z \right) \, {\hat k} - {\hat e}_z }{\vert  {\hat k} \times {\hat e}_z \vert}  \;. 
\label{uv-choice} 
\end{equation} 
The set $\{ {\hat k} ,\, {\hat u} ,\, {\hat v} \}$ forms an orthonormal basis. Note that ${\hat k}$ and ${\hat u}$ are odd under parity, while ${\hat v}$ is even. In standard  {spherical} 
coordinates
\begin{eqnarray}
{\hat k} = \left( \sin \theta \, \cos \phi ,\, \sin \theta \, \sin \phi ,\, \cos \theta \right)\,,
\end{eqnarray}
the unit vectors ${\hat u}$ and ${\hat v}$ acquire the form 
\begin{eqnarray} \label{eq:CanonicalOrthonormalBasis}
{\hat u  } = \left( \sin \phi ,\, - \cos \phi ,\, 0 \right) \;\;\;,\;\;\; 
{\hat v  } = {\hat k} \times {\hat u} = (\cos\theta \cos\phi, \cos\theta \sin\phi, -\sin\theta)\,. 
\end{eqnarray} 

Choosing an auxiliary unit vector different from ${\hat e}_z$ does not change any physical results, but induces a rotation of $\hat u, \hat v$ around $\hat k$, hence transforming the polarization operators and the GW mode functions~\cite{Maggiore:1900zz}. We discuss this in appendix~\ref{app:polarizationII}. In this appendix (like in the bulk of the text) we stick to identifying the auxiliary vector with ${\hat e}_z$, and then study how the polarization operators, and the mode functions, change under a rotation of the wave vector $\vec{k}$. These two different transformations, changing ${\hat e}_z$ vs. changing ${\hat k}$, produce a similar effect on the polarizations operators and on the mode function, as can be understood from the structure of (\ref{uv-choice}). 

Out of these vectors, it is customary to define 
\begin{equation}\label{eq:PlusCrossTensors}
e_{ab}^{(+)} = \frac{{\hat u}_a {\hat u}_b - {\hat v}_a {\hat v}_b }{\sqrt{2}}  \;\;,\;\; 
e_{ab}^{(\times)} = \frac{{\hat u}_a {\hat v}_b + {\hat v}_a {\hat u}_b }{\sqrt{2}} \;. 
\end{equation} 
These polarization operators are real, and they satisfy $e_{ab}^{(+)} e_{ab}^{(+)} = e_{ab}^{(\times)} e_{ab}^{(\times)} = 1 \;,\; e_{ab}^{(+)} e_{ab}^{(\times)} = 0$. We also note that $e_{ab}^{(+)}$ is even under parity, while $e_{ab}^{(\times)}$ is odd. 

The chiral polarization operators are then defined as 
\begin{eqnarray}\label{eq:ChiralTensors}
e_{ab}^{(R)} = \frac{e_{ab}^{(+)} + i \, e_{ab}^{(\times)}}{\sqrt{2}} \equiv e_{ab,+1} \;\;,\;\; 
e_{ab}^{(L)} = \frac{e_{ab}^{(+)} - i \, e_{ab}^{(\times)}}{\sqrt{2}} \equiv e_{ab,-1} \;, 
\end{eqnarray} 
and they satisfy the properties (\ref{e-properties}) given in the main text. 

We are interested in studying how the polarization operators change under a rotation. As we will see they acquire an additional factor with respect to the transformation of a rank 2 tensor, which is due to the fact that the fixed vector ${\hat e}_z$ used in (\ref{uv-choice}) does not change under the rotation.  To see this, it is convenient to rewrite 
\begin{eqnarray}
e_{ab}^{(R)} = 
\frac{ {\hat u}_a  + i \,  {\hat v}_a  }{\sqrt{2}} \, 
\frac{ {\hat u}_b  + i \,  {\hat v}_b  }{\sqrt{2}} 
\equiv \epsilon_{a,R} \,  \epsilon_{b,R} \;,  \nonumber\\ 
e_{ab}^{(L)}  = 
\frac{ {\hat u}_a  - i \,  {\hat v}_a  }{\sqrt{2}} \, 
\frac{ {\hat u}_b  - i \,  {\hat v}_b  }{\sqrt{2}} 
\equiv \epsilon_{a,L}  \epsilon_{b,L} \;.  
\label{u-v-oneindex} 
\end{eqnarray} 

To find how the polarization operators transform under a rotation, we will now study the transformation properties of $\epsilon_{a,L/R}$ (this is facilitated by the fact that they have one spacial index, rather than two). Specifically, we want to obtain the quantities $R^{-1} {\hat u} \left( R {\hat k} \right)$ and $R^{-1} {\hat v} \left( R {\hat k} \right)$, where $R$ represents a spatial rotation. To do so, we compute the coordinates of these two quantities in the  $\left\{ {\hat k} ,\, {\hat u} \left( {\hat k} \right) ,\, {\hat v} \left( {\hat k} \right) \right\}$ basis:  
\begin{eqnarray} 
R^{-1} {\hat u} \left( R {\hat k} \right) &=& c_{uu} \, {\hat u} \left( {\vec k} \right) +  c_{uv} \, {\hat v} \left( {\vec k} \right) + c_{uk} \, {\hat k} \;, \nonumber\\ 
R^{-1} {\hat v} \left( R {\hat k} \right) &=& c_{vu} \, {\hat u} \left( {\vec k} \right) +  c_{vv} \, {\hat v} \left( {\vec k} \right) + c_{vk} \, {\hat k} \;. 
\end{eqnarray} 

We note that 
\begin{equation}
{\hat u} \left( R {\hat k} \right) =  \frac{ \left( R {\hat k} \right) \times {\hat e}_z}{\left\vert  \left( R {\hat k} \right) \times {\hat e}_z \right\vert} \;\; \Rightarrow \;\; 
R^{-1} {\hat u} \left( R {\hat k} \right) = \frac{{\hat k} \times \left( R^{-1} \, {\hat e}_z \right)}{\vert {\hat k} \times \left( R^{-1} \, {\hat e}_z \right) \vert} \;. 
\end{equation} 
and analogously for ${\hat v}$. The various coordinates $c_{uu} ,\, \dots ,\, c_{vk}$ can be then computed via scalar products, such as 
\begin{equation} 
c_{uu} = {\hat u} \left( {\hat k} \right) \cdot R^{-1} {\hat u} \left( R {\hat k} \right) = 
\frac{ {\hat k} \times {\hat e}_z  }{\vert  {\hat k} \times {\hat e}_z \vert} \cdot 
\frac{ {\hat k} \times {\hat e}_z'  }{\vert  {\hat k} \times {\hat e}_z' \vert} \;\;\;,\;\;\; 
{\hat e}_z' \equiv R^{-1} \, {\hat e}_z \;. 
\label{cuu} 
\end{equation} 

It is convenient to define the metric 
\begin{equation}
\Pi_{ij} \left( {\hat k} \right) \equiv \delta_{ij} - {\hat k}_i {\hat k}_j \;, 
\end{equation} 
on the plane transverse to ${\hat k}$, and the product with respect to this metric 
\begin{eqnarray} 
\left( \vec{a} \cdot \vec{b} \right)_\Pi \equiv a_i \, \Pi_{ij} \, b_j \;. 
\end{eqnarray} 
In terms of this product, the expression (\ref{cuu}) acquires the compact form 
\begin{equation}
c_{uu} \equiv  \frac{\left( {\hat e}_z \cdot {\hat e}_z' \right)_\Pi }{\sqrt{ \left( {\hat e}_z \cdot {\hat e}_z \right)_\Pi } \; \sqrt{\left( {\hat e}_z' \cdot {\hat e}_z' \right)_\Pi }} \;. 
\end{equation} 

Computing the other coordinates in an analogous way, we find that $c_{uk} = c_{vk} = 0$ (namely, the rotated ${\hat u}$ and ${\hat v}$ remain orthogonal to the rotated ${\hat k}$, as one would expect), and that 
\begin{equation}
\left( \begin{array}{c}  R^{-1} {\hat u} \left( R {\hat k} \right) \\ R^{-1} {\hat v} \left( R {\hat k} \right) \end{array} \right) = 
\left( \begin{array}{cc} 
\cos \gamma  \left[ {\hat k} ,\, R \right]  & \sin \gamma  \left[ {\hat k} ,\, R \right] \\[8pt] 
- \sin \gamma  \left[ {\hat k} ,\, R \right]  & \cos \gamma  \left[ {\hat k} ,\, R \right] 
\end{array} \right) 
\left( \begin{array}{c}  {\hat u} \left( {\hat k} \right) \\[8pt]  {\hat v} \left(  {\hat k} \right) \end{array} \right) \;, 
\end{equation}  
where 
\begin{equation}
\cos \gamma  \left[ {\hat k} ,\, R \right]  \equiv \frac{\left( {\hat e}_z \cdot {\hat e}_z' \right)_\Pi }{\sqrt{ \left( {\hat e}_z \cdot {\hat e}_z \right)_\Pi } \; \sqrt{\left( {\hat e}_z' \cdot {\hat e}_z' \right)_\Pi }} \;\;\;,\;\;\; 
\sin \gamma  \left[ {\hat k} ,\, R \right] \equiv  \frac{{\hat k} \cdot \left( {\hat e}_z \times {\hat e}_z'  \right)}{\sqrt{ \left( {\hat e}_z \cdot {\hat e}_z \right)_\Pi } \; \sqrt{\left( {\hat e}_z' \cdot {\hat e}_z' \right)_\Pi }} \;. 
\label{gamma} 
\end{equation} 

Combining this with (\ref{u-v-oneindex}), we can write 
\begin{eqnarray}
R^{-1} \, \epsilon_{a,\lambda} \left( R {\hat k} \right) = \frac{
R^{-1} \, {\hat u}_{a} \left( R {\hat k} \right)  + i \lambda R^{-1} \, {\hat v}_{a} \left( R {\hat k} \right) }{\sqrt{2}} = {\rm e}^{-i \lambda \gamma}  \epsilon_{a,\lambda} \left( {\hat k} \right) \;, 
\end{eqnarray} 
where $\lambda = + 1$ denotes the $R$ helicity, while $\lambda = - 1$ denotes the $L$ one. This leads to 
\begin{eqnarray} 
e_{ab,\lambda} \left( R {\hat k} \right) =  {\rm e}^{-2i \lambda \gamma \left[ {\hat k} ,\, R \right]} \, R_{ac} \, R_{bd} \, e_{cd,\lambda} \left(  {\hat k} \right) \;. 
\label{rotate-e} 
\end{eqnarray} 

As it is clear from eq.~(\ref{GW-classicalv2}), the product between the polarization operator and the mode function is a rank two tensor. We then learn that, under the rotation, the GW mode function transforms as 
\begin{equation}
h_{\lambda} \left( t ,\, R \vec{k} \right) =  {\rm e}^{+2i \lambda \gamma \left[ {\hat k} ,\, R \right]} \, 
h_{\lambda} \left( t ,\,  \vec{k} \right)  \;. 
\label{rotate-h} 
\end{equation} 
We note from (\ref{gamma}) that $\gamma \left[ - {\hat k} ,\, R \right] = - \gamma \left[ {\hat k} ,\, R \right] $. As a consequence, the power spectrum $P_\lambda \left( k \right)$ is invariant under a rotation, since the two modes involved in the correlation have momenta $\vec{k}' = - \vec{k}$, cf. eq.~(\ref{PS}). This is not the case for the bispectrum. Denoting by $\vec{k}_i' = R \vec{k}_i$, we have 
\begin{eqnarray} 
{\cal B}_{\lambda_1,\lambda_2,\lambda_3} \left( \vec{k}_1' ,\, 
 \vec{k}_2' ,\,  \vec{k}_3' \right) &=& {\rm e}^{2 i \sum_{i=1}^3 \lambda_i \gamma \left[ {\hat k}_i , R \right]} \, 
{\cal B}_{\lambda_1,\lambda_2,\lambda_3} \left( \vec{k}_1 ,\, 
 \vec{k}_2 ,\,  \vec{k}_3 \right) \nonumber\\ 
 & \equiv & \Phi_{\lambda_i} \left( {\hat k}_i ,\, R \right) {\cal B}_{\lambda_1,\lambda_2,\lambda_3} \left( \vec{k}_1 ,\,  \vec{k}_2 ,\,  \vec{k}_3 \right)     \;. 
 \label{rotate-B} 
\end{eqnarray} 
An identical relation is satisfied by the Kernel function $K_{\lambda_1\lambda_2\lambda_2} \left( \vec{k}_1,\, \vec{k}_2,\,\vec{k}_3 \right)$.

\section{Transformation of polarization operators and  bispectrum under a rotation around wave vectors}
\label{app:polarizationII} 

As mentioned in the previous appendix, the decomposition in eq.~(\ref{GW-classicalv2}) [c.f.~eq.~(\ref{GW-classical})], depends on the choice of orientation of $\{ {\hat u} , {\hat v} \}$ within the plane transverse to $\hat k$. For a given GW mode propagating in the direction ${\hat k}$, our choice of the orthonormal basis $\lbrace \hat u, \hat v \rbrace$ in eq.~(\ref{eq:CanonicalOrthonormalBasis}), based on choosing $\hat e_z$ as an auxiliar fixed unit vector, was actually arbitrary (even though very convenient, as we will see). Our canonical choice of $\{{\hat u}, {\hat v}\}$ can be actually rotated by an arbitrary angle $\alpha$ around $\hat k$, into a new triad $\{ {\hat u}(\alpha), {\hat v}(\alpha), {\hat k}\}$, where ${\hat u}(0) \equiv {\hat u}$, ${\hat v}(0) \equiv {\hat v}$ are given by eq.~(\ref{eq:CanonicalOrthonormalBasis}). This can be implemented explicitly by a rotation matrix around $\hat k$ as
\begin{eqnarray} \label{eq:RotatedVectors}
\left(
\begin{array}{c}
\hat u(\alpha)\\
\hat v(\alpha)
\end{array}
\right) = 
R[\alpha]\cdot
\left(
\begin{array}{c}
\hat u\\
\hat v
\end{array}
\right)\,,~~~~~ R[\alpha] \equiv \left(\begin{array}{cc}
\cos\alpha & \sin\alpha\\
- \sin\alpha & \cos\alpha
\end{array}\right)\,,
\end{eqnarray}
from where it is easy to verify that $|\hat u(\alpha)|^2 = |\hat v(\alpha)|^2 = 1$ and $\hat u(\alpha) \times \hat v(\alpha) = \hat k$, as it should. The rotated triad $\{ {\hat u}(\alpha), {\hat v}(\alpha), {\hat k}\}$ form therefore an equally valid orthonormal basis, from which to build up new $+,\times$ polarizations tensors as
\begin{eqnarray}
e_{ij}^{(+)}(\alpha) = \frac{\hat u_i(\alpha) \hat u_j(\alpha) - \hat v_i(\alpha) \hat v_j(\alpha) }{\sqrt{2}}  \;\;,\;\; 
e_{ij}^{(\times)}(\alpha) = \frac{\hat u_i(\alpha) \hat v_j(\alpha) + \hat v_i(\alpha) \hat u_j(\alpha) }{\sqrt{2}} \;,
\end{eqnarray}
and from here a set of chiral polarization tensors in the rotated basis, 
\begin{eqnarray}\label{eq:RotatedChiralTensors}
e_{ab}^{(R)}(\alpha) = \frac{e_{ab}^{(+)}(\alpha) + i \, e_{ab}^{(\times)}(\alpha)}{\sqrt{2}} \equiv e_{ab,+1}(\alpha)\,,~ 
e_{ab}^{(L)}(\alpha) = \frac{e_{ab}^{(+)}(\alpha) - i \, e_{ab}^{(\times)}(\alpha)}{\sqrt{2}} \equiv e_{ab,-1}(\alpha).
\end{eqnarray}
Basic algebra leads to relate the new chiral polarization tensors with respect to the canonical `unrotated' basis $\{ \hat u, \hat v\}$  eq.~(\ref{eq:CanonicalOrthonormalBasis}), as
\begin{eqnarray}\label{eq:ChiralityRotation}
e_{ij,\pm 1}(\alpha) = e^{\mp i2\alpha}\,e_{ij,\pm 1}\,,
\end{eqnarray}
where the chiral polarization tensors on the right hand side are those from eq.~(\ref{eq:ChiralTensors}), and on the left hand side those from eq.~(\ref{eq:RotatedChiralTensors}). This implies that if we were to express the GW in eq.~(\ref{GW-classicalv2}) [c.f.~eq.~(\ref{GW-classical})] in terms of the new $\left\{e_{ij}^{(R)}(\alpha),\, e_{ij}^{(L)}(\alpha)\right\}$ tensors, the tensor modes should equally be re-scaled by an opposite phase as in eq.~(\ref{eq:ChiralityRotation}), i.e.
\begin{eqnarray}\label{eq:ChiralityRotationModes}
h_{\pm 1}[\alpha] = e^{\pm i2\alpha}h_{\pm 1}\,.
\end{eqnarray}
These transformation rules simply reflect, actually, the spin-2 nature of the tensor field $h_{ij}$ representing the GWs. Equations~(\ref{eq:ChiralityRotation}), (\ref{eq:ChiralityRotationModes}) are relevant as they determine the complex phase of the response function we have calculated in Section~\ref{sec-signal}. Let recall that the response functions presented along the main text were always computed for a tensor basis obeying the property expressed in eq.~(\ref{e-properties}), namely that under a ``parity'' tranformation $\vec k \longleftrightarrow - \vec k$, the chirality polarization tensors transform as $e_{ab,\lambda}(-\vec k) = e_{ab,\lambda}^*(\vec k)$. That property is certainly true when the polarization tensors are computed in the canonical basis $\{ \hat u, \hat v\}$ of eq.~(\ref{eq:CanonicalOrthonormalBasis}). However Eq.~(\ref{eq:ChiralityRotation}) implies that such relations are not universal: on the contrary, they are only verified by a very small subset of orthonormal basis $\{ \hat u(\alpha), \hat v(\alpha)\}$. In particular, from eq.~(\ref{eq:ChiralityRotation}), and using eq.~(\ref{e-properties}), we deduce immediately that, in general, under a flip of momentum, the rotated basis of polarization tensors transform as
\begin{eqnarray}\label{eq:ChiralRotatedTensorsFlipped}
e_{ab,\pm}[\alpha](-\vec k) = e^{\mp i2\alpha}e_{ab,\pm}^*(\vec k)\,.
\end{eqnarray}
As expected, for $\alpha = 0$ we recover the result expressed in eq.~(\ref{e-properties}) that that used in the main text, i.e. $e_{ab,\lambda}(-\vec k) = e_{ab,\lambda}^*(\vec k)$. We see that this result also holds for $\alpha = \pi$, which is nothing else but the flip of the canonical vectors as $\hat u \longrightarrow -\hat u$, $\hat v \longrightarrow -\hat v$. For any angle $\alpha$ for which $\alpha~{\rm mod}(\pi) \neq 0$, the property $e_{ab,\lambda}(-\vec k) = e_{ab,\lambda}^*(\vec k)$ will not hold. For instance for $\alpha = \pi/2$, this is equivalent to $\hat u \longrightarrow \hat v$ and $\hat v \longrightarrow -\hat u$, and hence it holds that $e_{ab,\lambda}(-\vec k) = -e_{ab,\lambda}^*(\vec k)$. In general, the result of flipping the momentum will introduce a phase, as indicated in eq.~(\ref{eq:ChiralRotatedTensorsFlipped}). 

It is clear then that the choice of a `rotated' vector basis other than the canonical one(s), will determine the complex phase of the polarization tensor via eq.~(\ref{eq:ChiralityRotation}), and hence this will have necessarily an impact on the final expression of the 3-point response functions that we have derived in Section~\ref{sec-signal}. The reason is that $\mathcal{R}_{\lambda \lambda' \lambda''}$ is always proportional to three polarization tensors, schematically $\sim e_{**}^{\lambda}e_{**}^{\lambda'}e_{**}^{\lambda''}$. Denoting as $\mathcal{R}_{\lambda \lambda' \lambda''}[\alpha_1,\alpha_2,\alpha_3]$ the response functions calculated in a orthonormal vector basis rotated an angle $\alpha_i$ around each vector $\vec k_i$, with respect the canonical basis $\{\hat u, \hat v\}$ eq.~(\ref{eq:CanonicalOrthonormalBasis}), we deduce the following relation(s)
\begin{eqnarray}\label{eq:RxxxRotationI}
\mathcal{R}_{\lambda_1 \lambda_2 \lambda_3}[\alpha_1,\alpha_2,\alpha_3] = e^{-i2\sum_{i=1}^3{\lambda_i\alpha_i}}\mathcal{R}_{\lambda_1 \lambda_2 \lambda_3}[0]\,,
\end{eqnarray}
where the response functions $\mathcal{R}_{\lambda \lambda' \lambda''}[0]$ on the right hand side should be identified with the response functions we presented in the main text. Due to eq.~(\ref{eq:RxxxRotationI}), we see that we can have pure real, imaginary or simply complex response functions, simply depending on the choice of the phases~$\alpha_i$. The fact that our 3-point response function is real applies because of the choice of polarization operators built from the canonical basis given in eq.~(\ref{eq:CanonicalOrthonormalBasis}). For example, let us take all equal angles, $\alpha_i = \alpha~\forall i = 1,2,3$, and consider $\alpha = 0$ or $\alpha = \pi$. Then we find that all 3-point response functions are real (like in the main text), while for $\alpha = \pi/6$ the response functions $\mathcal{R}_{RRR}, \mathcal{R}_{LLL}$ are still real (but different in a phase), while $\mathcal{R}_{LRR}, \mathcal{R}_{RLL}$ become complex (and differing in a phase). The condition to be purely real, purely imaginary, or simply complex, depends on the choice of $\alpha_i$'s according to eq.~(\ref{eq:RxxxRotationI}). Of course, the final 3-point signal Eq.~(\ref{forstar}) [c.f.~Eqs.~(\ref{3-point-start},\ref{3-point-middle})] in the GW detector does not depend on the choice of the $\hat u, \hat v$ basis. The left hand side of eq.~(\ref{forstar}) is the same no matter the choice of $\alpha$. However, depending on $\alpha$, one must modify our expressions for the response functions by the phases indicated in eq.~(\ref{eq:RxxxRotationI}), and at the same time correct the tensor bispectra $\mathcal{B}_{\lambda \lambda' \lambda''}$ by exactly the opposite phases, so that at the end the physical signal eq.~(\ref{forstar}) remains invariant. In order to use our formalism, special care must be put into what vector basis is being used in the computation of the tensor bispectrum, and hence what properties of the polarization operators and response functions hold.  

\section{Vanishing of the 3-point response function in the $k L \rightarrow 0$ limit}%
\label{app:R3k0} 
\setcounter{equation}{0}

In this Appendix we show that the LISA three-point response function vanishes in the long wavelength limit $L k \rightarrow 0$. In this limit, the three quantities in eq.~(\ref{calQ}) simplify to ${\cal Q}^1_{AB}   = 2  \left[   {\hat U}_1^A {\hat U}_1^B -    {\hat U}_3^A {\hat U}_3^B \right] $, ${\cal Q}^2_{AB} = 2  \,  \left[   {\hat U}_2^A {\hat U}_2^B -     {\hat U}_1^A {\hat U}_1^B \right] $, and ${\cal Q}^3_{AB} =  2 \, \left[   {\hat U}_3^A {\hat U}_3^B -     {\hat U}_2^A {\hat U}_2^B \right] $. We can then write the expressions for the three measurements in a compact way as 
\begin{eqnarray}
\Sigma_O& = & L^3 \, {\cal D}^O_{ab} \, \sum_\lambda \int d^3 k \;  h_\lambda \left( t - L ,\, \vec{k} \right)  \, {\rm e}_{ab,\lambda} \left( {\hat k} \right) \;, 
\end{eqnarray} 
with 
\begin{equation}
{\cal D}_{ab}^O =  \alpha^O  \, {\hat U}_1^a {\hat U}_1^b 
+ \beta^O \,  {\hat U}_2^a {\hat U}_2^b 
+ \gamma^O  \, {\hat U}_3^a {\hat U}_3^b  \;, 
\end{equation} 
and
\begin{equation}
\alpha^1 = 1 \;,\; \beta^1 = 0 \;,\; \gamma^1 = -1 \;\;\;,\;\;\; 
\alpha^2 = \frac{1}{\sqrt{3}} \;,\; \beta^2 = - \frac{2}{\sqrt{3}} \;,\; \gamma^2 =  \frac{1}{\sqrt{3}} \;. 
\label{alpha-beta-gamma} 
\end{equation} 
We recall that the index $O=1$ refers to the $A$ channel, while $O=2$ to the $E$ channel. We instead find $\Sigma_T=0$ (so $\alpha^3 =\beta^3 = \gamma^3 =0$). 

This gives the response function 
\begin{eqnarray} 
{\cal R}^{OO'O''}_{\lambda_1,\lambda_2\lambda_3} \left(  {\hat k}_1^* ,\,   {\hat k}_2^* ,\,   {\hat k}_3^* \right) &=&  
 {\cal D}_{ab}^O {\cal D}_{cd}^{O'} {\cal D}_{ef}^{O''} \; \times \; 
e_{AB,\lambda_1} \left( {\hat k}_1^* \right) \, 
e_{CD,\lambda_2} \left(  {\hat k}_2^* \right) \, 
e_{EF,\lambda_3} \left(  {\hat k}_3^* \right) \nonumber\\ 
& & \int_0^\pi d \theta_1 \, \sin \theta_1 \, \int_0^{2 \pi} d \phi_1 \,  \int_0^{2 \pi} d \phi_2 \, 
R_{aA} R_{bB} R_{cC} R_{dD} R_{eE} R_{fF} \;. \nonumber\\ 
\end{eqnarray} 

The quantity contracted with ${\cal D}_{ab}^O {\cal D}_{cd}^{O'} {\cal D}_{ef}^{O''}$ in the $rhs$ of the above expression, carries $6$ spatial indices, none of which can coincide with the spatial indices of any of the ${\hat k}_i$ directions (otherwise the response function could be different for different orientations of the LISA arms). In fact, the direct evaluations confirms that 
\begin{eqnarray} 
{\cal R}^{OO'O''}_{\lambda_1,\lambda_2\lambda_3} \left(  \vec{k}_1^* ,\,   \vec{k}_2^* ,\,   \vec{k}_3^* \right) &=&  
 {\cal D}_{ab}^O {\cal D}_{cd}^{O'} {\cal D}_{ef}^{O''} \;  {\cal N}_{\lambda_1,\lambda_2,\lambda_3} \left[ {\hat k}_1^* ,\,  {\hat k}_2^* ,\,  {\hat k}_3^* \right] 
\, \Bigg\{ \delta_{ab} \delta_{cd} \delta_{ef} \nonumber\\ 
& & - \frac{3}{4} \left[ 
\delta_{ab} \left( \delta_{ce} \delta_{df} + \delta_{cf} \delta_{de} \right) 
+  \delta_{cd} \left( \delta_{ae} \delta_{bf} + \delta_{af} \delta_{be} \right) 
+  \delta_{ef} \left( \delta_{ac} \delta_{bd} + \delta_{ad} \delta_{bc} \right) 
\right] \nonumber\\ 
& & + \frac{9}{16} \Big[ 
\delta_{ac} \left( \delta_{de} \delta_{fb} +  \delta_{df} \delta_{eb} \right) 
+ \delta_{ad} \left( \delta_{ce} \delta_{fb} +  \delta_{cf} \delta_{eb} \right) \nonumber\\ 
& & \quad\quad + 
\delta_{ae} \left( \delta_{fc} \delta_{db} +  \delta_{fd} \delta_{cb} \right) 
+ \delta_{af} \left( \delta_{ec} \delta_{db} +  \delta_{ed} \delta_{cb} \right) 
\Big] \Bigg\} \;. 
\end{eqnarray} 
where only the number $ {\cal N}_{\lambda_1,\lambda_2,\lambda_3} $ depends on the orientation of the reference plane used to evaluate the bispectrum. For the choice given by eqs.~(\ref{starframe}) and (\ref{v1hat-v2hat}) we obtain ${\cal N}_{RRR} = - \frac{9 \pi^2}{35} $ and ${\cal N}_{RRL} = - \frac{ \pi^2}{35} $. The structure in the curly parenthesis is compatible with the GW origin of the signal, and it vanishes when we trace over the $ab$, the $cd$, or the $ef$ pair. 
Taking into account the fact that ${\cal D}_{aa}^O = 0$, this expression simplifies to 
\begin{eqnarray} 
{\cal R}^{OO'O''}_{\lambda_1,\lambda_2\lambda_3} \left(  \vec{k}_1^* ,\,   \vec{k}_2^* ,\,   \vec{k}_3^* \right) &=&  \frac{9}{4} 
 \;  {\cal N}_{\lambda_1,\lambda_2,\lambda_3} \left[ {\hat k}_1^* ,\,  {\hat k}_2^* ,\,  {\hat k}_3^* \right] \, 
 \left[  {\rm Tr} \left(  {\cal D}^O {\cal D}^{O'} {\cal D}^{O''} \right) +  {\rm Tr} \left(  {\cal D}^O {\cal D}^{O''} {\cal D}^{O'} \right) \right] \;. \nonumber\\ 
\end{eqnarray} 

For brevity, we denote the sum of the two traces as   ${\tilde T}^{OO'O''}$. Using the values in (\ref{alpha-beta-gamma}), this expression gives 
\begin{eqnarray}
{\tilde T}^{AAA} &=& {\tilde T}^{AEE} = 0 \;, \nonumber\\ 
{\tilde T}^{AAE} &=&  - {\tilde T}^{AAE}  \nonumber \\
&=& \frac{4}{\sqrt{3}} \left[ 1 - \left( {\hat U}_1 \cdot {\hat U}_2 \right)^2  - \left( {\hat U}_1 \cdot {\hat U}_3 \right)^2  - \left( {\hat U}_2 \cdot {\hat U}_3 \right)^2 + 2 \left( {\hat U}_1 \cdot {\hat U}_2 \right) \left( {\hat U}_1 \cdot {\hat U}_3 \right) \left( {\hat U}_2 \cdot {\hat U}_3 \right) \right] \;. 
 \nonumber\\ 
\end{eqnarray} 
Finally, recalling the definitions (\ref{LISA-arms}) of the arm directions, we have ${\hat U}_1 \cdot {\hat U}_2 =  {\hat U}_1 \cdot {\hat U}_3 =  {\hat U}_2 \cdot {\hat U}_3 = - \frac{1}{2}$. This gives 
\begin{equation} 
{\tilde T}^{AAE} = {\tilde T}^{EEE} = 0 \;. 
\end{equation} 

As we anticipated, this shows (through a fully analytical computation) that the three-point response function vanishes in the  large wavelength limit, in agreement with what obtained numerically in Figures \ref{fig:BS-eq} and \ref{fig:BS-sq}.

\section{Comparison with GW decomposition in the frequency basis} 
\label{app:f-basis} 
\setcounter{equation}{0}

It is instructive to compare the GW decomposition (\ref{GW-classical}) used in this paper with the decomposition in terms of positive and negative frequencies that is often encountered in the literature of the stochastic GW background \cite{Allen:1997ad}
\begin{equation}
h_{ab} \left( t ,\, \vec{x} \right) = \int_{-\infty}^{+\infty} d f \, \int d^2 {\hat n} \, \sum_\lambda \, {\rm e}_{ab,\lambda} \left( {\hat n} \right) \, 
{\rm e}^{2 \pi i f \left( t - {\hat n} \cdot \vec{x} \right)} \, {h}_\lambda \left( f ,\, {\hat n} \right) \,, 
\label{GW-f}
\end{equation}
where we are considering left and right polarizations, and where the requirement of a real GW background is used to define the negative frequency field, $h_\lambda \left( - f ,\, {\hat n} \right) = h_{-\lambda}^* \left(  f ,\, {\hat n} \right)$ (in the $\left\{ + ,\, \times \right\}$ basis, one has instead $h_\sigma \left( - f ,\, {\hat n} \right) = h_{\sigma}^* \left(  f ,\, {\hat n} \right)$). Using this property, the decomposition (\ref{GW-f}) can be rewritten as 
\begin{align} 
h_{ab} \left( t ,\, \vec{x} \right) =& \int_0^{+\infty} d f \,  \int d^2 {\hat n} \, \sum_\lambda 
{\rm e}_{ab,\lambda} \left( {\hat n} \right)  \left[  {\rm e}^{2 \pi i f \left( t - {\hat n} \cdot \vec{x} \right)} {h}_\lambda \left( f ,\, {\hat n} \right) 
+  {\rm e}^{-2 \pi i f \left( t - {\hat n} \cdot \vec{x} \right)} {h}_\lambda \left( -f ,\, {\hat n} \right) 
\right] \nonumber\\ 
=& \int_0^{+\infty} d f \,  \int d^2 {\hat n} \, {\rm e}^{-2 \pi i f  {\hat n} \cdot \vec{x} } 
 \sum_\lambda {\rm e}_{ab,\lambda} \left( {\hat n} \right)  \left[ {\rm e}^{2 \pi i f  t } \, {h}_\lambda \left( f ,\, {\hat n} \right) +  {\rm e}^{-2 \pi i f  t }  \, {h}_\lambda^* \left( f ,\, - {\hat n} \right) \right] \,, 
\end{align} 
where we note that we also sent $\left\{ {\hat n} ,\, \lambda \right\} \rightarrow \left\{ -{\hat n} ,\, -\lambda \right\} $ in the second term of the second line. Comparing this with the first line of (\ref{GW-classical}) allows to relate the operators in the two decompositions as 
\begin{equation}
{h}_\lambda \left( f ,\, {\hat n} \right) = k^2 \left[ h_\lambda \left( \vec{k} = f \, {\hat n} \right) \theta \left( f \right) + h_{-\lambda}^* \left( \vec{k} = - f \, {\hat n} \right) \theta \left( - f \right) \right] \,. 
\end{equation} 

Using the relation between the two lines of  (\ref{GW-classical}) we can also relate the operator in (\ref{GW-f}) to the full GW Fourier transform as 
\begin{equation}
{h}_\lambda \left( f ,\, {\hat n} \right) = \frac{{\rm e}^{-2 \pi i f t}}{2} \, f^2 
\left[ h_{{\rm sign} \, f \times \lambda} \left( t ,\, \vec{k} = f {\hat n} \right) - \frac{i}{2 \pi f}  \dot{h}_{{\rm sign} \, f \times \lambda} \left( t ,\, \vec{k} = f {\hat n} \right) 
\right] \,,  
\end{equation}
where dot denotes time differentiation. 

We can use this relation to also relate correlators between modes in the two basis. For instance, for the $2-$point correlator we have 
\begin{eqnarray} 
&& \!\!\!\!\!\!\!\!  \!\!\!\!\!\!\!\! 
\left\langle {h}_\lambda \left( f ,\, {\hat n} \right)  {h}_{\lambda'} \left( f' ,\, {\hat n}' \right) \right\rangle 
= \frac{{\rm e}^{-2 \pi i \left( f t + f' t' \right)}}{4} \, f^2 \, f^{' 2} \left[ 1 - \frac{i}{2 \pi f} \, \frac{\partial}{\partial t} \right] 
 \left[ 1 - \frac{i}{2 \pi f'} \, \frac{\partial}{\partial t'} \right] \, \nonumber\\ 
& & \quad\quad\quad\quad   \quad\quad\quad\quad   \quad\quad\quad\quad   \quad\quad\quad\quad  
\times  \left\langle h_{{\rm sign} \,  f \times \lambda} \left( t ,\, f {\hat n} \right) \,  h_{{\rm sign} \,  f' \times \lambda'} \left( t' ,\, f' {\hat n}' \right) \right\rangle \,.  
\label{hfhf-hkhk}
\end{eqnarray} 

Eq. (\ref{PS}) in the main text provides the equal time correlator in the Fourier basis. This is obtained from 
\begin{equation}
\left\langle h_\lambda \left( \vec{k} \right)  h_{\lambda'} \left( \vec{k}' \right) \right\rangle = 0 \;\;,\;\; 
\left\langle h_\lambda \left( \vec{k} \right)  h_{\lambda'}^* \left( \vec{k}' \right) \right\rangle = \frac{P_\lambda \left( k \right)}{8 \pi k^3} \, 
\delta_{\lambda \lambda'} \, \delta^{(3)} \left( \vec{k} - \vec{k}' \right) \;. 
\end{equation} 
These relations lead to the unequal time correlator 
\begin{equation}
\left\langle h_\lambda \left( t ,\, \vec{k} \right)  h_{\lambda'} \left( t' ,\, \vec{k}' \right) \right\rangle  = 
\cos \left[ 2 \pi k \left( t - t' \right) \right] \, \frac{P_\lambda \left( k \right)}{4 \pi k^3} \, \delta_{\lambda \lambda'} \, \delta^{(3)} \left( \vec{k} + \vec{k}' \right) \,, 
\end{equation} 
(which immediately reduces to (\ref{PS}) at equal times). Inserting this relation in (\ref{hfhf-hkhk}) leads to  
\begin{eqnarray} 
&& \left\langle h_L \left( f ,\, {\hat n} \right)  h_L \left( f' ,\, {\hat n}' \right) \right\rangle = 
\left\langle h_R \left( f ,\, {\hat n} \right)  h_R \left( f' ,\, {\hat n}' \right) \right\rangle = 0 \;, \nonumber\\ 
&& \left\langle h_L \left( f ,\, {\hat n} \right)  h_R \left( f' ,\, {\hat n}' \right) \right\rangle = 
\frac{P_L \left(  f  \right) \theta \left( f \right) + P_R \left(  f'  \right) \theta \left(  f' \right) }{8 \pi \vert f \vert} \, 
\delta \left( f + f' \right) \delta^{(2)} \left( {\hat n} - {\hat n}' \right) \;. 
\end{eqnarray} 

As a check on our algebra, we verified that combining this result with the decomposition (\ref{GW-f}) leads again to the equal-time 
real-space correlator (\ref{hx-hy}).

\section{Technical calculations of Section \ref{section-bumpy}}
\label{app:tech} 
\setcounter{equation}{0}

In this Appendix we provide some technical steps that were omitted in the computation of Section \ref{section-bumpy}. 
Inserting the kernel of eq \eqref{bump-ansatz} relation into eqs.~(\ref{B-tilde}) and  (\ref{S-ijk}),  we obtain
%
\begin{align} 
S_s^{ijk} \left( f_1, f_2, f_3 \right) =& \frac{ L^3 \, f_{\rm NL}^{LLL} }{ 32 \, \pi^2 } 
\int \frac{dk_1}{k_1}  \frac{dk_2}{k_2}  \frac{dk_3}{k_3} \, k_1^2 k_2^2 k_3^2 \, {\cal R}_{LLL}^{ijk} \left( \vec k^*_1, \vec k^*_2, \vec k^*_3 \right) \nonumber \\
& \times \exp\left(-\frac{1}{2 \sigma^2} \left[ \left( | \vec k_1^*| - \bar k \right)^2 +  \left( ( | \vec k_2^*| - \bar k  \right)^2 +  \left( ( | \vec k_3^*| - \bar k  \right)^2 \right] \right)    \nonumber\\ 
&\times \left\{  
\frac{P_L \left( k_2 \right)}{k_2^3} \, \frac{P_L \left( k_3 \right)}{k_3^3} \, 
\delta \left( k_2 - \vert f_2 \vert \right) \delta \left( k_3 - \vert f_3 \vert \right)  + {\rm 2 \; permutations} \right\} \;. 
\end{align} 


The curly bracket has three terms. In each term, two $d k_i$ integrals are immediately performed thanks to the $\delta-$functions. 
The third integral is performed in the limit of narrow width of the bump, in which the functions multiplying the exponential term are evaluated at $\bar k$. For instance, in the first term $k_1 \, {\cal R}_{LLL}^{ink} \left(\vec k^*_1 ,\, \vert f_2 \vert ,\, \vert f_3 \vert \right) \, {\rm e}^{- \frac{\left(| \vec k^*_1| - k_* \right)^2}{2 \sigma^2}} \simeq \bar  k \, {\cal R}_{LLL}^{ink} \left( \bar k \hat k_1 ,\, \vert f_2 \vert ,\, \vert f_3 \vert \right) 
 $. 
The third integral is then also immediately done by extending it from $-\infty$ to $\infty$ (which is also appropriate in the narrow width approximation). We obtain 
\begin{align} 
S_s^{ijk} \left( f_1,f_2, f_3 \right) \simeq&  \frac{ L^3 \, f_{\rm NL}^{LLL} }{ 32 \, \pi^2 } \, \sqrt{2 \pi} \, \sigma \, \bar k \nonumber \\ 
& \times \Bigg\{ 
\frac{P_L \left( \vert f_2 \vert \right)}{f_2^2} \, \frac{P_L \left( \vert f_3 \vert \right)}{f_3^2} \, {\rm e}^{-\frac{1}{2 \sigma^2} \left[ \left( \vert f_2 \vert - \bar k \right)^2 +  \left( \vert f_3 \vert - \bar k \right)^2 \right]}  \; {\cal R}_{LLL}^{ijk} \left(\bar k, \vert f_2 \vert , \vert f_3 \vert \right) \nonumber\\ 
& \quad
+ 2 \; {\rm permutations}  \Bigg\} \;. 
\end{align} 


We need to square this quantity, and insert it in (\ref{SNR}). The mixed products between the three different terms in the curly bracket provide a negligible contribution to this result. Let us for instance discuss the product between the first and the second term. In the narrow peak approximation, the first term is mostly supported at $\vert f_2 \vert = \vert f _3 \vert = \bar k$, while the second term at  $\vert f_1 \vert = \vert f _3 \vert = \bar k$. So the mixed product is supported only around the points where the magnitude of all the frequencies is equal to $k_*$. But this is incompatible with the $\delta-$function $\delta \left( f_1 + f_2 + f_3 \right)$ present in  (\ref{SNR}). 

Therefore, for the purpose of computing the signal to noise ratio, only the sum of the square of the three terms in the curly bracket is relevant. In this sum we also approximate all functions multiplying the exponential terms by their value at the center of the bump. This leads to 
\begin{align} 
S_s^{ijk} \left( f_1,f_2, f_3 \right)^2 \cong& \left[ \frac{ L^3 \, f_{\rm NL}^{LLL} }{ 32 \, \pi^2 } \, \sqrt{2 \pi} \, \sigma 
\, \frac{P_L^2 \left( k_* \right)}{k_*^3}  {\cal R}_{LLL}^{ijk} (\bar k\,\hat k^*)\right]^2 
 \Bigg\{ 
 {\rm e}^{-\frac{1}{\sigma^2} \left[ \left( \vert f_2 \vert - \bar k \right)^2 +  \left( \vert f_3 \vert - \bar k \right)^2 \right]} +  \nonumber \\ 
& \quad +  {\rm e}^{-\frac{1}{\sigma^2} \left[ \left( \vert f_1 \vert - \bar k \right)^2 +  \left( \vert f_3 \vert - \bar k \right)^2 \right]}  
+  {\rm e}^{-\frac{1}{\sigma^2} \left[ \left( \vert f_1 \vert - \bar k \right)^2 +  \left( \vert f_2 \vert - \bar k \right)^2 \right]} 
\Bigg\} \;.  
\end{align} 

The three terms in the curly bracket provide the same contribution to the signal to noise ratio 
\begin{align} 
{\rm SNR } &\simeq  \frac{ L^3 \, f_{\rm NL}^{LLL} }{ 32 \, \pi^2 } \, \sqrt{2 \pi} \, \sigma 
\, \frac{P_L^2 \left( k_* \right)}{k_*^3} \Bigg[ 
\frac{T}{6} \, \frac{1}{\left( 4 L^2 \right)^3 }  \,  \sum_{ijk} {\cal R}_{LLL}^{ijk} \left(  \bar k\,\hat k^* \right)^2 \nonumber\\ 
 \times & \,  3 \, \int_{-\infty}^{+\infty} \frac{d f_1}{P_n \left( \vert f_1 \vert \right)} \,  \int_{-\infty}^{+\infty}  \frac{d f_2}{P_n \left( \vert f_2 \vert \right)}  \, {\rm e}^{-\frac{\left( \vert f_2 \vert - \bar k \right)^2}{\sigma^2}} \int_{-\infty}^{+\infty}  \frac{d f_3}{P_n \left( \vert f_3 \vert \right)}  \, {\rm e}^{-\frac{\left( \vert f_3 \vert -\bar k \right)^2}{\sigma^2}} \delta \left( f_1 + f_2 + f_3 \right)  \Bigg]^{1/2} . 
\end{align} 

We perform the $d f_2$ and the $d f_3$ integrals in the narrow width limit. The second line of this expression then becomes 
\begin{equation}
3 \frac{\pi \sigma^2}{P_n^2 \left( \bar k \right)} 
\int_{-\infty}^{+\infty} \frac{d f_1}{P_n \left( \vert f_1 \vert \right)} 
\left[ \delta \left( f_1 - 2\bar k \right) +  \delta \left( f_1 \right) +   \delta \left( f_1 \right)  +   \delta \left( f_1 + 2 \bar k\right) \right] \;. 
\end{equation} 

The contribution proportional to $\delta \left( f_1 \right)$ can be disregarded, since the noise diverges in that limit. This leads to eq.~(\ref{SNR-bump}) written in the main text.

\section{Cosmological scaling of the non-linear parameter} 
\label{app:astar} 
\setcounter{equation}{0}

The parameterization (\ref{f-NL}) of non-Gaussianity in terms of a nonlinear parameter is subject to the cosmological evolution, giving rise to the relation (\ref{f-NL-primordial}). When the momenta in the convolution have equal magnitude, we have a single rescaling, that can be roughly approximated as 
\begin{equation}
f_{\rm NL} \simeq \frac{f_{\rm NL}^{\rm primordial}}{a_{k_*}} \;, 
\end{equation} 
where $k_*$ is the magnitude of the momentum and $a_{k_*}$ denotes the value of the scale factor when the modes re-entered the horizon. For GW generated after inflation inside the horizon, $a_{k_*}$ denotes instead the value of the scale factor at the moment of the GW production. In this relation, the scale factor is normalized to one at the present time. In this appendix we estimate the value of $a_{k_*}$ for both cases. 

We start from GW generated during inflation. In this case, the modes re-enter the horizon during radiation domination. We can write 
\begin{eqnarray} 
\frac{1}{a_{k_*}} &=& \frac{1}{a_{\rm eq}} \, \frac{a_{\rm eq}}{a_{k_*}} = \frac{1}{a_{\rm eq}}  \, \sqrt{\frac{t_{\rm eq}}{t_*}} = 
\frac{1}{a_{\rm eq}} \, \frac{a_* H_*}{a_{\rm eq} \, H_{\rm eq}} \;, 
\end{eqnarray} 
where we have used the fact that $a \propto t^{1/2}$ and $H \propto \frac{1}{t}$ during radiation domination. In this relation, the suffix `eq' refers to the moment of matter-radiation equality, while the star denotes the moment at which the mode $k_*$ re-enters the horizon. 

At horizon re-entry, $a_* H_* = 2 \pi k_*$, where the $2 \pi$ is a consequence of the Fourier transform convention adopted in this work (see for instance eq.~(\ref{GW-classical})). The Hubble rate scales as the square root of the energy density. At equality, the energy density was twice that of radiation, $\rho_{\rm eq} = 2 \times \rho_{\rm rad,eq} = 2 \times \, a_{\rm eq}^{-4} \, \rho_{\rm rad,0} =  2 \times \, a_{\rm eq}^{-4} \, \Omega_{\rm rad,0} \, \rho_0$, where $\rho_0$ is the present value of the energy density. Therefore, 
\begin{eqnarray} 
\frac{1}{a_{k_*}} &=& \frac{2 \pi k_*}{a_{\rm eq}^2} \, \frac{1}{H_0} \, \frac{1}{\sqrt{2 \, a_{\rm eq}^{-4} \, \Omega_{\rm rad,0} }}  \;. 
\end{eqnarray} 

In scaling the energy density of radiation  as $a^{-4}$ we have disregarded the fact that massive neutrinos become non-relativistic, so we consistently evaluate the present fractional density in radiation as if neutrinos were massless, leading to $\Omega_{\rm rad,0} \simeq 4.18 \times 10^{-5} h^{-2}$. This leads to 
\begin{eqnarray} 
\frac{1}{a_{k_*}} &\simeq&   2 \times 10^{17} \,  \frac{k_*}{10^{-3} \, {\rm Hz}}  \;\;,\;\; {\rm production \; from \; inflation} \;. 
\end{eqnarray} 
This corresponds to a temperature  of $T \simeq \frac{T_0}{a_*} \sim 50 \, {\rm TeV}$ (disregarding the variation of relativistic degrees of freedom). 
 
Let us instead assume that the GW are produced inside the horizon, when the temperature of the universe was $T_*$. For convenience, we normalize the temperature at $ 100 \, {\rm GeV}$. We then have (disregarding the variation of relativistic degrees of freedom for the purpose of our estimate) 
\begin{equation}
\frac{1}{a_{k_*}} \sim \frac{T_*}{T_0} \simeq 4 \times 10^{14} \, \frac{T_*}{100 \, {\rm GeV}}  \;\;,\;\; {\rm production \; inside \; horizon \; at \; temperature \; } T_*  \;. 
\end{equation}  
 
\newpage
 
\bibliography{GWbispectrum} 

\begin{thebibliography}{100}

\bibitem{Maggiore:1900zz}
M.~Maggiore,
\newblock {\em {Gravitational Waves. Vol. 1: Theory and Experiments}} (Oxford
  University Press, 2007).

\bibitem{Maggiore:2018sht}
M.~Maggiore,
\newblock {\em {Gravitational Waves. Vol. 2: Astrophysics and Cosmology}}
  (Oxford University Press, 2018).

\bibitem{Caprini:2018mtu}
C.~Caprini and D.~G. Figueroa,
\newblock Class. Quant. Grav. {\bf 35}, 163001 (2018), [1801.04268],
\newblock 10.1088/1361-6382/aac608.

\bibitem{Romano:2016dpx}
J.~D. Romano and N.~J. Cornish,
\newblock Living Rev. Rel. {\bf 20}, 2 (2017), [1608.06889],
\newblock 10.1007/s41114-017-0004-1.

\bibitem{Audley:2017drz}
LISA, P.~Amaro-Seoane {\em et~al.},
\newblock 1702.00786.

\bibitem{Ade:2015ava}
Planck, P.~A.~R. Ade {\em et~al.},
\newblock Astron. Astrophys. {\bf 594}, A17 (2016), [1502.01592],
\newblock 10.1051/0004-6361/201525836.

\bibitem{Smith:2016jqs}
T.~L. Smith and R.~Caldwell,
\newblock Phys. Rev. {\bf D95}, 044036 (2017), [1609.05901],
\newblock 10.1103/PhysRevD.95.044036.

\bibitem{Salopek:1990jq}
D.~S. Salopek and J.~R. Bond,
\newblock Phys. Rev. {\bf D42}, 3936 (1990),
\newblock 10.1103/PhysRevD.42.3936.

\bibitem{Gangui:1993tt}
A.~Gangui, F.~Lucchin, S.~Matarrese and S.~Mollerach,
\newblock Astrophys. J. {\bf 430}, 447 (1994), [astro-ph/9312033],
\newblock 10.1086/174421.

\bibitem{Komatsu:2001rj}
E.~Komatsu and D.~N. Spergel,
\newblock Phys. Rev. {\bf D63}, 063002 (2001), [astro-ph/0005036],
\newblock 10.1103/PhysRevD.63.063002.

\bibitem{Babich:2004gb}
D.~Babich, P.~Creminelli and M.~Zaldarriaga,
\newblock JCAP {\bf 0408}, 009 (2004), [astro-ph/0405356],
\newblock 10.1088/1475-7516/2004/08/009.

\bibitem{Maldacena:2002vr}
J.~M. Maldacena,
\newblock JHEP {\bf 05}, 013 (2003), [astro-ph/0210603],
\newblock 10.1088/1126-6708/2003/05/013.

\bibitem{Acquaviva:2002ud}
V.~Acquaviva, N.~Bartolo, S.~Matarrese and A.~Riotto,
\newblock Nucl. Phys. {\bf B667}, 119 (2003), [astro-ph/0209156],
\newblock 10.1016/S0550-3213(03)00550-9.

\bibitem{Scoccimarro:2011pz}
R.~Scoccimarro, L.~Hui, M.~Manera and K.~C. Chan,
\newblock Phys. Rev. {\bf D85}, 083002 (2012), [1108.5512],
\newblock 10.1103/PhysRevD.85.083002.

\bibitem{Wagner:2010me}
C.~Wagner, L.~Verde and L.~Boubekeur,
\newblock JCAP {\bf 1010}, 022 (2010), [1006.5793],
\newblock 10.1088/1475-7516/2010/10/022.

\bibitem{Liguori:2003mb}
M.~Liguori, S.~Matarrese and L.~Moscardini,
\newblock Astrophys. J. {\bf 597}, 57 (2003), [astro-ph/0306248],
\newblock 10.1086/378394.

\bibitem{Finn:2008np}
L.~S. Finn,
\newblock Phys. Rev. {\bf D79}, 022002 (2009), [0810.4529],
\newblock 10.1103/PhysRevD.79.022002.

\bibitem{Adams:2010vc}
M.~R. Adams and N.~J. Cornish,
\newblock Phys. Rev. {\bf D82}, 022002 (2010), [1002.1291],
\newblock 10.1103/PhysRevD.82.022002.

\bibitem{Cornish:2001bb}
N.~J. Cornish,
\newblock Phys. Rev. {\bf D65}, 022004 (2002), [gr-qc/0106058],
\newblock 10.1103/PhysRevD.65.022004.

\bibitem{Cornish:2018dyw}
N.~Cornish and T.~Robson,
\newblock 1803.01944.

\bibitem{Namba:2015gja}
R.~Namba, M.~Peloso, M.~Shiraishi, L.~Sorbo and C.~Unal,
\newblock JCAP {\bf 1601}, 041 (2016), [1509.07521],
\newblock 10.1088/1475-7516/2016/01/041.

\bibitem{Kuroyanagi:2014nba}
S.~Kuroyanagi, T.~Takahashi and S.~Yokoyama,
\newblock JCAP {\bf 1502}, 003 (2015), [1407.4785],
\newblock 10.1088/1475-7516/2015/02/003.

\bibitem{Array:2015xqh}
BICEP2, Keck Array, P.~A.~R. Ade {\em et~al.},
\newblock Phys. Rev. Lett. {\bf 116}, 031302 (2016), [1510.09217],
\newblock 10.1103/PhysRevLett.116.031302.

\bibitem{Ade:2015lrj}
Planck, P.~A.~R. Ade {\em et~al.},
\newblock Astron. Astrophys. {\bf 594}, A20 (2016), [1502.02114],
\newblock 10.1051/0004-6361/201525898.

\bibitem{Hazumi:2012gjy}
M.~Hazumi {\em et~al.},
\newblock Proc. SPIE Int. Soc. Opt. Eng. {\bf 8442}, 844219 (2012),
\newblock 10.1117/12.926743.

\bibitem{Finelli:2016cyd}
CORE, F.~Finelli {\em et~al.},
\newblock JCAP {\bf 1804}, 016 (2018), [1612.08270],
\newblock 10.1088/1475-7516/2018/04/016.

\bibitem{Abazajian:2016yjj}
CMB-S4, K.~N. Abazajian {\em et~al.},
\newblock 1610.02743.

\bibitem{Maldacena:2011nz}
J.~M. Maldacena and G.~L. Pimentel,
\newblock JHEP {\bf 09}, 045 (2011), [1104.2846],
\newblock 10.1007/JHEP09(2011)045.

\bibitem{Agrawal:2018gzp}
A.~Agrawal,
\newblock 1804.01481.

\bibitem{Gao:2011vs}
X.~Gao, T.~Kobayashi, M.~Yamaguchi and J.~Yokoyama,
\newblock Phys. Rev. Lett. {\bf 107}, 211301 (2011), [1108.3513],
\newblock 10.1103/PhysRevLett.107.211301.

\bibitem{Soda:2011am}
J.~Soda, H.~Kodama and M.~Nozawa,
\newblock JHEP {\bf 08}, 067 (2011), [1106.3228],
\newblock 10.1007/JHEP08(2011)067.

\bibitem{Shiraishi:2011st}
M.~Shiraishi, D.~Nitta and S.~Yokoyama,
\newblock Prog. Theor. Phys. {\bf 126}, 937 (2011), [1108.0175],
\newblock 10.1143/PTP.126.937.

\bibitem{Guzzetti:2016mkm}
M.~C. Guzzetti, N.~Bartolo, M.~Liguori and S.~Matarrese,
\newblock Riv. Nuovo Cim. {\bf 39}, 399 (2016), [1605.01615],
\newblock 10.1393/ncr/i2016-10127-1.

\bibitem{Kobayashi:2011nu}
T.~Kobayashi, M.~Yamaguchi and J.~Yokoyama,
\newblock Prog. Theor. Phys. {\bf 126}, 511 (2011), [1105.5723],
\newblock 10.1143/PTP.126.511.

\bibitem{Zumalacarregui:2013pma}
M.~Zumalacarregui and J.~Garcia-Bellido,
\newblock Phys. Rev. {\bf D89}, 064046 (2014), [1308.4685],
\newblock 10.1103/PhysRevD.89.064046.

\bibitem{Gleyzes:2014dya}
J.~Gleyzes, D.~Langlois, F.~Piazza and F.~Vernizzi,
\newblock Phys. Rev. Lett. {\bf 114}, 211101 (2015), [1404.6495],
\newblock 10.1103/PhysRevLett.114.211101.

\bibitem{Langlois:2015cwa}
D.~Langlois and K.~Noui,
\newblock JCAP {\bf 1602}, 034 (2016), [1510.06930],
\newblock 10.1088/1475-7516/2016/02/034.

\bibitem{Langlois:2015skt}
D.~Langlois and K.~Noui,
\newblock JCAP {\bf 1607}, 016 (2016), [1512.06820],
\newblock 10.1088/1475-7516/2016/07/016.

\bibitem{Crisostomi:2016czh}
M.~Crisostomi, K.~Koyama and G.~Tasinato,
\newblock JCAP {\bf 1604}, 044 (2016), [1602.03119],
\newblock 10.1088/1475-7516/2016/04/044.

\bibitem{BenAchour:2016fzp}
J.~Ben~Achour {\em et~al.},
\newblock JHEP {\bf 12}, 100 (2016), [1608.08135],
\newblock 10.1007/JHEP12(2016)100.

\bibitem{Akita:2015mho}
Y.~Akita and T.~Kobayashi,
\newblock Phys. Rev. {\bf D93}, 043519 (2016), [1512.01380],
\newblock 10.1103/PhysRevD.93.043519.

\bibitem{Creminelli:2014wna}
P.~Creminelli, J.~Gleyzes, J.~Nore{\~n}a and F.~Vernizzi,
\newblock Phys. Rev. Lett. {\bf 113}, 231301 (2014), [1407.8439],
\newblock 10.1103/PhysRevLett.113.231301.

\bibitem{Bordin:2017hal}
L.~Bordin, G.~Cabass, P.~Creminelli and F.~Vernizzi,
\newblock JCAP {\bf 1709}, 043 (2017), [1706.03758],
\newblock 10.1088/1475-7516/2017/09/043.

\bibitem{Lue:1998mq}
A.~Lue, L.-M. Wang and M.~Kamionkowski,
\newblock Phys. Rev. Lett. {\bf 83}, 1506 (1999), [astro-ph/9812088],
\newblock 10.1103/PhysRevLett.83.1506.

\bibitem{Jackiw:2003pm}
R.~Jackiw and S.~Y. Pi,
\newblock Phys. Rev. {\bf D68}, 104012 (2003), [gr-qc/0308071],
\newblock 10.1103/PhysRevD.68.104012.

\bibitem{Crisostomi:2017ugk}
M.~Crisostomi, K.~Noui, C.~Charmousis and D.~Langlois,
\newblock Phys. Rev. {\bf D97}, 044034 (2018), [1710.04531],
\newblock 10.1103/PhysRevD.97.044034.

\bibitem{Bartolo:2017szm}
N.~Bartolo and G.~Orlando,
\newblock JCAP {\bf 1707}, 034 (2017), [1706.04627],
\newblock 10.1088/1475-7516/2017/07/034.

\bibitem{Mollerach:2003nq}
S.~Mollerach, D.~Harari and S.~Matarrese,
\newblock Phys. Rev. {\bf D69}, 063002 (2004), [astro-ph/0310711],
\newblock 10.1103/PhysRevD.69.063002.

\bibitem{Saito:2008jc}
R.~Saito and J.~Yokoyama,
\newblock Phys. Rev. Lett. {\bf 102}, 161101 (2009), [0812.4339],
\newblock 10.1103/PhysRevLett.102.161101, 10.1103/PhysRevLett.107.069901.

\bibitem{Baumann:2007zm}
D.~Baumann, P.~J. Steinhardt, K.~Takahashi and K.~Ichiki,
\newblock Phys. Rev. {\bf D76}, 084019 (2007), [hep-th/0703290],
\newblock 10.1103/PhysRevD.76.084019.

\bibitem{Ananda:2006af}
K.~N. Ananda, C.~Clarkson and D.~Wands,
\newblock Phys. Rev. {\bf D75}, 123518 (2007), [gr-qc/0612013],
\newblock 10.1103/PhysRevD.75.123518.

\bibitem{Bartolo:2016ami}
N.~Bartolo {\em et~al.},
\newblock JCAP {\bf 1612}, 026 (2016), [1610.06481],
\newblock 10.1088/1475-7516/2016/12/026.

\bibitem{Cook:2011hg}
J.~L. Cook and L.~Sorbo,
\newblock Phys. Rev. {\bf D85}, 023534 (2012), [1109.0022],
\newblock 10.1103/PhysRevD.86.069901, 10.1103/PhysRevD.85.023534.

\bibitem{Senatore:2011sp}
L.~Senatore, E.~Silverstein and M.~Zaldarriaga,
\newblock JCAP {\bf 1408}, 016 (2014), [1109.0542],
\newblock 10.1088/1475-7516/2014/08/016.

\bibitem{Carney:2012pk}
D.~Carney, W.~Fischler, E.~D. Kovetz, D.~Lorshbough and S.~Paban,
\newblock JHEP {\bf 11}, 042 (2012), [1209.3848],
\newblock 10.1007/JHEP11(2012)042.

\bibitem{Biagetti:2013kwa}
M.~Biagetti, M.~Fasiello and A.~Riotto,
\newblock Phys. Rev. {\bf D88}, 103518 (2013), [1305.7241],
\newblock 10.1103/PhysRevD.88.103518.

\bibitem{Biagetti:2014asa}
M.~Biagetti, E.~Dimastrogiovanni, M.~Fasiello and M.~Peloso,
\newblock JCAP {\bf 1504}, 011 (2015), [1411.3029],
\newblock 10.1088/1475-7516/2015/04/011.

\bibitem{Goolsby-Cole:2017hod}
C.~Goolsby-Cole and L.~Sorbo,
\newblock JCAP {\bf 1708}, 005 (2017), [1705.03755],
\newblock 10.1088/1475-7516/2017/08/005.

\bibitem{Sorbo:2011rz}
L.~Sorbo,
\newblock JCAP {\bf 1106}, 003 (2011), [1101.1525],
\newblock 10.1088/1475-7516/2011/06/003.

\bibitem{Anber:2012du}
M.~M. Anber and L.~Sorbo,
\newblock Phys. Rev. {\bf D85}, 123537 (2012), [1203.5849],
\newblock 10.1103/PhysRevD.85.123537.

\bibitem{Barnaby:2010vf}
N.~Barnaby and M.~Peloso,
\newblock Phys. Rev. Lett. {\bf 106}, 181301 (2011), [1011.1500],
\newblock 10.1103/PhysRevLett.106.181301.

\bibitem{Barnaby:2012xt}
N.~Barnaby {\em et~al.},
\newblock Phys. Rev. {\bf D86}, 103508 (2012), [1206.6117],
\newblock 10.1103/PhysRevD.86.103508.

\bibitem{Maleknejad:2011jw}
A.~Maleknejad and M.~M. Sheikh-Jabbari,
\newblock Phys. Lett. {\bf B723}, 224 (2013), [1102.1513],
\newblock 10.1016/j.physletb.2013.05.001.

\bibitem{Dimastrogiovanni:2012ew}
E.~Dimastrogiovanni and M.~Peloso,
\newblock Phys. Rev. {\bf D87}, 103501 (2013), [1212.5184],
\newblock 10.1103/PhysRevD.87.103501.

\bibitem{Adshead:2013qp}
P.~Adshead, E.~Martinec and M.~Wyman,
\newblock Phys. Rev. {\bf D88}, 021302 (2013), [1301.2598],
\newblock 10.1103/PhysRevD.88.021302.

\bibitem{Adshead:2013nka}
P.~Adshead, E.~Martinec and M.~Wyman,
\newblock JHEP {\bf 09}, 087 (2013), [1305.2930],
\newblock 10.1007/JHEP09(2013)087.

\bibitem{Obata:2014loa}
I.~Obata, T.~Miura and J.~Soda,
\newblock Phys. Rev. {\bf D92}, 063516 (2015), [1412.7620],
\newblock 10.1103/PhysRevD.95.109902, 10.1103/PhysRevD.92.063516.

\bibitem{Maleknejad:2016qjz}
A.~Maleknejad,
\newblock JHEP {\bf 07}, 104 (2016), [1604.03327],
\newblock 10.1007/JHEP07(2016)104.

\bibitem{Dimastrogiovanni:2016fuu}
E.~Dimastrogiovanni, M.~Fasiello and T.~Fujita,
\newblock JCAP {\bf 1701}, 019 (2017), [1608.04216],
\newblock 10.1088/1475-7516/2017/01/019.

\bibitem{Agrawal:2017awz}
A.~Agrawal, T.~Fujita and E.~Komatsu,
\newblock 1707.03023.

\bibitem{Adshead:2017hnc}
P.~Adshead and E.~I. Sfakianakis,
\newblock JHEP {\bf 08}, 130 (2017), [1705.03024],
\newblock 10.1007/JHEP08(2017)130.

\bibitem{Caldwell:2017chz}
R.~R. Caldwell and C.~Devulder,
\newblock Phys. Rev. {\bf D97}, 023532 (2018), [1706.03765],
\newblock 10.1103/PhysRevD.97.023532.

\bibitem{Agrawal:2018mrg}
A.~Agrawal, T.~Fujita and E.~Komatsu,
\newblock 1802.09284.

\bibitem{Espinosa:2018eve}
J.~R. Espinosa, D.~Racco and A.~Riotto,
\newblock 1804.07732.

\bibitem{Peloso:2016gqs}
M.~Peloso, L.~Sorbo and C.~Unal,
\newblock JCAP {\bf 1609}, 001 (2016), [1606.00459],
\newblock 10.1088/1475-7516/2016/09/001.

\bibitem{Garcia-Bellido:2016dkw}
J.~Garcia-Bellido, M.~Peloso and C.~Unal,
\newblock JCAP {\bf 1612}, 031 (2016), [1610.03763],
\newblock 10.1088/1475-7516/2016/12/031.

\bibitem{Shiraishi:2016yun}
M.~Shiraishi, C.~Hikage, R.~Namba, T.~Namikawa and M.~Hazumi,
\newblock Phys. Rev. {\bf D94}, 043506 (2016), [1606.06082],
\newblock 10.1103/PhysRevD.94.043506.

\bibitem{Thorne:2017jft}
B.~Thorne {\em et~al.},
\newblock Phys. Rev. {\bf D97}, 043506 (2018), [1707.03240],
\newblock 10.1103/PhysRevD.97.043506.

\bibitem{Endlich:2012pz}
S.~Endlich, A.~Nicolis and J.~Wang,
\newblock JCAP {\bf 1310}, 011 (2013), [1210.0569],
\newblock 10.1088/1475-7516/2013/10/011.

\bibitem{Bartolo:2015qvr}
N.~Bartolo, D.~Cannone, A.~Ricciardone and G.~Tasinato,
\newblock JCAP {\bf 1603}, 044 (2016), [1511.07414],
\newblock 10.1088/1475-7516/2016/03/044.

\bibitem{Ricciardone:2016lym}
A.~Ricciardone and G.~Tasinato,
\newblock Phys. Rev. {\bf D96}, 023508 (2017), [1611.04516],
\newblock 10.1103/PhysRevD.96.023508.

\bibitem{Ricciardone:2017kre}
A.~Ricciardone and G.~Tasinato,
\newblock JCAP {\bf 1802}, 011 (2018), [1711.02635],
\newblock 10.1088/1475-7516/2018/02/011.

\bibitem{Domenech:2017kno}
G.~Dom{\`e}nech {\em et~al.},
\newblock JCAP {\bf 1705}, 034 (2017), [1701.05554],
\newblock 10.1088/1475-7516/2017/05/034.

\bibitem{Ballesteros:2016gwc}
G.~Ballesteros, D.~Comelli and L.~Pilo,
\newblock Phys. Rev. {\bf D94}, 124023 (2016), [1603.02956],
\newblock 10.1103/PhysRevD.94.124023.

\bibitem{Cannone:2015rra}
D.~Cannone, J.-O. Gong and G.~Tasinato,
\newblock JCAP {\bf 1508}, 003 (2015), [1505.05773],
\newblock 10.1088/1475-7516/2015/08/003.

\bibitem{Lin:2015cqa}
C.~Lin and L.~Z. Labun,
\newblock JHEP {\bf 03}, 128 (2016), [1501.07160],
\newblock 10.1007/JHEP03(2016)128.

\bibitem{Cannone:2014uqa}
D.~Cannone, G.~Tasinato and D.~Wands,
\newblock JCAP {\bf 1501}, 029 (2015), [1409.6568],
\newblock 10.1088/1475-7516/2015/01/029.

\bibitem{Akhshik:2014bla}
M.~Akhshik,
\newblock JCAP {\bf 1505}, 043 (2015), [1409.3004],
\newblock 10.1088/1475-7516/2015/05/043.

\bibitem{Takahashi:2009wc}
T.~Takahashi and J.~Soda,
\newblock Phys. Rev. Lett. {\bf 102}, 231301 (2009), [0904.0554],
\newblock 10.1103/PhysRevLett.102.231301.

\bibitem{Zhu:2013fja}
T.~Zhu, W.~Zhao, Y.~Huang, A.~Wang and Q.~Wu,
\newblock Phys. Rev. {\bf D88}, 063508 (2013), [1305.0600],
\newblock 10.1103/PhysRevD.88.063508.

\bibitem{Huang:2013epa}
Y.~Huang, A.~Wang, R.~Yousefi and T.~Zhu,
\newblock Phys. Rev. {\bf D88}, 023523 (2013), [1304.1556],
\newblock 10.1103/PhysRevD.88.023523.

\bibitem{Cook:2013xea}
J.~L. Cook and L.~Sorbo,
\newblock JCAP {\bf 1311}, 047 (2013), [1307.7077],
\newblock 10.1088/1475-7516/2013/11/047.

\bibitem{Garcia-Bellido:2017aan}
J.~Garcia-Bellido, M.~Peloso and C.~Unal,
\newblock JCAP {\bf 1709}, 013 (2017), [1707.02441],
\newblock 10.1088/1475-7516/2017/09/013.

\bibitem{Shiraishi:2013kxa}
M.~Shiraishi, A.~Ricciardone and S.~Saga,
\newblock JCAP {\bf 1311}, 051 (2013), [1308.6769],
\newblock 10.1088/1475-7516/2013/11/051.

\bibitem{Shiraishi:2014ila}
M.~Shiraishi, M.~Liguori and J.~R. Fergusson,
\newblock JCAP {\bf 1501}, 007 (2015), [1409.0265],
\newblock 10.1088/1475-7516/2015/01/007.

\bibitem{Flauger:2014qra}
R.~Flauger, J.~C. Hill and D.~N. Spergel,
\newblock JCAP {\bf 1408}, 039 (2014), [1405.7351],
\newblock 10.1088/1475-7516/2014/08/039.

\bibitem{Seljak:2003pn}
U.~Seljak and C.~M. Hirata,
\newblock Phys. Rev. {\bf D69}, 043005 (2004), [astro-ph/0310163],
\newblock 10.1103/PhysRevD.69.043005.

\bibitem{Meerburg:2016ecv}
P.~D. Meerburg, J.~Meyers, A.~van Engelen and Y.~Ali-Ha{\"\i}moud,
\newblock Phys. Rev. {\bf D93}, 123511 (2016), [1603.02243],
\newblock 10.1103/PhysRevD.93.123511.

\bibitem{Shiraishi:2017yrq}
M.~Shiraishi, M.~Liguori and J.~R. Fergusson,
\newblock JCAP {\bf 1801}, 016 (2018), [1710.06778],
\newblock 10.1088/1475-7516/2018/01/016.

\bibitem{Easther:2006gt}
R.~Easther and E.~A. Lim,
\newblock JCAP {\bf 0604}, 010 (2006), [astro-ph/0601617],
\newblock 10.1088/1475-7516/2006/04/010.

\bibitem{GarciaBellido:2007dg}
J.~Garcia-Bellido and D.~G. Figueroa,
\newblock Phys. Rev. Lett. {\bf 98}, 061302 (2007), [astro-ph/0701014],
\newblock 10.1103/PhysRevLett.98.061302.

\bibitem{GarciaBellido:2007af}
J.~Garcia-Bellido, D.~G. Figueroa and A.~Sastre,
\newblock Phys. Rev. {\bf D77}, 043517 (2008), [0707.0839],
\newblock 10.1103/PhysRevD.77.043517.

\bibitem{Dufaux:2007pt}
J.~F. Dufaux, A.~Bergman, G.~N. Felder, L.~Kofman and J.-P. Uzan,
\newblock Phys. Rev. {\bf D76}, 123517 (2007), [0707.0875],
\newblock 10.1103/PhysRevD.76.123517.

\bibitem{Dufaux:2008dn}
J.-F. Dufaux, G.~Felder, L.~Kofman and O.~Navros,
\newblock JCAP {\bf 0903}, 001 (2009), [0812.2917],
\newblock 10.1088/1475-7516/2009/03/001.

\bibitem{Dufaux:2010cf}
J.-F. Dufaux, D.~G. Figueroa and J.~Garcia-Bellido,
\newblock Phys. Rev. {\bf D82}, 083518 (2010), [1006.0217],
\newblock 10.1103/PhysRevD.82.083518.

\bibitem{Figueroa:2017vfa}
D.~G. Figueroa and F.~Torrenti,
\newblock JCAP {\bf 1710}, 057 (2017), [1707.04533],
\newblock 10.1088/1475-7516/2017/10/057.

\bibitem{Adshead:2018doq}
P.~Adshead, J.~T. Giblin and Z.~J. Weiner,
\newblock 1805.04550.

\bibitem{Kamionkowski:1993fg}
M.~Kamionkowski, A.~Kosowsky and M.~S. Turner,
\newblock Phys. Rev. {\bf D49}, 2837 (1994), [astro-ph/9310044],
\newblock 10.1103/PhysRevD.49.2837.

\bibitem{Caprini:2007xq}
C.~Caprini, R.~Durrer and G.~Servant,
\newblock Phys. Rev. {\bf D77}, 124015 (2008), [0711.2593],
\newblock 10.1103/PhysRevD.77.124015.

\bibitem{Huber:2008hg}
S.~J. Huber and T.~Konstandin,
\newblock JCAP {\bf 0809}, 022 (2008), [0806.1828],
\newblock 10.1088/1475-7516/2008/09/022.

\bibitem{Hindmarsh:2013xza}
M.~Hindmarsh, S.~J. Huber, K.~Rummukainen and D.~J. Weir,
\newblock Phys. Rev. Lett. {\bf 112}, 041301 (2014), [1304.2433],
\newblock 10.1103/PhysRevLett.112.041301.

\bibitem{Hindmarsh:2015qta}
M.~Hindmarsh, S.~J. Huber, K.~Rummukainen and D.~J. Weir,
\newblock Phys. Rev. {\bf D92}, 123009 (2015), [1504.03291],
\newblock 10.1103/PhysRevD.92.123009.

\bibitem{Caprini:2015zlo}
C.~Caprini {\em et~al.},
\newblock JCAP {\bf 1604}, 001 (2016), [1512.06239],
\newblock 10.1088/1475-7516/2016/04/001.

\bibitem{Cutting:2018tjt}
D.~Cutting, M.~Hindmarsh and D.~J. Weir,
\newblock 1802.05712.

\bibitem{Vachaspati:1984gt}
T.~Vachaspati and A.~Vilenkin,
\newblock Phys. Rev. {\bf D31}, 3052 (1985),
\newblock 10.1103/PhysRevD.31.3052.

\bibitem{Sakellariadou:1990ne}
M.~Sakellariadou,
\newblock Phys. Rev. {\bf D42}, 354 (1990),
\newblock 10.1103/PhysRevD.42.354, 10.1103/PhysRevD.43.4150.2.

\bibitem{Damour:2000wa}
T.~Damour and A.~Vilenkin,
\newblock Phys. Rev. Lett. {\bf 85}, 3761 (2000), [gr-qc/0004075],
\newblock 10.1103/PhysRevLett.85.3761.

\bibitem{Damour:2001bk}
T.~Damour and A.~Vilenkin,
\newblock Phys. Rev. {\bf D64}, 064008 (2001), [gr-qc/0104026],
\newblock 10.1103/PhysRevD.64.064008.

\bibitem{Damour:2004kw}
T.~Damour and A.~Vilenkin,
\newblock Phys. Rev. {\bf D71}, 063510 (2005), [hep-th/0410222],
\newblock 10.1103/PhysRevD.71.063510.

\bibitem{Figueroa:2012kw}
D.~G. Figueroa, M.~Hindmarsh and J.~Urrestilla,
\newblock Phys. Rev. Lett. {\bf 110}, 101302 (2013), [1212.5458],
\newblock 10.1103/PhysRevLett.110.101302.

\bibitem{Blanco-Pillado:2017oxo}
J.~J. Blanco-Pillado and K.~D. Olum,
\newblock Phys. Rev. {\bf D96}, 104046 (2017), [1709.02693],
\newblock 10.1103/PhysRevD.96.104046.

\bibitem{Figueroa:2013vif}
D.~G. Figueroa and T.~Meriniemi,
\newblock JHEP {\bf 10}, 101 (2013), [1306.6911],
\newblock 10.1007/JHEP10(2013)101.

\bibitem{Dufaux:2009wn}
J.-F. Dufaux,
\newblock Phys. Rev. Lett. {\bf 103}, 041301 (2009), [0902.2574],
\newblock 10.1103/PhysRevLett.103.041301.

\bibitem{Tranberg:2017lrx}
A.~Tranberg, S.~T{\"a}htinen and D.~J. Weir,
\newblock JCAP {\bf 1804}, 012 (2018), [1706.02365],
\newblock 10.1088/1475-7516/2018/04/012.

\bibitem{Krauss:1991qu}
L.~M. Krauss,
\newblock Phys. Lett. {\bf B284}, 229 (1992),
\newblock 10.1016/0370-2693(92)90425-4.

\bibitem{JonesSmith:2007ne}
K.~Jones-Smith, L.~M. Krauss and H.~Mathur,
\newblock Phys. Rev. Lett. {\bf 100}, 131302 (2008), [0712.0778],
\newblock 10.1103/PhysRevLett.100.131302.

\bibitem{Fenu:2009qf}
E.~Fenu, D.~G. Figueroa, R.~Durrer and J.~Garcia-Bellido,
\newblock JCAP {\bf 0910}, 005 (2009), [0908.0425],
\newblock 10.1088/1475-7516/2009/10/005.

\bibitem{Adshead:2009bz}
P.~Adshead and E.~A. Lim,
\newblock Phys. Rev. {\bf D82}, 024023 (2010), [0912.1615],
\newblock 10.1103/PhysRevD.82.024023.

\bibitem{Regimbau:2011bm}
T.~Regimbau, S.~Giampanis, X.~Siemens and V.~Mandic,
\newblock Phys. Rev. {\bf D85}, 066001 (2012), [1111.6638],
\newblock 10.1103/PhysRevD.85.066001, 10.1103/PhysRevD.85.069902.

\bibitem{Vilenkin:2000jqa}
A.~Vilenkin and E.~P.~S. Shellard,
\newblock {\em {Cosmic Strings and Other Topological Defects}} (Cambridge
  University Press, 2000).

\bibitem{Brandenberger:1988aj}
R.~H. Brandenberger and C.~Vafa,
\newblock Nucl. Phys. {\bf B316}, 391 (1989),
\newblock 10.1016/0550-3213(89)90037-0.

\bibitem{Sakellariadou:1995vk}
M.~Sakellariadou,
\newblock Nucl. Phys. {\bf B468}, 319 (1996), [hep-th/9511075],
\newblock 10.1016/0550-3213(96)00123-X.

\bibitem{Nayeri:2005ck}
A.~Nayeri, R.~H. Brandenberger and C.~Vafa,
\newblock Phys. Rev. Lett. {\bf 97}, 021302 (2006), [hep-th/0511140],
\newblock 10.1103/PhysRevLett.97.021302.

\bibitem{Brandenberger:2006xi}
R.~H. Brandenberger, A.~Nayeri, S.~P. Patil and C.~Vafa,
\newblock Phys. Rev. Lett. {\bf 98}, 231302 (2007), [hep-th/0604126],
\newblock 10.1103/PhysRevLett.98.231302.

\bibitem{Chen:2007js}
B.~Chen, Y.~Wang, W.~Xue and R.~Brandenberger,
\newblock The Universe {\bf 3}, 2 (2015), [0712.2477].

\bibitem{Feldstein:2006hm}
B.~Feldstein and B.~Tweedie,
\newblock JCAP {\bf 0704}, 020 (2007), [hep-ph/0611286],
\newblock 10.1088/1475-7516/2007/04/020.

\bibitem{Berera:1995wh}
A.~Berera and L.-Z. Fang,
\newblock Phys. Rev. Lett. {\bf 74}, 1912 (1995), [astro-ph/9501024],
\newblock 10.1103/PhysRevLett.74.1912.

\bibitem{Bartolo:2018evs}
N.~Bartolo {\em et~al.},
\newblock Phys. Rev. Lett. {\bf 122}, 211301 (2019), [1810.12218],
\newblock 10.1103/PhysRevLett.122.211301.

\bibitem{Bartolo:2018rku}
N.~Bartolo {\em et~al.},
\newblock Phys. Rev. {\bf D99}, 103521 (2019), [1810.12224],
\newblock 10.1103/PhysRevD.99.103521.

\bibitem{Bartolo:2019oiq}
N.~Bartolo {\em et~al.},
\newblock 1908.00527.

\bibitem{Dimastrogiovanni:2019bfl}
E.~Dimastrogiovanni, M.~Fasiello and G.~Tasinato,
\newblock 1906.07204.

\bibitem{Powell:2019kid}
C.~Powell and G.~Tasinato,
\newblock 1910.04758.

\bibitem{Allen:1997ad}
B.~Allen and J.~D. Romano,
\newblock Phys. Rev. {\bf D59}, 102001 (1999), [gr-qc/9710117],
\newblock 10.1103/PhysRevD.59.102001.

\end{thebibliography}
\bibliographystyle{h-physrev4}
\end{document}